\documentclass[pdflatex,sn-basic]{sn-jnl}
\setcitestyle{aysep={}} 

\usepackage{graphicx}%
\usepackage{multirow}%
\usepackage{amsmath,amssymb,amsfonts}%
\usepackage{amsthm}%
\usepackage{mathrsfs}%
\usepackage[title]{appendix}%
\usepackage{xcolor}%
\usepackage{textcomp}%
\usepackage{manyfoot}%
\usepackage{booktabs}%
\usepackage{algorithm}%
\usepackage{algorithmicx}%
\usepackage{algpseudocode}%
\usepackage{listings}%

\makeatletter
\input{aas_macros.sty}
\def\ref@jnl#1{{\jnl@style#1\ }}
\makeatother

\newcommand{\orcit}[1]{\protect\href{https://orcid.org/#1}{\protect\includegraphics[width=8pt]{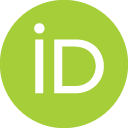}}}

\begin{document}

\title[The variability of blazars throughout the electromagnetic spectrum]{The variability of blazars throughout the electromagnetic spectrum}


\author[]{\fnm{Claudia M.} \sur{Raiteri\orcit{0000-0003-1784-2784}}}\email{claudia.raiteri@inaf.it}

\affil[]{\orgname{}{INAF-Osservatorio Astrofisico di Torino}, \orgaddress{\street{Via Osservatorio 20}, \city{Pino Torinese (TO)}, \postcode{I-10025}, \country{Italy}}}

\abstract{With their jet pointing towards us, blazars are ideal tools to study the physics and structure of extragalactic jets. Their powerful jets are cosmic particle accelerators and are alleged to be one of the production sites of the high-energy neutrinos detected by the IceCube Observatory. Doppler beaming of the jet nonthermal radiation increases blazar brightness, blue-shifts their emission, and shortens their variability time scales, which are observed to range from years down to minutes. This review will focus on blazar flux, spectral, and polarization variability across the electromagnetic spectrum. After introducing blazars and their peculiarities, we will consider 
the statistical tools that are used to characterize the variability and to reveal correlations and time delays between flux variations at different frequencies. Then we will outline the main observed properties of the blazar multiwavelength behaviour. Interpretation of blazar variability calls into question both intrinsic and extrinsic mechanisms. Shock waves, magnetic reconnection, and turbulence can accelerate particles inside the jet, while jet precession, rotation, and twisting can produce variations in Doppler beaming. Changes in the broad-band spectral energy distribution have commonly been explained by variations in the jet physical parameters in one- or two-zone models. However, microvariability observed at all wavelengths puts strong constraints on the size of the emitting regions, suggesting a multizone emitting jet. Twisting jets have been proposed to explain the long-term multiwavelength variability. They are supported by radio observations of bent or helical jets, and by results of relativistic magnetohydrodynamics simulations of plasma jets. Detection of (quasi)periodic behaviour at all frequencies and on all time scales has been ascribed to orbital motion in black hole binary systems, jet precession, kink instabilities developing inside the jet, or perturbations in the accretion disc. Gravitational microlensing has been suggested to explain blazar behaviour in some cases.
Polarization is another important ingredient in blazar variability studies, providing information on the structure and behaviour of the magnetic field in the emission zones. Both the degree and angle of polarization can show strong and fast variability, which is sometimes correlated with flux. Overall, polarimetric observations indicate that turbulence must play an important role in the emitting regions. Recent results obtained by the Imaging X-ray Polarimetric Explorer (IXPE) satellite have revealed some unexpected behaviour favouring a multizone emitting jet model. The interpretation of flux, spectral, and polarization variability within a consistent picture challenges current models of blazar variability and tells us that we may still miss some tiles of the puzzle.}


\maketitle

\setcounter{tocdepth}{3} 
\tableofcontents

\section{Introduction}\label{intro}

According to the commonly accepted paradigm for the active galactic nuclei (AGNs), these are powerful and long-standing sources of energy at the centre of some galaxies. The engine is a supermassive black hole (SMBH), with millions to billions of solar masses, fed by matter falling from a surrounding accretion disc. 
Rapidly moving gas clouds in the disc neighbourhood form the broad-line region (BLR), while more externally slowly moving gas clouds constitute the narrow-line region (NLR). BLR and NLR are responsible respectively for the broad and narrow emission lines that are visible in the AGN spectra. Depending on the AGN inclination to the line of sight, the BLR can be hidden by a dusty torus.
The typical length scale of the accretion disc is of the order of light days, while that of the BLR is light months; the torus begins approximately at this distance, while the length scale of the NLR ranges from a few to several light years.

A fraction of AGNs, which depends on redshift and luminosity, \citep[][and references therein]{kratzer2015} 
is made up of radio-loud objects. They possess two plasma jets that are launched roughly perpendicularly to the accretion disc in both directions.
These jets are likely formed through the extraction of energy and angular momentum from either the rotating SMBH \citep{blandford1977} or the accretion disc \citep{blandford1982}.
Since blazars are hosted by massive early-type galaxies that were likely formed through mergers, the blazar core may actually contain a binary black hole system \citep[BBHS,][]{begelman1980}, where one or both SMBHs carry a jet. 
The possible presence of a BBHS must be inferred indirectly, looking for (quasi)periodic behaviour in the light curves.

\subsection{Blazars as beamed sources}
Blazars form a heterogeneous class of jetted active galactic nuclei, which share the common geometric property of having one of their plasma jets closely aligned with the line of sight \citep{blandford1979}. 
Inside the jet, particles are flowing at relativistic speeds along the magnetic field lines, emitting nonthermal radiation.
Because of the jet orientation, this radiation undergoes Doppler beaming, which produces a series of effects that can be expressed in terms of the Doppler factor 
\begin{equation}
    \delta=[\Gamma (1- \beta \cos\theta)]^{-1},
    \label{df}
    \end{equation}
where $\Gamma={(1-\beta^2)}^{-1/2}$ is the Lorentz factor, $\beta$ is the bulk velocity of the plasma in units of the speed of light, and $\theta$ is the viewing angle, i.e.\ the angle between the velocity vector and the line of sight.
The parameters $\delta$ and $\Gamma$ in blazar jets are often assumed to have typical values of 10, but actually their indirect estimates through theory \citep[e.g.][]{ghisellini2011} and especially from observations \citep[e.g.][]{kellermann2007,hovatta2009,liodakis2018,homan2021,weaver2022}
show a broad distribution, reaching much higher values.

The Doppler factor increases when $\theta$ decreases. In the extreme case where $\theta=0^{\circ}$, it is $\delta \approx 2 \, \Gamma$, while when $\theta=90^{\circ}$, then $\delta=\Gamma^{-1}$. The value of the critical viewing angle below which Doppler beaming is effective ($\delta > 1$) depends on the bulk velocity of the plasma: for $\Gamma=5$, 10, 15, and 20 this occurs for $\theta \approx 35^{\circ}, \, 25^{\circ}, \, 21^{\circ}$, and $18^{\circ}$, respectively.

Aberration causes the viewing angle $\theta$ in the observer's frame usually to be much smaller than that in the source rest frame $\theta'$
\begin{equation}
    \theta'=\rm arccos \left( {\frac{\cos\theta - \beta}{1-\beta \,\cos\theta}} \right) .
\end{equation}
For example, for $\Gamma=10$ and $\theta=1^{\circ}$ the rest-frame viewing angle is $\theta' \sim 20^{\circ}$.
In the following, primed quantities will refer to the rest frame.

Doppler beaming produces an enhancement of the observed flux.
For an isotropic emitting source with a power-law dependence of the flux density on frequency, $F'_{\nu'} \propto (\nu')^{-\alpha}$, 
it is 
\begin{equation}
F_\nu (\nu) = \delta^{n+\alpha} F'_{\nu'} (\nu),
\end{equation}
where $n=2$ for a continuous jet and $n=3$ for a moving discrete source \citep{ghisellini1993}.
The dependence of $\delta$ on the viewing angle (Eq. \ref{df}) then implies that we can observe variability even when the rest-frame flux does not change, if the jet emitting region changes its orientation with respect to the line of sight. Variability due to changes in the Doppler factor can also be produced by variations in the Lorentz factor, i.e.\ in the plasma velocity.

Doppler beaming modifies time intervals:
\begin{equation}
\Delta t = \Delta t' \frac{(1+z)}{\delta}
\end{equation}
and frequencies 
\begin{equation}
\nu = \nu' \frac{\delta}{(1+z)} \,,
\label{eq:nu}
\end{equation}
where $z$ is the source redshift.
Since beaming is usually stronger than the effect of the cosmological expansion of the Universe, i.e.\ $\delta > (1+z)$, time intervals become shorter and the jet radiation is blue-shifted. The decrease in the variability time scales together with flux boosting make blazars appear as very variable luminous sources.


Relativistic motion towards the observer also explains the apparent superluminal motion \citep{rees1966} estimated by observing moving features in radio images obtained by interferometric techniques \citep[e.g.][]{whitney1971,cohen1976,kellermann2007,jorstad2017}.

\subsection{Spectral energy distribution}
\label{sec:sed}

Blazars are observed at all wavelengths, from the radio band to $\gamma$-rays with very high energies (VHE) of the order of a few TeV. The broad-band spectral energy distribution (SED) is dominated by the beamed jet emission, which is thought to occur at distances of less than 1 pc from the SMBH. In small frequency windows, the jet broad-band spectrum can be approximated by a power law $F_\nu \propto \nu^{-\alpha}$. In the radio and optical bands, values of the spectral index $\alpha \leq 0.5$ define a ``flat" spectrum, while for $\alpha > 0.5$ the spectrum is known as ``steep".
In the X-ray and $\gamma$-ray bands, values $\alpha \leq 1$ describe a ``hard" spectrum, while for $\alpha > 1$ the spectrum is named ``soft".\footnote{At high energies it is common to consider the photonic spectrum, which in the case of a power-law distribution is $dN/dE\propto E^{-\Gamma}$, where $E$ is the energy and $\Gamma=\alpha+1$ is the photon index.}

In the $\log (\nu F_\nu)$ versus $\log (\nu)$ diagram, the broad-band SED shows two bumps. The low-energy bump covers from radio to optical--UV frequencies, but can extend up to X-rays in some sources. It is due to synchrotron radiation emitted by relativistic electrons that spiralise around the lines of the magnetic field in the jet.

The nature of the high-energy bump, ranging from X-rays to $\gamma$-rays, has been actively discussed. According to leptonic models, it comes from inverse-Compton scattering of seed photons off the same relativistic electrons that produce the synchrotron radiation. 
If the energy of the seed photons is similar or greater than the electron rest-mass energy $m_e c^2$, the classical Thomson cross section is replaced by the Klein-Nishina cross section, which decreases with photon energy.
The origin of the seed photons is also debated. A synchrotron-self-Compton (SSC) mechanism \citep{blandford1978,marscher1983} 
implies that they are the same photons produced by the synchrotron process. The difficulty in reproducing the SEDs of the most powerful blazars with the SSC model led to the introduction of an external Compton (EC) mechanism, where seed photons come from an external radiation field, such as disc, BLR or dusty torus emission \citep{dermer1992,sikora1994,blazejowski2000}.

Both SSC and EC models predict that high-energy emission is correlated with low-energy emission, in particular $\gamma$-ray with optical radiation. 
The relationship between the high-energy flux and the low-energy flux depends on the cause of the variability. An increase in the number density of electrons is expected to produce a linear correlation in the EC process and a quadratic relationship in the SSC process \citep{maraschi1994}. 
Changes in the magnetic field strength will affect the synchrotron emission and produce a linear dependence of the $\gamma$-ray flux on the optical flux in the SSC regime, and no changes in the $\gamma$-ray emission in the case of EC, possibly leading to optical flares without $\gamma$-ray counterpart (``sterile flares", see Sect.~\ref{section:mw}).
If instead the variability is due to orientation changes, a linear correlation is predicted \citep{larionov2016a,raiteri2024}.
A ``mirror" mechanism has also been proposed, where photons produced in the jet during a flaring event photoionise the closest BLR gas clouds, which then provide enhanced seed photons for the EC process \citep{ghisellini1996}. In this case, the relationship between the optical and $\gamma$-ray fluxes is the same as in the SSC case, but the brightness changes observed at $\gamma$-rays should lag those in the optical.

Hadronic models have also been developed, where the high-energy emission is produced by synchrotron emission from protons and/or particle cascades \citep{mannheim1992,mucke2003}. These models can in particular account for the emission of very high-energy neutrinos detected by the IceCube Observatory that have been associated with blazars (see Sect.~\ref{sec:neu}), and also the connection with the ultra high-energy cosmic rays detected by the Pierre Auger Observatory and the Telescope Array \citep{resconi2017}.
Moreover, hadronic models may explain the lack of correlation between optical and $\gamma$-ray flux variations that have sometimes been observed (see Sect.~\ref{section:mw}).

The leptonic versus hadronic debate also calls into question the particle content of blazar jets. 
Protons should be present, but they must be much less numerous than leptons \citep[see][and references therein]{madejski2016_rev}.
Moreover, the non detection of circular polarisation was interpreted as an indication of low magnetic field strengths in a jet mostly composed of electron-positron pairs rather than electron-proton pairs \citep{liodakis2023}. 
This would disfavour a hadronic process as the main mechanism for the origin of high-energy photons.

In addition to the jet radiation, other emission contributions can appear in the blazar SED (see also next section). They can come from the accretion disc, BLR, NLR, dusty torus, and host galaxy.
The contribution from less variable emission components, such as the accretion disc and BLR, or from a steady component, like the host galaxy, can sometimes overwhelm the strong variability of the beamed jet emission. This makes it very difficult to identify low-power blazars in the near-infrared-to-UV frequency range, where these unbeamed emission contributions are present.

\subsection{Blazar types}
\label{sec:blatyp}
\begin{figure}[htbp]
\centering
\includegraphics[width=0.66\textwidth]{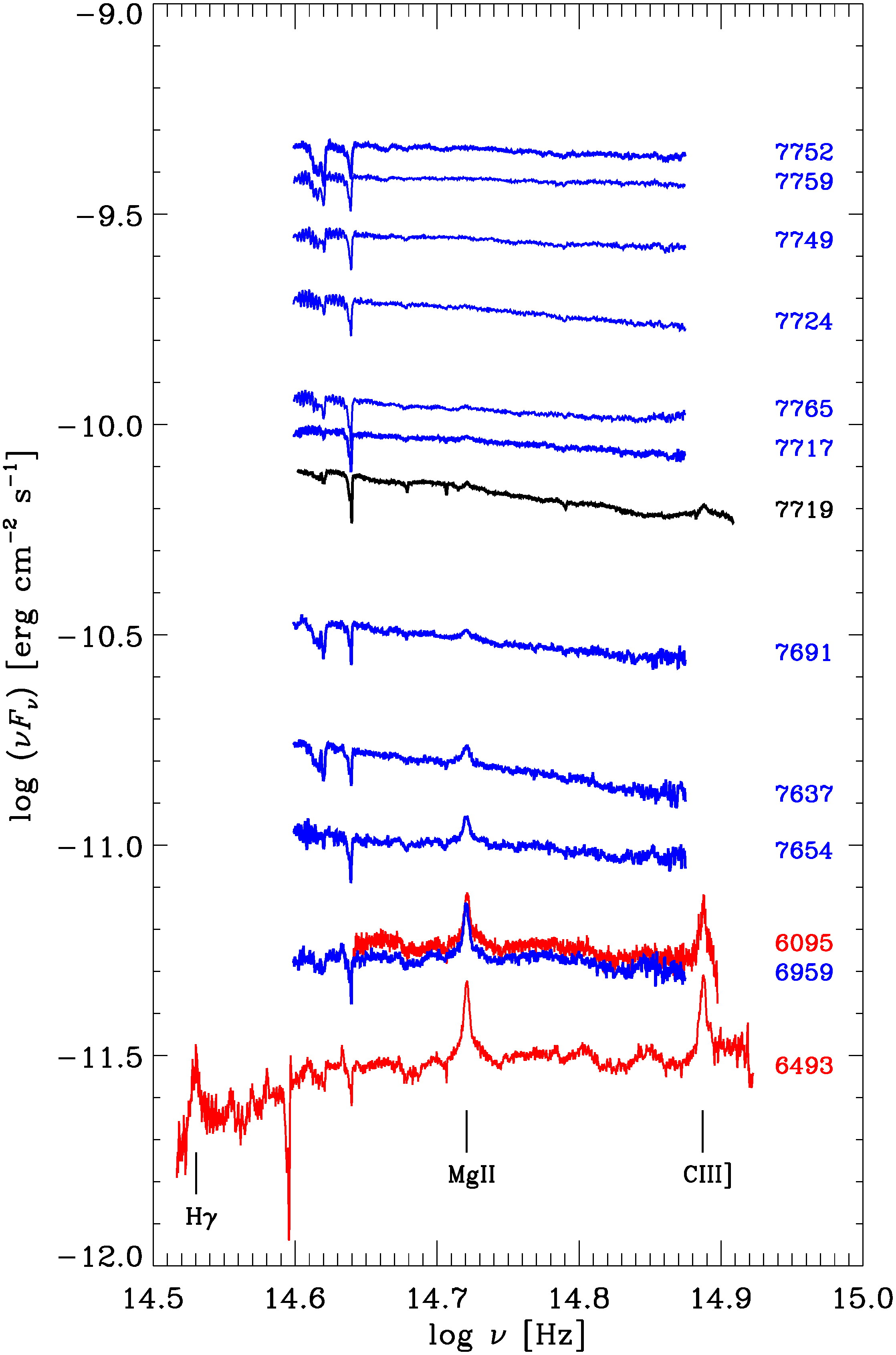}
\caption{Spectra of the FSRQ CTA~102 in different brightness states. When the flux is low the spectra show strong emission lines from the BLR and the ``big blue bump" due to the contribution of thermal radiation from the accretion disc. As the jet continuum increases, the lines gradually disappear and in the highest flux levels the object displays a featureless spectrum, which is typical of BL Lac objects. Image reproduced with permission from \citet{raiteri2017_nature}, copyright by Macmillan.}
\label{fig:cta_spectra}
\end{figure}

The blazar class includes two main subclasses, that of flat-spectrum radio quasars (FSRQs) and that of BL Lacertae-type objects, usually abbreviated into BL Lacs. The divide between these two blazar types was established more than 30 years ago by \citet{stickel1991} and \citet{stocke1991} based on the spectroscopic behaviour of the sources, namely on the strength of the emission lines. They set a limit of 5 \AA\ on the line equivalent width (EW) in the rest frame: BL Lacs are sources with (almost) featureless spectra, while FSRQs have quasar-like lines whose EW exceeds the limit. However, the line EW, being the ratio between the line flux and the continuum flux density, in the case of blazars is a misleading quantity, because the continuum flux density is generally dominated by the jet emission, which is not the main ionising source of the line (see also Sect.~\ref{sec:lines}). Therefore, when the jet flux increases, a source that usually appears as a FSRQ can look like a BL Lac \citep{carnerero2015,raiteri2017_nature}. 
Likewise, a BL Lac object can show emission lines as strong as to be classified as a FSRQ when the jet contribution becomes weak. This is even the case for BL Lacertae, the prototype of the BL Lac subclass \citep{vermeulen1995}.
These changing-type objects in the literature are sometimes called ``changing-look blazars". However, this term should be avoided, because it reminds the term ``changing-look AGNs'' that is used for those AGNs where the BLR contribution appears or disappears for some reasons connected to physical phenomena in the nuclear zones of the AGN. 
In contrast, in ``changing-type blazars" the lines disappear when the source is bright and appear when the source is faint just because of the variations in the jet emission.
An example is given in Fig.~\ref{fig:cta_spectra}, showing that the broad emission lines in the optical spectra of the FSRQ CTA~102 gradually disappear when the jet continuum increases towards an outburst state \citep{raiteri2017_nature}.

Other, more physical distinctions between the two blazar types have been proposed, based on the accretion rate or luminosity of the BLR \citep{ghisellini2011}.
According to \citet{giommi2012} FSRQs and BL Lacs correspond to intrinsically high- and low-ionisation sources, respectively.

Considering the shape of the SED, FSRQs are defined as low-energy peaked sources, because the peak frequencies of the two bumps are in the mm--near-IR and MeV--GeV ranges. Their SED often shows an additional bump peaking in the rest-frame UV. This emission component is called the ``big blue bump" and is ascribed to thermal emission from the accretion disc. Moreover, sometimes it is also possible to recognise a ``little blue bump" due to the contribution of Balmer line and continuum emission and Mg\,II and Fe\,II lines at about 3000 \AA\ rest frame \citep{bregman1986,raiteri2007}.
Both features appear, or are more evident, in faint states, i.e.\ when the jet contribution is weaker, and reveal a quasar-like core. In some FSRQs, the optical--UV part of the SED can even be dominated by the big blue bump. 

The SEDs of BL Lacs are more heterogeneous and, depending on the frequency at which the bumps peak, these objects are further subdivided into low-energy cutoff BL Lacs (LBLs) and high-energy cutoff BL Lacs (HBLs), a definition that replaced the previous classification into radio-selected and X-ray-selected BL Lacs \citep{padovani1995}.
The SEDs of LBLs exhibit maxima in the IR-optical and MeV--GeV bands, whereas those of HBLs peak in the X-ray and VHE domains. A further class is that of the extremely high-energy peaked BL Lacs (EHBL), with the synchrotron peak in the medium-hard X-ray band and the high-energy peak above 1 TeV.
BL Lacs usually do not show disc or BLR signatures in their SEDs, but the emission of some close objects can receive an important and even prevailing contribution from the light of the stars in the host galaxy.

Moreover, infrared and sub-mm observations by several space observatories
revealed thermal dust emission in several blazars, both FSRQs and BL Lacs, which has been ascribed to either interstellar matter in the host galaxy (for low-redshift objects) or the AGN torus \citep[e.g.][]{malmrose2011}.

Blazars are also classified as low-, intermediate-, or high-synchrotron peaked (LSP, ISP, and HSP) objects if the rest-frame frequency of their SED synchrotron peak is lower than $10^{14} \rm \, Hz$, between $10^{14} \rm \, Hz$ and $10^{15} \rm \, Hz$, or greater than $10^{15} \rm \, Hz$, respectively \citep{abdo2010_sed}. 
In general, variability is more pronounced at frequencies close or just beyond the peaks of the SED bumps. This means that LSP blazars show more variability in the optical and GeV bands, while HSP sources are very active in X-rays and TeV bands.

\begin{figure}[htbp]
\centering
\includegraphics[width=\textwidth]{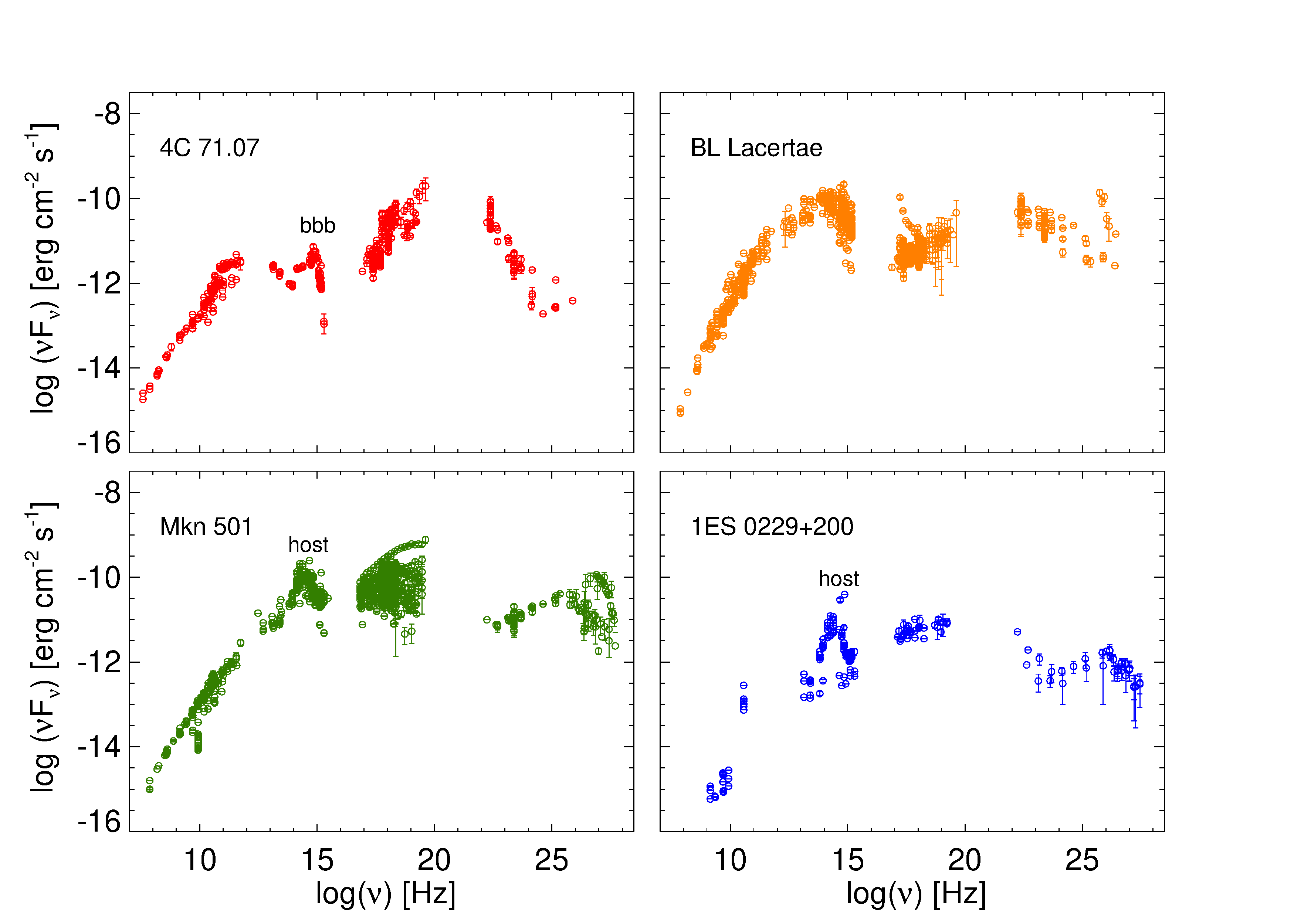}
\caption{Spectral energy distribution of different blazar types obtained with data acquired at different epochs. Top-left: 4C 71.07, a FSRQ with a prominent contribution from the big blue bump (bbb) in the optical--UV spectral range, and a strong Compton dominance. Top-right: BL Lacertae, an LBL characterised by strong variability at all wavelengths. Bottom-left: Mkn~501, an HBL with a prominent contribution from the host galaxy (host) in the near-IR to optical spectral range, a strong variability in the X-rays and at TeV energies, and a high-energy bump peak lower than the low-energy bump peak. Bottom-right: 1ES~0229+200, an EHBL with a strong host-galaxy contribution and a low peak of the high-energy bump. Data for the plots were obtained through the ASI-SSDC SED builder tool (\url{https://tools.ssdc.asi.it/SED/}), including multiwavelength data from various observing facilities. A few strong outliers have been removed.}
\label{fig:seds}
\end{figure}

Figure \ref{fig:seds} shows four SEDs corresponding to blazars of different types: the FSRQ 4C 71.07, the LBL BL Lacertae, the HBL Mkn~501, and the EHBL 1ES~0229+200. Typical features of the 4C 71.07 SED are the prominent big blue bump in the optical--UV energy range and the strong energy output in $\gamma$-rays (``Compton dominance", see Sect.~\ref{sec:blase}). The SED of BL Lacertae displays intense variability at all wavelengths and almost the same level of the low- and high-energy bumps. The SEDs of Mkn~501 and 1ES~0229+200 show a prominent contribution from the host galaxy in the near-IR to optical bands and a peak of the high-energy bump much lower than the low-energy one. Mkn~501 also show extreme variability in X-rays and at TeV energies.

\subsection{The blazar sequence}
\label{sec:blase}

\begin{figure}[ht]
\centering
\includegraphics[width=0.6\textwidth]{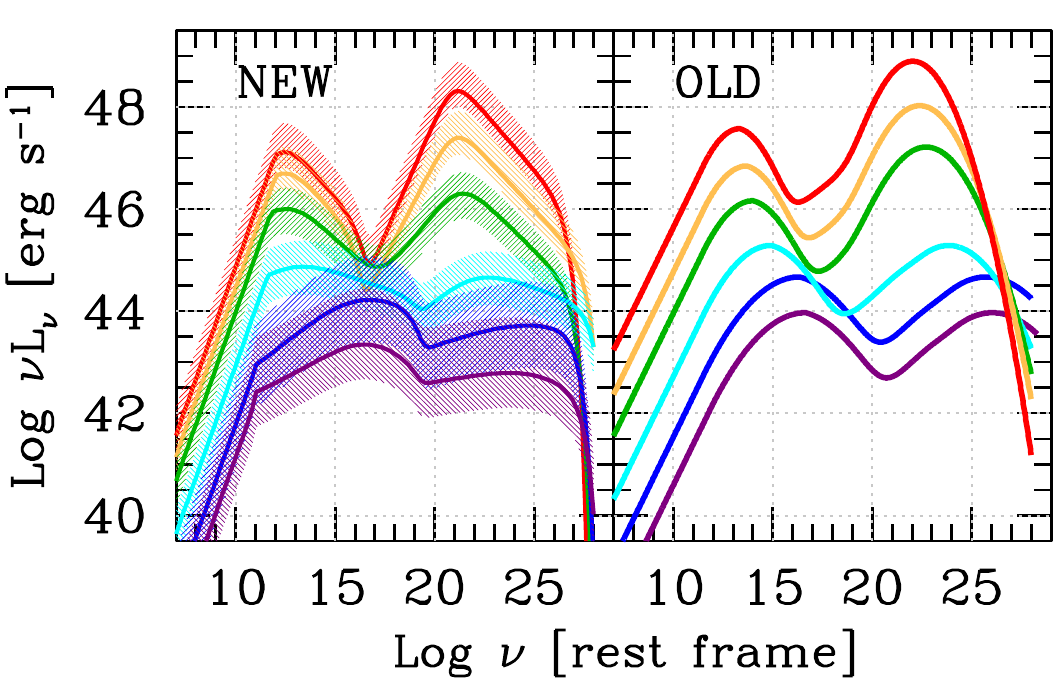}
\caption{The blazar sequence by \citet{ghisellini2017} (labeled ``NEW", left), which is based on bins in $\gamma$-ray luminosities, compared to the original blazar sequence by \citet{fossati1998} (labeled ``OLD", right), which considers bins in radio luminosity. FSRQs present the largest luminosities and lowest synchrotron (and inverse-Compton) peak frequencies, together with high Compton dominance, while HBLs display the lowest luminosities and the highest $\nu_{\rm S}$ and $\nu_{\rm IC}$. LBLs are located at intermediate positions. Image reproduced with permission from \citet{ghisellini2017}, copyright by the author(s).}
\label{fig:bs}
\end{figure}

\citet{fossati1998} analysed the SEDs of a sample of blazars in the $\log (\nu L_\nu)$ versus $\log (\nu)$ diagram and showed that these SEDs, grouped by radio luminiosities, produce a sequence connecting the highest luminosity sources with the lowest frequencies of the synchrotron peak $\nu_{\rm S}$, to the lowest luminosity objects with the highest $\nu_{\rm S}$.
This means that the synchrotron peak moves from the top left to the bottom right in the diagram, leading to an anticorrelation between the luminosity of the source and $\nu_{\rm S}$, and also between the luminosity and the frequency of the peak of the SED high-energy bump $\nu_{\rm IC}$, which correlates with $\nu_{\rm S}$. Moreover, the higher the luminosity, the higher the Compton dominance, i.e.\ the ratio between the peak luminosities of the high-energy and low-energy bumps.
The blazar sequence also seems to separate blazar types, as FSRQs are found in the top left, and HBL in the bottom right, while LBL occupy intermediate positions.
The blazar sequence was later revised by \citet{ghisellini2017}. They grouped the sources according to their $\gamma$-ray luminosities instead of their radio luminosities, and essentially confirmed the results of \citet{fossati1998}.
A comparison between the original and revised blazar sequence is shown in Fig.~\ref{fig:bs}.

The transition from HBLs, IBLs, LBLs and FSRQs along the blazar sequence was interpreted 
as due to an increasing importance of the BLR radiation field and thus of the electron cooling due to inverse-Compton scattering off these external photons (EC process), leading to $\gamma$-ray emission dominating the SED (Compton dominance). 
In HBLs, which have a weak BLR, if any, cooling would be less effective, and particles would be able to reach high energies and produce high-energy synchrotron and inverse-Compton photons by SSC.
The origin of $\gamma$-ray radiation by SSC in HBLs, with an increasing EC contribution from IBLs to LBLs, and a major EC contribution in FSRQs, has been confirmed by other studies \citep[e.g.][]{abdo2010_sed,hovatta2014,liodakis2019}.

However, the discovery of sources that do not fit into the blazar sequence, either FSRQs with high $\nu_{\rm S}$ \citep[e.g.][]{padovani2003},
or LBL with low luminosities \citep[e.g.][]{caccianiga2004} 
indicates that this blazar unification scheme may work in general, but cannot describe the variety of cases observed.
According to some authors \citep[e.g.][]{caccianiga2004,giommi2012}, the blazar sequence is at least partly produced by selection effects, due to flux-limited surveys and the lack of redshift estimates for many BL Lacs, which may be sources with high luminosities and high synchrotron peaks.
Moreover, \citet{nieppola2008} showed that the sequence disappears when intrinsic luminosities and $\nu_{\rm S}$ are considered, which are obtained by correcting the observed values for the Doppler factors derived from variability.
From the theoretical point of view, \citet{ghisellini2008} revisited the blazar sequence assuming that the two main parameters are the mass of the SMBH and the accretion rate, and predicted the existence of blazars outside the original sequence.
In the interpretation by \citet{meyer2011}, the blazar sequence is actually an envelope in which different subclasses are formed by progressive misalignment of two intrinsically different populations.


\subsection{Blazar census}
In population studies, it is common to define a blazar as a jetted AGN with a viewing angle of the order of $1/\Gamma \, \rm rad$. Simple geometrical considerations predict that for each blazar in the sky with Lorentz factor $\Gamma$, there should be about $\Gamma^2$ misaligned jetted AGNs.
The fact that blazars have been observed up to redshift $z \sim 7$ then may imply the presence of hundreds to thousands of quasars powered by SMBHs with billion solar masses in the very early Universe \citep{banados2025}.
However, \citet{lister2019} showed that the $\Gamma^2$ rule is incorrect for flux-limited blazar samples and that the expected number of misaligned AGNs has a complex dependence on $\Gamma$.

Vice versa, the number of blazars that can be detected in the sky can be estimated starting from the number of expected observable radio-loud quasars and then considering those with the appropriate jet orientation. The result is that several tens of thousands of blazars will likely be observed with deep surveys like the incoming Legacy Survey of Space and Time at the Vera C. Rubin Observatory (Rubin-LSST), which will be able to provide light curves for most of them over its ten-year sky monitoring \citep{raiteri2022}.

Several blazar catalogues exist, including up to many tens of thousands of objects, most of which are candidates selected in different ways out of the Galactic Plane.
The BZCAT5 catalogue\footnote{\url{https://www.ssdc.asi.it/bzcat5/}} 
\citep{massaro2015} 
contains 1909 FSRQs, 1059 BL Lacs, 92 BL Lac candidates, 274 galaxy-dominated BL Lacs, and 227 blazars of uncertain type. This catalogue has been extensively used for statistical studies on blazars.
The CRATES \citep{healey2007} and BROS \citep{itoh2020} catalogues include $\sim 11000$ and 88211 flat-spectrum radio sources, respectively.
3980 blazar candidates were selected by \citet{everett2020} in the South Pole Telescope (SPT)  Sunyaev-–Zel'dovich (SZ) survey.
\citet{chang2019} adopted a series of multiwavelength criteria to select the 2013 HBL objects in the 3HSP catalogue\footnote{\url{https://www.asdc.asi.it/3hsp}}.
The WIBRaLS2 and KDEBLLACS catalogues contain blazar candidates selected mainly based on their infrared colours \citep{dabrusco2019}.
1580 blazar candidates were identified by \citet{paggi2020} among the ALMA calibrators, some of which located at low Galactic latitude. 
The 4LAC catalogue \citep{ajello2020} includes 2863 $\gamma$-ray objects detected by the \textit{Fermi} satellite, most of which are blazars. A revised catalogue of the $\gamma$-ray emitting jetted AGNs from 4LAC was published by \citet{foschini2022}. Another recent catalogue dominated by blazars is the Radio Fundamental Catalogue based on VLBI observations \citep{petrov2025}.

\section{Time series analysis}

Blazars are characterised by variable emission on a variety of time scales, which is likely due to the overlapping of different mechanisms. 
Like the other AGNs, blazar light curves are characterised by ``red noise".
This implies random variability, whose amplitude increases with time scale.
Without going into many details, in this section we summarise the main methods to asses if variability is present in a light curve and what are its properties.

\subsection{Variability tests}
A quantitative estimate of source variability is made difficult by both measurement errors and the presence of outliers in the light curves. These represent a big problem, especially for time series obtained from wide surveys, where data are processed with automated procedures. Indeed, it is very hard to fine-tune an algorithm that can automatically remove outliers without the risk of deleting genuine data. And sometimes the identification of outliers is difficult even when a visual inspection of the light curves is performed.

One of the classical ways to assess variability is to use the reduced $\chi^2$ for a constant model:

\begin{equation}
\frac{\chi^2}{\rm d.o.f.}= \frac{1}{N-1} \, \sum_{i=1}^N \left( \frac{x_i- \langle x \rangle}{\xi_i} \right)^2
\end{equation}
where $x_i$ represents a single measurement in either flux or magnitude at time $t_i$ of a time series with $N$ elements, $<x>$ is the arithmetic mean value, and $\xi_i$ the photometric error on $x_i$.

The amount of variability can be defined through the \textit{excess variance} $\sigma^2_{\rm XS}$, which corrects the \textit{sample variance} $\sigma^2$  for the contribution of the measurement errors
\begin{equation}
    \sigma^2_{\rm XS}=\sigma^2 - \langle \xi^2 \rangle
\end{equation}
where
$
    \sigma^2=\frac{1}{N-1} \sum _{i=1}^N (x_i-\langle x \rangle)^2
$
and
$
\langle \xi^2 \rangle = \frac{1}{N} \sum _{i=1}^N \xi_i^2
$

\medskip
The \emph{normalised} or \emph{fractional excess variance} is $\sigma^2_{\rm NXS}=\sigma^2_{\rm XS}/ \langle x \rangle^2$, and its square root is also known as the \emph{fractional variability}:
\begin{equation}
F_{\rm var}=\sqrt{\frac{\sigma^2 - \langle \xi^2 \rangle}{\langle x \rangle^2}}
\end{equation}

Uncertainties on these parameters were estimated by \citet{edelson2002} and \citet{vaughan2003}. 

The fact that the errors on the data points are not always reliable can make it preferable to use the \emph{intrinsic variability} parameter adopted by \citet{sesar2007}:
\begin{equation}
\sigma_{\rm IV}=\sqrt{\sigma^2- \langle \xi(m) \rangle^2}
\end{equation}
where $\langle \xi(m) \rangle$ is the magnitude-dependent mean photometric error.
Another widely used parameter is the \emph{variability amplitude} \citep{heidt1996}:
\begin{equation}
    A=\sqrt{(x_{\max}-x_{\min})^2-2 \langle \xi \rangle^2}
\end{equation}
where $x_{\max}$ and $x_{\min}$ represent the maximum and minimum values in the light curve and $\langle \xi \rangle$ is the average measurement error.

Sometimes the strength of the observed flux variations is measured through the \emph{modulation index} \citep{quirrenbach2000}:
\begin{equation}
    m[\%]=100 \, \sigma/ \langle x \rangle
\end{equation}

To test variability on short time scales, in particular the flux variations detected by a satellite over subsequent orbits, one can calculate the point-to-point fractional variability \citep{edelson2002}:
\begin{equation}
F_{\rm pp}= \frac{1}{\langle x \rangle} \, \sqrt{ \frac{1}{2(N-1)} \sum_{i=1}^{N-1} (x_{i+1}-x_i)^2- \langle \xi^2 \rangle}
\end{equation}

\citet{dediego1998} proposed to use the \emph{analysis of variance} (ANOVA) to test optical microvariability in quasars. This implies splitting the data train into several segments, evaluating their mean and variance, and then testing the null hypothesis, i.e.\ the equality of the means, which would imply no variability.

\subsection{Variability characterization}
Variability can be characterised through a variety of metrics, often called \emph{variability features} \citep[see e.g.][]{richards2011}.
These features were proposed especially in view of the need to classify sources in wide time-domain optical surveys through machine learning techniques. 
The application of variability-based methods to identify blazars is expected to be very successful, because the emission is in general dominated by the very variable beamed radiation from the jet. However, in the near-IR to UV band, the possible prevailing presence of additional, less variable, or even steady emission components (see Sect.~\ref{sec:blatyp}) can hinder an effective identification of blazars based on variability.

An important aspect of variability characterization is the search for characteristic time scales and periodicities, which can provide insights into the variability mechanisms. The main techniques to reach this goal are the {\it Structure Function} (SF), the {\it Autocorrelation Function} (ACF), and the {\it Power Spectral Density} (PSD).

The primary tool to search for periodic signals in time series is the periodogram, which is 
a data-based estimator of the PSD \citep[e.g.][]{vanderplas2018}. 
There are several implementations of the periodogram, which all work fine if the periodic signal in the time series has a sinusoidal shape and the period is regular. The most famous is the Lomb-Scargle periodogram \citep{scargle1982}, which was developed for unevenly-sampled data.


One critical point is that blazar light curves, like those of AGNs in general, are affected by red noise, which makes the evaluation of the period significance quite delicate \citep{vaughan2005}. 
An often used method to search for periodicities and estimate their significance in unevenly sampled time series affected by red noise is REDFIT\footnote{\url {https://www.manfredmudelsee.com/soft/redfit/index.htm}}, which was originally developed to analyse paleoclimatic time series \citep{schulz2002}.

A particularly useful tool to explore the presence of periodic signals is wavelet analysis\footnote{\url{ http://paos.colorado.edu/research/wavelets/}} \citep[see e.g.][]{torrence1998}, which gives information not only in the space of frequencies, but also in the space of times.

The SF essentially measures the mean variability amplitude of a time series as a function of the time lag, $\tau$, between the data points.
In its classical formulation by \citet{simonetti1985}, the first-order SF Function is given by
\begin{equation}
{\rm SF}(\tau)=\frac{1}{N(\tau)}\sum (x_i-x_j)^2
\end{equation}
For each bin of the time lag centred on $\tau$, $N(\tau)$ is the number of data pairs $\{x_i,x_j\}$ whose time separation $t_j-t_i$ falls into the bin, and the sum runs on these pairs.
Several other definitions of the SF have been proposed, which also take measurement errors into account \citep[see e.g.][]{graham2014}.

In the $\log ({\rm SF})$ versus $\log(\tau)$ diagram, the SF ideally shows a first plateau, then a linear increase, and then a second plateau \citep[e.g.][]{hughes1992}. The first plateau is due to the measurement noise. The slope of the linear increase depends on the type of noise process that causes variability and is related to the index of the PSD: when
$\rm SF (\tau) \propto \tau^\alpha$, then ${\rm PSD}(f) \propto f^{-(\alpha+1)}$, where $f=1/t$ is the frequency.
In particular, if $\alpha$=0 the process is called \emph{flicker noise}, while if $\alpha = 1$ it is known as \emph{random-walk} or \emph{red} noise.

The study of blazar variability benefits greatly from a multiwavelength context, where the flux changes detected in one band are compared to those observed at another frequency. 
The \emph{Discrete Correlation Function} (DCF) is a method that was specifically designed to analyse the correlation between two time series $x$ and $y$ that are unevenly sampled \citep{edelson1988,hufnagel1992}. It is defined as
\begin{equation}
    {\rm DCF}(\tau)=\frac{1}{N(\tau)} \sum {\rm UDCF_{ij} (\tau)}
    \label{eq:dcf}
\end{equation}
with
\begin{equation}
    {\rm UDCF}_{ij}=\frac{(x_i-\langle x \rangle)(y_j-\langle y \rangle)}{\sigma_x \sigma_y}
\end{equation}
where the sum in Eq.~\eqref{eq:dcf} runs over all the $N(\tau)$ data pairs $\{x_i,x_j\}$ whose time separation $t_j-t_i$ falls into the $\tau$ bin, and $\sigma_x$ and $\sigma_y$ are the standard deviations of each data set.
The standard error on the DCF is given by
\begin{equation}
\sigma_{\rm DCF}(\tau)=\frac{1}{N(\tau)-1} \left\{ \sum [ {\rm UDCF}_{ij} - {\rm DCF}(\tau) ]^2 \right\}^{1/2}
\end{equation}
and depends on the number of data pairs in the bin. Larger $\tau$ bins decrease the uncertainty, but also decrease the time resolution, so a compromise must be reached between the two needs. To alleviate the problem, \citet{alexander1997} proposed an adaptive-bin version of the DCF, called ZDCF.

A positive (negative) peak of the DCF indicates correlation (anticorrelation), and the corresponding value of $\tau$ gives the time delay with which the variations in the second time series follow those in the first one.
In order to take possible asymmetry of the peaks into account and to increase time resolution, a better estimate of the time lag is obtained by considering centroids rather than peaks. The uncertainty on $\tau$ can be calculated through a ``flux-randomization, random subset selection" Monte Carlo technique \citep{peterson1998}. The strength of the (anti)correlation increases with the DCF absolute value, and a value of the DCF around $(-)1$ means strong (anti)correlation. A more precise estimate of the signal significance requires Monte Carlo simulations of the time series \citep{emma2013,max2014}.  

If $y=x$, the \emph{Auto-Correlation Function} (ACF) is obtained, which can be used like the SF to identify characteristic time scales and periodicities. 
Note that characteristic time scales produce maxima in the SF and minima in the ACF, while periodicites yield minima in the SF and maxima in the ACF.
As shown by \citet{smith1993}, removal of long-term linear trends from the light curve before calculating the SF helps to recover variability time scales. This is true in general for both SF and ACF when light curves are corrected for long-term brightness oscillations, a procedure that is known as ``detrending" \citep[][]{raiteri2021b}.
Warnings have been raised by \citet{emma2010} on the use of the SF to study blazar variability, since spurious signals can appear in some cases. 

Another popular technique to identify and characterise local episodes of variability in light curves is the Bayesian Block analysis
\citep{scargle2013}.

Time scales of specific events can be calculated as $T=F/(|\Delta F/\Delta t|)$ \citep{wagner1995}.
In case $F$ shows a constant growth rate during the flare, the time scale can be measured by the \emph{doubling time} $T_d=|\Delta t/ \Delta \ln F| \, \ln(2)$, while if the flux grows exponentially, the time scale can be estimated by the \emph{e-folding time} $T_e=T_d/\ln(2)$. The rise and fall phases can have different time scales. 
From the time scales of the fastest flares, together with their maximum flux amplitude, it is possible to calculate the observed variability brightness temperature and then, from a comparison with the assumed intrinsic value, the Doppler factor \citep[e.g.][]{hovatta2009}.

The \emph{duty cycle} is the time that a source is seen in a certain state divided by the total observing time.
It can be used to determine the percentage of time spent in flaring states, i.e.\ with a brightness level higher than some threshold,
or to establish the percentage of time during which a source shows microvariability.


\subsection{Flux distributions}
Many authors have tried to extract information on the processes that cause the observed variability by analysing the flux distributions, with the idea that normal distributions are expected from additive processes, such as multiple independent emitting zones, while lognormal distributions arise from multiplicative processes like a variable accretion rate \citep{uttley2005}.

Several works have found lognormal flux distributions in the light curves of blazars at different wavelengths.
This was the case with the X-ray light curve of BL Lacertae \citep{giebels2009}, whose variability was interpreted as the imprint of activity in the accretion disc. However, a lognormal distribution of the optical flux densities of BL Lacertae was found to be consistent with changes in the jet orientation \citep[][]{raiteri2024}.
\citet{kushwaha2016} recognized a lognormal profile in the X-ray flux distribution of PKS~1510--089, and bi-lognormal profiles in its near-infrared, optical, and $\gamma$-ray light curves. They discussed the possibility that, besides disc modulation, lognormality in blazars can also be the result of magnetic reconnection in the jet. 
The analysis of the $\gamma$-ray light curves of several blazars 
mostly revealed lognormal flux distributions \citep{shah2018,bhatta2020}.
\citet{tavecchio2020} showed that a stochastic model applied to the $\gamma$-ray light curves of some powerful FSRQs can reproduce the observed flux distribution, including the power-law decrease at high fluxes.

The works mentioned above indicate that it is hard to unveil the variability process from the profile of the flux distribution, since different models for blazar variability can produce the same flux distribution. Moreover, \citet{scargle2020} demonstrated that linear additive processes can produce both normal and lognormal flux distributions, so that lognormal distributions do not necessarily imply nonlinear multiplicative processes.
The author also warned against using the statistical properties of a time series alone to infer the type of the underlying physical process.

\section{Blazar monitoring programs}
\label{blamo}
Understanding blazar variability on all time scales requires an enormous monitoring effort. 
In the radio band, some long-standing monitoring programs have collected huge datasets for large samples of blazars. In particular, the Blazar Monitoring Program at the University of Michigan Radio Astronomy Observatory (UMRAO) in US was active from the mid-1960s until 2012, measuring flux density and polarisation at 14.5, 8.0, and 4.8 GHz \citep{aller2017}. 
Other single-dish long-term radio programs have been running at the Mets\"ahovi Radio Observatory in Finland \citep[37 GHz and formerly also 22 GHz,][]{terasranta2005}, at the Medicina and Noto radio telescopes in Italy\footnote{\url{https://radio.oato.inaf.it/}} \citep[5, 8.4, 22, and 43 GHz,][]{bach2007},
and at the Owens Valley Radio Observatory\footnote{\url{https://sites.astro.caltech.edu/ovroblazars/}} \citep[OVRO, 15 GHz,][]{richards2011_ovro} in US.
The F-GAMMA project, which was in operation between 2007 and 2015, involved observations at both the 100~m Effelsberg telescope in Germany (8 frequencies between $\sim 2$ and 43 GHz) and the 30~m IRAM telescope in Spain \citep[86, 142, and 229 GHz,][]{angelakis2019}. The IRAM telescope also provides data for the Polarimetric Monitoring of AGN with Millimetre Wavelength \citep[POLAMI,][]{agudo2018} project, which acquires data at 86 and 230 GHz. 
Many blazars are monitored as calibrators of the Submillimeter Array\footnote{\url{http://sma1.sma.hawaii.edu/callist/callist.html}} \citep[SMA,][]{gurwell2007} at 230 and 345 GHz.

High-resolution observations with the Very Long Baseline Array (VLBA) are performed mostly at 15 GHz by the Monitoring Of Jets in Active galactic nuclei with VLBA Experiments\footnote{\url{https://www.cv.nrao.edu/MOJAVE/index.html}} \citep[MOJAVE,][]{lister2018} program. 
The Boston University blazar team carries on a monitoring program with the VLBA at 43 and 86 GHz\footnote{\url{https://www.bu.edu/blazars/BEAM-ME.html}} \citep{jorstad2017}. 

Blazar optical monitoring programs have been run in several observatories.
Some of them also provide polarimetry.
The oldest observations were recorded on plates, while from the late 1980s these were replaced by CCD cameras.
Among the historical optical monitoring programs, we must mention
the photographic monitoring of more than 200 AGN run at the Rosemary Hill Observatory starting from 1969 \citep[e.g.][]{smith1993}, as well as the Hamburg Quasar Monitoring Program (HQM), which was born in 1988 with the main goal of searching for flares that may be caused by microlensing events in the light curves of about 100 quasars \citep[e.g.][]{schramm1994}.
More recently, the ``Ground-based Observational Support of the Fermi Gamma-ray Space Telescope at the University of Arizona"\footnote{\url{https://james.as.arizona.edu/\~psmith/Fermi/}} \citep{smith2016}, and the Small and Medium Aperture Research Telescope System \citep[SMARTS,][]{bonning2012} optical/near-IR observations of \textit{Fermi} blazars\footnote{\url{http://www.astro.yale.edu/smarts/glast/home.php}} projects were triggered by the launch of the \textit{Fermi} $\gamma$-ray satellite and included spectropolarimetric and photometric plus spectroscopic observations, respectively.
Other blazar monitoring programs are the ``Monitoring AGN with Polarimetry at the Calar Alto Telescopes" (MAPCAT\footnote{\url{https://home.iaa.csic.es/~iagudo/research/MAPCAT/}}), and those running at the St.~Petersburg\footnote{\url{https://vo.astro.spbu.ru/program_all/}} and Tuorla\footnote{\url{https://tuorlablazar.utu.fi}} observatories.
 Optical polarimetric monitoring for a large sample of blazars has been carried out by the RoboPol program at the Skinakas Observatory\footnote{\url{https://robopol.physics.uoc.gr/}} \citep{angelakis2016,kiehlmann2017,blinov2018}.



A robust interpretation of blazar variability requires well-sampled multifrequency continuous light curves, and this is feasible only through the coordinated monitoring effort of many astronomers and observing facilities. 
One of the largest and longest-running international collaborations for the study of blazar multiwavelength variability is the Whole Earth Blazar Telescope\footnote{\url{https://www.oato.inaf/blazars/webt/}} \citep[WEBT, e.g.][]{villata2002,raiteri2017_nature,larionov2020,jorstad2022}. 
Born in 1997,
the aim of the WEBT is to observe blazars in a continuous way to understand their emission and variability properties. The many tens of members, together with the distribution in longitude of their telescopes, allow the WEBT to obtain very well-sampled light curves in the optical filters. Observations in the radio and near-IR bands often lead to a broad-band coverage, which is further extended by high-energy data from satellites and Cherenkov telescopes.
WEBT observations also include polarimetry and sometimes spectroscopy.

\section{Blazar multiwavelength behaviour}
\label{section:mw}

Blazars show flux variability at all frequencies on all time scales and dramatic outbursts with an apparent huge release of energy. 
These outbursts capture the attention of astronomers and trigger follow-up observations at all accessible wavelengths.
Sometimes VLBA observations analysed together with photometric monitoring data have shown the emergence of a bright superluminal radio knot from the core at the same time as an outburst observed in the radio and/or other bands \citep[e.g][]{vicente1996,savolainen2002,wehrle2012,jorstad2013,abe2025}.

Insight on the variability mechanisms is more effectively obtained in a multiwavelength context, which allows the analysis of correlation between the flux changes at different frequencies and the identification of possible time lags. 
In this section, we review the main features of the blazar behaviour in the various observing bands, from radio to $\gamma$-rays, and the relationship between the brightness variations across the electromagnetic spectrum.

\subsection{Radio}
Observations in the radio band usually show correlated events at individual frequencies, with an increasing time delay of the flux changes, together with decreasing variability amplitude, with increasing wavelength.
These correlated radio events are often preceded by an optical counterpart
\citep[e.g][]{bregman1986,bregman1990b,clements1995,villata2009b,jorstad2013}. 
However, sometimes correlated delayed radio flares that grow with wavelength and lack a clear optical precursor are observed, and both types of correlated radio events can be present in the light curves of the same object \citep[][]{villata2004b,larionov2020}.
In correlated optical-radio events, flares at 230 GHz can occur almost simultaneously with the optical ones, while flares at centimetre wavelengths can follow the optical ones by several weeks or months.
The time delays between the optical and radio flux variations, and among those at different radio frequencies, vary from source to source and even in the same source at different epochs. The delay is generally ascribed to synchrotron-self-absorption opacity, which increases with wavelength and likely decreases going from the inner, denser jet regions closer to the SMBH, towards the outer, more transparent jet zones \citep{blandford1979,bregman1990a,lobanov1998}.
In general, radio oscillations are smoother than those observed at higher frequencies, which suggests a larger emitting zone. Therefore, it is generally accepted that the emission that we observe in the radio bands comes from wider jet regions downstream with respect to the radiation that we see at higher frequencies. Characteristic variability time scales in radio bands show a wide distribution with mean values around 2 yr \citep{hughes1992}, but very fast radio variability has been observed too (see Sect.~\ref{sec:vfv}).

\subsection{Infrared}
The \textit{Herschel} satellite allowed blazar monitoring in the far-infrared for a few sources. In the FSRQ 3C~454.3, flux variations in this band were found to be almost simultaneous with those in both mm and $\gamma$-rays, indicating that the emissions at all these frequencies come from the same jet region \citep{wehrle2012}. Differences in the short-term variability were ascribed to inhomogeneities that arise in a turbulent plasma. \textit{Herschel} observations of BL~Lacertae showed spectral variability also on day time scales, suggesting that the source emission comes from multiple jet regions with distinct physical properties \citep{wehrle2016}.
Blazar variability behaviour in the near-IR is commonly well correlated with the optical one without significant time delay \citep{bonning2012,dammando2013}. 

\subsection{Optical}
The literature on the optical variability of blazars is very wide. In a few cases, light curves extending for about a century or even more are available.
Some blazars, often referred to as ``Optically Violent Variable" (OVV) sources, have shown huge outbusts with dramatic brightness changes of several magnitudes, implying the release of large amount of energy.
This is the case of the FSRQ 3C~279, whose optical light curve was observed to span 8 mag in brightness, with an estimated maximum luminosity\footnote{Rescaled to a flat universe with an Hubble constant $H_0=70 \, \rm km \, s^{-1} Mpc^{-1}$} of  $\approx 1.0 \times 10^{48} \, \rm erg \, s^{-1}$ during the 1988 outburst \citep{webb1990}.
Similar power characterized the peak of the 2005 optical outburst of the FSRQ 3C~454.3 \citep{villata2006}.
An even more dramatic event was observed in the FSRQ CTA~102, which underwent a brightness increase of $\sim 6$ mag during the 2016--2017 outburst, reaching a peak luminosity of $\approx 1.3 \times 10^{48} \, \rm erg \, s^{-1}$ \citep{raiteri2017_nature}.
Another well-known OVV is the BL Lac object AO~0235+16, whose historical optical light curve covers an amplitude range of about 6 mag, including extreme episodes of $\sim 2 \, \rm mag$ brightness changes in 4 days \citep{raiteri2008_0235}.

Very long-term trends in blazar optical light curves covering about 20 years were identified by \citet{smith1993}. These authors estimated variability time scales trough visual inspection, structure function, and power spectral density, finding values mostly ranging from 3 to 15 yr.
Short-term variability will be treated in Sect.~\ref{sec:vfv}.



\subsection{Ultraviolet}
Blazar variability in UV usually appears to be well correlated with the flux changes observed in the optical with no time delay, suggesting that the emission in these two bands is produced by the same mechanism in the same jet region.
In general, if the emission in these bands is pure synchrotron, the observed variability in UV is stronger than in the optical \citep[e.g.][]{raiteri2013}, whereas the reverse is true when there is an important contribution from the big blue bump radiation in the UV part of the SED (see Sect.~\ref{sec:blatyp}).

\subsection{X-rays}
The X-ray emission is mostly inverse-Compton radiation in LSP sources, it receives contribution from both synchrotron and inverse-Compton in ISP objects, and it is dominated by synchrotron radiation in HSPs. Therefore, the X-ray behaviour of the various types of blazars is different: in general, HSP sources show intense flux and spectral changes, even on short time scales, whereas LSP display much less variability.
In the HBL Mkn~421, flux variations at hard X-rays were observed to both lag and precede those at soft X-rays by a few minutes, while sometimes the correlation is lost; the change in behaviour occurs on a few ksec time scales and confirms emission from multiple jet emission regions in a complex hydrodynamic scenario \citep{brinkmann2005}. 
In HBLs, the X-ray variability does not seem to correlate with the optical flux changes \citep[e.g.][]{carnerero2017,ahnen2016,abe2025}.
In LSP sources, the X-ray emission usually lags the optical one and correlates with the radio-mm behaviour, and it is likely produced by inverse-Compton scattering in a larger, possibly downstream, jet region  \citep[e.g.][]{bregman1984,marscher2010,raiteri2011,jorstad2013,raiteri2013}. 

\subsection{Gamma-rays}
The \textit{Compton Gamma Ray Observatory} (CGRO) satellite, in orbit from 1991 to 2000, discovered that blazars can be strong $\gamma$-ray emitters and opened the $\gamma$-ray sky to the investigation of blazar variability \citep[e.g.][]{wehrle1998,hartman2001}. 

In 2007 the small Italian satellite Astro-rivelatore Gamma a Immagini Leggero (AGILE) 
was launched, which allowed the study of the $\gamma$-ray behaviour of a number of blazars. 
The FSRQ~3C~454.3 was one of the most-frequently pointed sources because of its strong flaring activity in the first years of satellite life, which earned the source the nickname ``The Crazy Diamond"  \citep{vercellone2010,vercellone2011}.
The AGILE observations were supported by 
monitoring at low frequencies by the WEBT Collaboration
\citep[see e.g.][and references therein]{raiteri2011}.
The flux variations observed at $\gamma$-rays were found to be strongly correlated with those in the optical band, with no significant time lag. 
This is indeed expected if the high-energy photons are produced through an inverse-Compton process (see Sect.~\ref{sec:sed}). 

The study of blazar variability in the $\gamma$-rays has been revolutionised by the Large Area Telescope (LAT)
onboard the \textit{Fermi} satellite,
which was launched in 2008.
\textit{Fermi} can monitor the whole sky every about 3 hours, providing continuous light curves of objects above its detection threshold.
Optical variability was found to be stronger in $\gamma$-ray detected sources \citep{hovatta2014}.

In some FSRQs, the optical flux variations were observed to lag after the $\gamma$-ray ones by several days, as in PKS~1510--089 \citep{abdo2010} and 3C~279 \citep{hayashida2012}, or by several weeks, as in OJ~248 \citep{carnerero2015}. One possible explanation was given in terms of a faster drop along the jet of the external radiation energy density profile, from which the $\gamma$-ray radiation would depend in an EC scenario, with respect to the magnetic energy density profile, which determines the production of synchrotron photons \citep{hayashida2012}. Further theoretical work on this line demonstrated that $\gamma$-ray flux variations can both precede and follow the optical ones, and the lags can potentially shed light on the jet structure and the external radiation field \citep{mateusz2012}. 
Cross-correlation studies on large samples of objects yielded a variety of results \citep{liodakis2019,dejaeger2023,mccall2024}.
In most cases, a strong correlation was identified, with a range of positive and negative lags, but mainly consistent with no delay. This would favour the leptonic hypothesis for the origin of the high-energy radiation.

In general, flux changes in $\gamma$-rays appear faster than in the optical band, suggesting that they are produced in smaller emitting regions \citep[e.g.][]{jorstad2013}. Moreover, $\gamma$-ray flares usually have larger amplitude. This is expected if the flares are due to an increase in the number of emitting electrons and the $\gamma$-ray photons are produced by an SSC or ``mirror" EC processes (see Sect.~\ref{sec:sed}). 
Otherwise, larger amplitude of the $\gamma$-ray variations may indicate that the optical radiation also receives less variable contributions from other jet zones, as can happen in a spine-sheath model \citep[][]{sikora2016}.

Sometimes $\gamma$-ray flares are found that lack an optical counterpart (``orphan flares"), and vice versa there are optical flares with no $\gamma$-ray counterpart (``sterile flares"). These uncorrelated flares are difficult to explain in the context of pure leptonic models, but may naturally arise in lepto-hadronic models, where low-energy photons are produced by accelerated electrons, while high-energy photons are obtained by accelerated protons (see also Sect.~\ref{sec:vhe}).


\subsection{Very high energies}
\label{sec:vhe}
The most extreme photon-observing window is that of $\gamma$-rays at very high energies reaching the TeV band, which is accessible through Cherenkov telescopes.
The detected targets are mostly HBLs, but some FSRQs with redshift below 1 have also been observed\footnote{\url{http://tevcat.uchicago.edu/}}. 
In the future, the Cherenkov Telescope Array Observatory\footnote{\url{https://www.ctao.org/}} (CTAO) will explore the VHE sky with unprecedented resolution. 

The variability of HBLs at VHE is generally strongly correlated with that in X-rays with no time delay, as expected if TeV photons in these sources are obtained through SSC from X-ray photons \citep{ahnen2016,abe2025}.
However, ``orphan" TeV flares without X-ray counterparts have also been observed, which in leptonic models could be explained by an additional electron population, or by a variation of the external photon field in an EC process, or by a magnetic field oriented along the line of sight. Alternatively, the orphan TeV flares can be ascribed to a lepto-hadronic scenario, where the physical conditions for electrons and protons may be different \citep{krawczynski2004}.

\begin{figure}[ht]
\includegraphics[width=\textwidth]{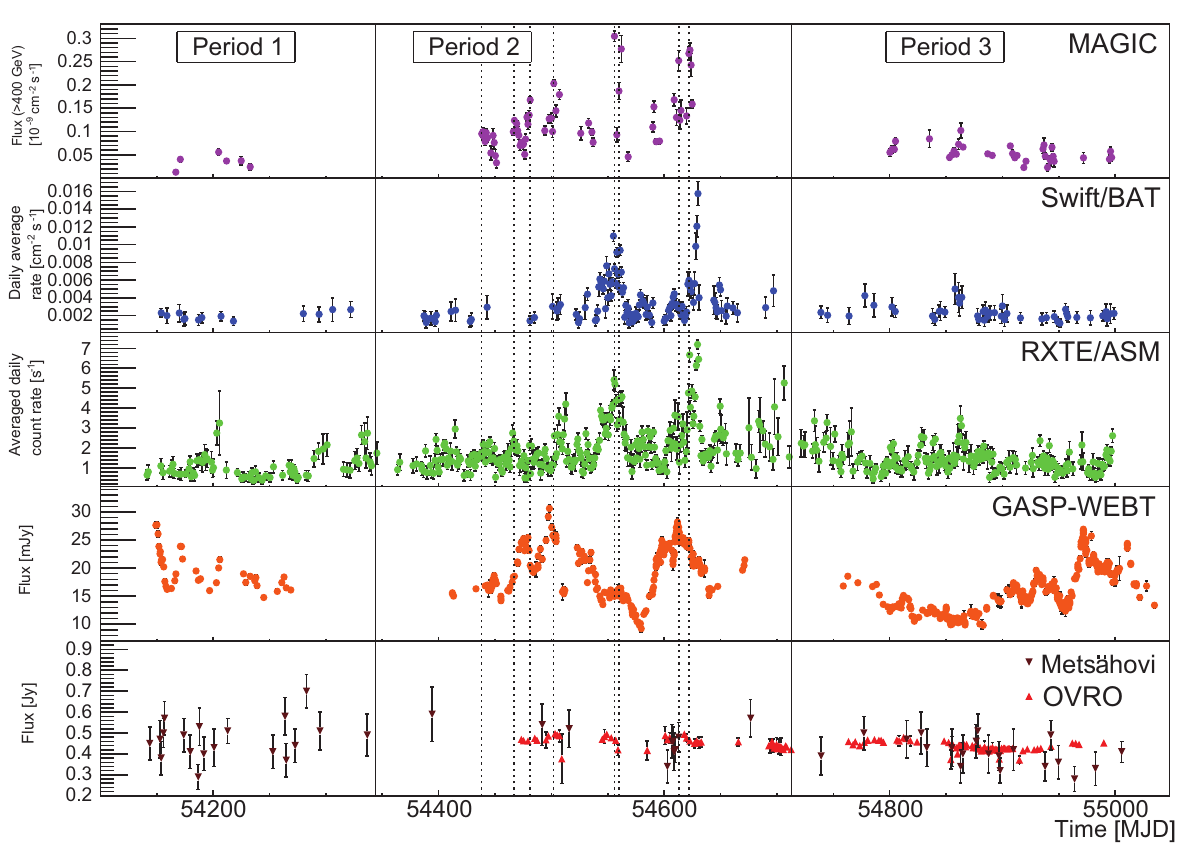}
\caption{Multiwavelength behaviour of the HBL Mkn~421 in 2007--2009. From top to bottom: light curves i) at TeV energies from the MAGIC telescope, ii) in the hard X-rays from the BAT instrument onboard the \textit{Swift} satellite, iii) in the soft X-rays from the ASM detector onboard the \textit{RXTE} satellite, iv) in the optical $R$-band from the WEBT, and v) at 37 and 15 GHz radio frequencies from the Mets\"ahovi Observatory and OVRO, respectively. Image reproduced with permission from \citet{ahnen2016}, copyright by ESO}
\label{fig:ahnen2016}
\end{figure}

Fig.~\ref{fig:ahnen2016} shows an example of multiwavelength behaviour of the HBL source Mkn~421 from TeV energies to radio frequencies \citep[from][]{ahnen2016}.
The flux at VHE clearly correlates with both the soft and the hard X-ray fluxes, but does not correlate with the optical and radio flux densities. Moreover, the variability time scales at VHE and X-rays appear shorter. The lack of correlation between high- and low-energy emission, together with the different variability time scales, may be explained in terms of multiple emitting regions.

\section{Very fast variability}
\label{sec:vfv}

Blazars show both long-term variability with time scales of the order of weeks to years, and short-term variability, whose time scales are measured in days, hours (intra-day variability, IDV), down to minutes (microvariability).
For causality arguments, the observed variability time scale $\Delta t$ implies an upper limit on the size $R$ of the region from which the radiation comes from:
\begin{equation}
    R < \frac{c \, \Delta t \, \delta}{(1+z)}
\end{equation}
The most stringent constraints on the size of the jet emitting region come from the fastest events. Microvariability on time scales of a few minutes then implies either very high and probably unrealistic $\delta$ values for the jet, even greater than 50, or a size much smaller than the typical jet dimension estimated chiefly from SED modelling. Therefore, the detection of microvariability in very well-sampled light curves with very good quality data most likely indicates that the blazar nonthermal emission comes from jet subregions.

Moreover, the size-time scale connection leads to the following expression for the brightness temperature of a source \citep{wagner1995,quirrenbach2000}:
\begin{equation}
    T_B[{\rm K}]=4.5 \times 10^{10} \, F[{\rm Jy}] \, \left( \frac{\lambda[{\rm cm}] \, D_L [{\rm Mpc}]}{\Delta t [{\rm d}] \, (1+z)} \right)^2 \, ,
\end{equation}
where $D_L$ is the luminosity distance.
IDV observed at radio wavelengths $\lambda$ translates into values of $T_B$ by far exceeding the inverse-Compton limit of $10^{12} \, \rm K$ \citep{kellermann1969}. This would imply the so-called ``Compton catastrophe", i.e.\ a dramatic rise in the luminosity of inverse-Compton radiation, which is not observed.
Beaming alleviates the problem, because it lowers the temperature by a factor $\delta^3$.

Radio IDV can also have an extrinsic origin, being due to interstellar scintillation produced by inhomogeneities in the Galactic interstellar medium \citep[ISM, e.g.][]{rickett2006}. In this case, scattering modelling is required to constrain the intrinsic source size. Extreme scattering events can also occur, due to lensing by intervening compact plasma structures in the ISM \citep[e.g.][]{vedantham2017}.
Scintillation can be ruled out if radio flux changes at centimetre wavelengths are found to correlate with flux variations at higher frequencies, where this phenomenon does not occur \citep[e.g.][]{quirrenbach1991}.

IDV in the radio and optical bands has been known for a long time \citep[e.g.][and references therein]{quirrenbach1989a,miller1989,wagner1995}, and
there are many examples of very fast variability detected in several sources at various wavelengths.
In the optical band, it has been observed in densely sampled light curves provided by both ground and space observations. 
During the first massive WEBT campaign on BL Lacertae in 2000, more than 15000 observations obtained by 24 telescopes in 11 countries were collected in the $R$ band \citep{villata2002}. Because of the distribution in longitude of the participating observatories, these data produced a nearly continuous light curve during the nearly one-month core campaign, which revealed intense rapid variability.

The exceptionally sampled monochromatic light curves provided by the Transiting Exoplanet Survey Satellite (\textit{TESS}), complemented by multiband observations by the WEBT, were used to perform detailed variability analyses on individual sources.
In the case of BL Lacertae \citep{weaver2020} and 3C~371 \citep{otero2024}, a minimum variability time scale of about half an hour was found, implying an upper limit for the emitting zone of the order of $5 \times 10^{13} \, \delta \, \rm cm$.

\citet{wehrle2023} analysed the light curves of the \textit{Kepler} K2 mission of a sample of BL Lacs and FSRQs. The time series were about 80 days long and sampled every 29.4 min.  The authors found a significant difference in the mean slope of the periodograms of FSRQs and BL Lacs. This may indicate that they belong to distinct populations, characterised by a different origin of the jet, or by different size or location of the emission regions. In contrast, no significant difference in short-term variability behaviour was found between FSRQs and BL Lacs when considering the densely-sampled light curves of 67 blazars provided by \textit{TESS} \citep{dingler2024}. 

\begin{figure}[htbp]
\centering
\includegraphics[width=0.66\textwidth]{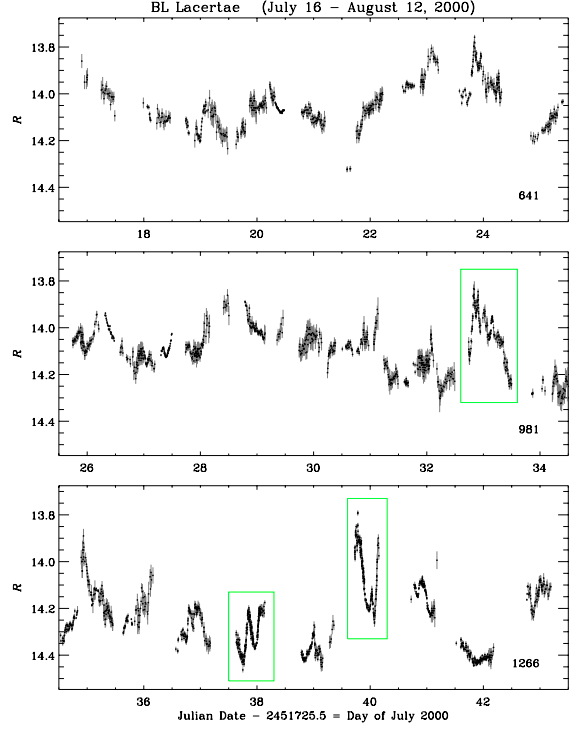}
\includegraphics[width=0.66\textwidth]{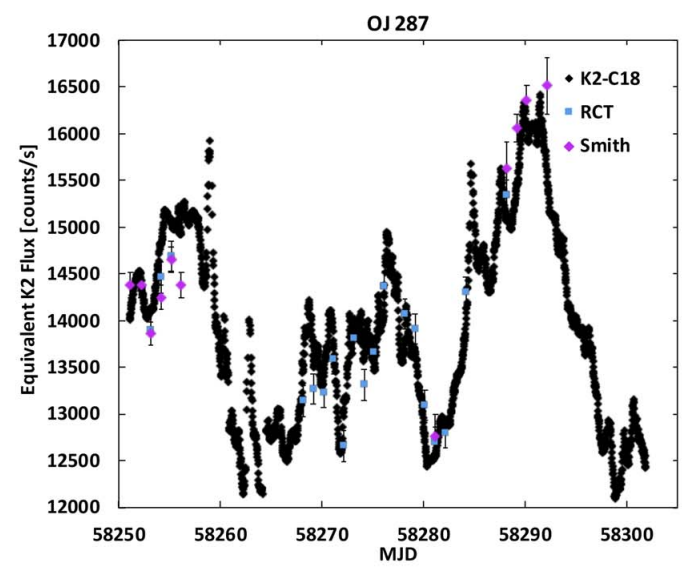}
\caption{Examples of short-term optical variability in blazars. Top: $R$-band light curve of BL~Lacertae in 2000 obtained from ground observations by the WEBT Collaboration \citep[from][]{villata2002}. The green boxes highlight some of the most significant periods. Bottom: \textit{Kepler}-K2 (black symbols) light curve of OJ~287 in 2018; additional ground-based data are shown in blue and magenta \citep[from][]{wehrle2023}.}
\label{fig:micro}
\end{figure}

Figure \ref{fig:micro} shows two examples of extremely well-sampled blazar optical light curves that were obtained from ground and space and were characterised by very fast variability.

Microvariability was also observed at high energies.
Extremely rapid variability in X-rays on a time scale of 30 seconds was detected in the BL Lac object H0323+022 by \citet{feigelson1986}, who estimated the size of the compact emitting region to be $\sim 10^{12} \, \delta \, \rm cm$.
\cite{shukla2020} identified variability on time scale of minutes in the \textit{Fermi} $\gamma$-ray light curve of 3C~279 during a giant flare. They suggested relativistic magnetic reconnection in a small jet region, a plasmoid of about $8 \times 10^{14} \, {\rm cm}$, as the most likely particle acceleration mechanism. This would happen far enough from the SMBH to make kink instabilities produce turbulence that leads to the formation of plasmoids.
Flux variations on minute time scales were also observed in the VHE data obtained by Cherenkov satellites, in particular by the Whipple Observatory on Mkn~421 \citep{gaidos1996}, by the High Energy Stereoscopic System (HESS) on PKS~2155--304 \citep{aharonian2007} and by the Major Atmospheric Gamma-ray Imaging Cherenkov Telescopes (MAGIC) on Mkn~501 \citep{albert2007}.
This very rapid TeV variability  has been interpreted in various ways \citep[e.g.][and references therein]{rieger2019}, including $\gamma$-ray emission in the SMBH magnetosphere, but more likely implies enhanced emission in very small jet regions and high bulk Lorentz factors. This is required to allow TeV photons to escape without being absorbed because of $\gamma$-$\gamma$ pair production \citep[][]{begelman2008}. However, high Lorentz factors seem to be in contradiction with the typically low jet speeds of TeV sources measured from VLBI observations.
In the jets-in-a-jet model by \citet{giannios2009} the observed TeV fast flares are due to small blobs accelerated by magnetic reconnection, which are moving with Lorentz factors of 100 in a strongly magnetized jet with a Lorentz factor of 10.

An extreme case is that of a jet emitting region made up of many subzones that are largely independent one from the others. This is how turbulence is usually treated in blazar emission models (see Sect.~\ref{sec:acc}).

\section{Spectral variability}

The flux variations at different frequencies are not always correlated, and even when they are, usually do not have the same brightness amplitude and may not occur at the same time. The result is that blazar SEDs can show changes in their shape. 
As a consequence, when modelling the blazar spectral behaviour at various epochs, it is of utmost importance to build the SEDs with simultaneous data, especially at those frequencies where flux variations are more intense.

In the optical band, spectral variability can be studied by analysing changes in the colour indices. In general, the synchrotron emission is characterized by a steeper-when-fainter (bluer-when-brighter) behaviour 
\citep[e.g.][]{gear1986,brown1989}. 
This is expected from the acceleration of particles, with faster cooling times for the more energetic ones. 
In some cases, the variation of the colour index on long time scales appears less chromatic than the short-term spectral changes. The mild bluer-when-brighter chromatism of the spectral behaviour on long time scales can be explained as the result of Doppler factor changes on a curved, convex spectrum \citep{villata2004a}. Therefore, while short-term spectral chromatism likely comes from energetic processes in the jet, the mild long-term chromatism 
is consistent with a geometric interpretation of the blazar variability on long time scales \citep[][]{villata2004a,papadakis2007,raiteri2021b}.

The presence of the big blue bump in FSRQs implies a redder-when-brighter trend, which by itself can reveal the presence of a thermal contribution from the nuclear region. The redder-when-brighter behaviour can turn into a bluer-when-brighter one when the brightness increases beyond a certain level at which the thermal component gives a negligible contribution to the total emission \citep{villata2006}. 

In LBLs and especially IBLs, the X-ray band represents the transition region between low- and high-energy bumps in the SED, and noticeable spectral variations can be observed when the relative contributions from synchrotron and inverse-Compton vary \citep[e.g.][]{madejski1996}.

Dramatic changes in the X-ray spectrum can also be detected in HBLs. One example is Mkn~501, where a shift of at least a factor 100 in the synchrotron peak frequency was observed during a very high activity phase in 1997. This was accompanied by a corresponding harder-when-brighter behaviour in X-rays and correlated variability in X-rays and TeV energies \citep[][]{pian1998}. The strong spectral variability of Mkn~501 in X-rays can be recognised in Fig.~\ref{fig:seds}.
Extreme shifts in the synchrotron peak frequency according to a harder-when-brighter trend were also detected in another HBL, Mkn~421 \citep{takahashi1996,carnerero2017,acciari2021}. 
A harder-when-brighter behaviour was sometimes seen also at GeV \citep{dammando2013} and TeV $\gamma$-ray energies \citep{acciari2021}. 

It was shown that spectral hardening in rising flux phases, and spectral softening in dimming phases, occurs when the electron accelerating time is shorter than their cooling time. In contrast, the opposite trend is seen when the two time scales are comparable \citep{kirk1998}.

\section{Variability signatures associated with neutrino emission}
\label{sec:neu}
Blazar jets are formidable particle accelerators, and this makes them good candidates for the origin of at least some of the high-energy neutrinos detected by the IceCube Observatory and other neutrino telescopes. As already mentioned, hadronic processes are needed for neutrino emission, and these produce high-energy photons too, so a flare should be observed
in the $\gamma$-ray light curve of a blazar associated with a neutrino event at the time of neutrino detection,  unless $\gamma$-rays get absorbed.
The $\gamma$-rays produced together with the neutrinos can cascade to lower energies, and enhanced emission may also be expected in X-rays, but not necessarily at optical or longer wavelengths.
The observational scenario is actually not well defined, especially because of the large uncertainty on the neutrino arrival direction,
which makes a secure association with astrophysical sources very hard to obtain.

One case of robust connection was found between the high-energy neutrino event IC-170922A detected by IceCube and the positionally-consistent IBL TXS~0506+056, which was flaring at $\gamma$-rays at the time of the neutrino arrival \citep{ice2018a}. In that period the source was in a high optical and radio state. TXS~0506+056 was also recognised to be responsible for a lower-energy neutrino flare occurred in 2014--2015, during a quiet radio-to-$\gamma$-ray state \citep{ice2018b}. 

\citet{righi2019} looked at the \textit{Fermi}-LAT light curves of eight BL Lacs of different types that were possible emitters of IceCube high-energy neutrinos. They did not find $\gamma$-ray flares around the times corresponding to neutrino detections, apart from the case of TXS~0506+056, whose hadronic nature of the flaring state in 2017 was questioned.

\citet{giommi2020} searched for $\gamma$-ray blazar counterparts to neutrino events reported by IceCube and found that they are mostly IBL and HBL sources. Yet, the availability of external photon fields in FSRQs should enhance neutrino production in these objects through interactions with protons (photomeson reactions) and make them better candidates. According to \citet{padovani2022}, objects like TXS~0506+056 and PKS 1424+240, another blazar identified as a neutrino source, are actually masquerading BL Lacs, i.e.\ they are FSRQs with a very bright jet continuum that overwhelms the emission lines. 
This would also be the case of PKS~0735+178, which was associated with multiple neutrino events detected in 2021 by various neutrino telescopes while the source was undergoing a major flaring episode from optical to $\gamma$-ray energies \citep{sahakyan2023}.

\citet{franckowiak2020} analysed $\gamma$-ray blazars associated with neutrino detections. They found no clear correlation between neutrino emission and multiwavelength behaviour of the sources.

Statistically significant association of 71 IceCube neutrino events with radio-bright blazars was obtained by \citet{plavin2023}. The authors also discussed individual associations. 
Two neutrino events in 2011 and 2022 were detected close in time to radio outbursts of the positionally-consistent FSRQ PKS~1741--038. 
In 2021 IceCube, together with other neutrino observatories, detected high-energy neutrinos from the direction of the LBL PKS~0735+178, which was undergoing a strong multiwavelength flare observed in the radio, optical, X-rays and $\gamma$-rays bands.
The very variable LBL TXS 1749+096 is likely responsible for the emission of a high-energy neutrino detected by IceCube in 2022.
Flaring multiwavelength activity in the FSRQ PKS 1502+106 was ongoing at the time of a neutrino arrival in 2019.
These associations seem to indicate LSPs rather than HSPs as the most favourable blazar sources for neutrino emission.
This finding is supported by the lack of neutrinos associated with the two brightest HBLs, i.e.\ Mkn~421 and Mkn~501 \citep{righi2019}.
According to \citet{righi2020}, neutrinos can be effectively produced by FSRQs, but mostly with energies that exceed the PeV energies characterising the detected events. This could justify the scarcity of robust associations.


\section{Interpretation of blazar variability}
\label{sec:int}
The observed variability of blazars can be explained by intrinsic physical mechanisms, i.e.\ energetic processes occurring inside the jet, and by extrinsic mechanisms, involving changes in the jet orientation.
Inside the jet, particles can be injected or accelerated by different mechanisms. 
On the other hand, there are several observational and theoretical results that indicate that the jet can rotate or precess or even twist, with consequent variation in time of the viewing angle and hence of the Doppler beaming.
Rotation of the jet around its axis is expected, since the jet is launched from the SMBH or the accretion disc, both of which rotate. 
If the jet is curved, as in the case of a helical pattern, rotation around the helix axis would produce changes in the viewing angle of the jet emitting region(s).
Precession can happen in situations such as when the jet axis is not well aligned with the SMBH or accretion disc rotation axis, or with the normal to the orbital plane in a BBHS.
Plasma instabilities developing inside the jet can distort the flow and produce  jet twisting with consequent orientation changes of the emitting jet region(s).
In general, it seems reasonable to consider that both intrinsic and geometric mechanisms are likely contributing to blazar variability, as will be evident from the arguments given below.


\subsection{Particle acceleration mechanisms}
\label{sec:acc}

Different mechanisms of particle acceleration can work in blazar jets, including shock waves propagating along the jet, magnetic reconnection, and turbulence.

Shock waves can arise in blazar jets because of an increase in flow velocity or pressure \citep{rees1978,marscher1985}.
The shock-in-jet model has been widely applied to explain blazar variability \citep{blandford1979,marscher1985,hughes2011,boettcher2019}.

Magnetic reconnection occurs when magnetic field lines of opposite polarity meet and annihilate, releasing magnetic energy. In some circumstances, the reconnection layer can fragment in many plasmoids that can grow through mergers, further accelerating particles \citep[e.g][and references therein]{giannios2013}. In particular, as mentioned in Sect.~\ref{sec:vfv}, ``monster" plasmoids could be responsible for the fastest blazar flares observed at high energies \citep{shukla2020}.
Moreover, magnetic reconnection has been shown to be much more powerful than shocks in accelerating particles in magnetically dominated flows \citep{sironi2015}.
Turbulence implies stochastic behaviour and is a less efficient acceleration mechanism.
However, relativistic turbulence, possibly coupled to reconnection, can play a non-negligible role \citep[see e.g.][and references therein]{sciaccaluga2022,sobacchi2023}.

\begin{figure}[ht]
\centering
\includegraphics[width=0.8\textwidth]{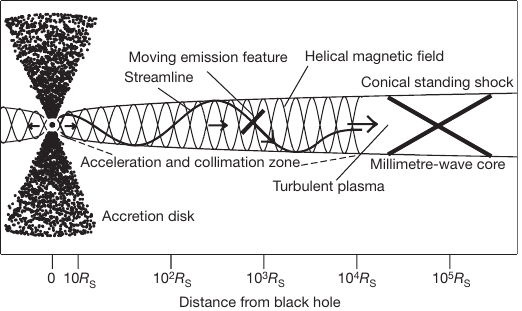}
\caption{The model for the inner jet of BL Lacertae proposed to explain its multiwavelength behaviour. A shock propagates in the acceleration and collimation zone of a jet with a helical magnetic field following a spiral streamline. The flow then becomes turbulent. Image reproduced with permission from \citet{marscher2008}, copyright by Nature.}
\label{fig:marscher2008}
\end{figure}

Indeed, sometimes different mechanisms are involved in the same scenario.
The multiwavelength flaring behaviour of BL Lacertae in 2005--2006 was explained by \citet{marscher2008} in terms of a shock that first crosses the jet acceleration and collimation zone along a spiral streamline in a helical magnetic field, and then meets a turbulent plasma zone (see Fig.~\ref{fig:marscher2008}).
A model based on shocks in a turbulent plasma jet was shown to reproduce the basic features of the blazar multiwavelength light curves, SEDs, and polarization behaviour \citep{marscher2014}.

\subsection{Energetic models based on SED reproduction}

The location and structure of the dissipation zone in blazar jets is not known. A classical, schematic representation involves a single homogeneous emitting zone described by a series of physical parameters, which are changed to reproduce the source broad-band SEDs at different epochs. 
These one-zone homogeneous models have been extensively used in blazar studies, and there are several implementations described in the literature \citep[e.g.][]{mastichiadis1997}.
In general, they involve about ten free parameters, while the viewing angle is assumed to be fixed.
A python code named JetSet is publicly available\footnote{\url{https://github.com/andreatramacere/jetset}} to fit broad-band SEDs with a one-zone model that includes both SSC and EC processes \citep{tramacere2020}; an earlier version of the model 
is implemented in the SED builder tool of the Agenzia Spaziale Italiana (ASI) Space Science Data Center\footnote{\url{https://www.ssdc.asi.it/}} (SSDC), which also allows to retrieve multiwavelength data from a number of catalogues and projects.

However, these models are known to be highly degenerate. Moreover, they usually cannot reproduce the radio emission, which is ascribed to an outer jet region, so they are actually two-zone models. 
In any case, as more broad-band simultaneous data become available, one-zone models appear no longer adequate to describe the source behaviour, and multizone models have been developed. One of them is the spine-sheath model, where an inner flow, the spine, moves faster than the outer flow, the sheath \citep[e.g][]{sikora2016}.

One-zone models usually do not include a self-consistent treatment of the time evolution of the physical parameters. An exception is the one-zone model with time-dependent electron injection rate developed by \citet{boettcher2002} to analyse the X-ray variability of BL Lac objects.

On the other hand, inhomogeneous models have been proposed since long time ago \citep{ghisellini1985,maraschi1992}. 
They parametrize the main physical quantities as a function of the distance along the jet. 
A class of inhomogeneous jet models also assumes jet curvature and twisting \citep{villata1999,villata2009a,raiteri2010,raiteri2017_nature},
and their presentation is deferred to Sect.~\ref{sec:geo}, after introducing what we know about the jet structure and motion from both the observational and theoretical point of view.

\subsection{Jet structure and motion: observations}

Advancements in radio-interferometric techniques have provided increasingly detailed information on the structure of extragalactic jets in general and on blazar jets in particular. 
They have revealed complex morphologies and motions at (sub)pc scales, where the multiwavelength emission presented in the previous sections comes from. 

VLBI images show a shift in the position of the core at different radio frequencies (core-shift effect) mostly along the jet direction \citep{lobanov1998,pushkarev2012}, which supports a scenario where the jet is made inhomogeneous by synchrotron opacity \citep{blandford1979}. An offset is also found with the optical position provided by the Gaia satellite in both directions \citep{kovalev2017}. In the case where the Gaia position is downstream of the VLBI one, the shift is likely due to an extended optical jet, while in the opposite case it may be due to synchrotron opacity as in the radio core-shift effect.

Both standing and propagating features are recognizable, the latter often showing superluminal motion and producing flares in the multiwavelength light curves when they emerge from the core \citep[e.g.][]{marscher2010}. 

Observing evidence of jet rotation around its axis has been found \citep{mertens2016,fuentes2023}. Moreover, changes in the jet direction have been observed.
In particular, \citet{kostrichkin2025} analysed  VLBI data of 317 AGNs at frequencies from 2 to 43 GHz. They found variations in the parsec-scale jet direction in those objects for which high-quality data with high temporal coverage were available, so they suggested that the property is likely ubiquitous. Apparent rotation speeds are frequency dependent, ranging from $0.21 \, \rm deg \, yr^{-1}$ at 2 GHz to $1.04 \, \rm deg \, yr^{-1}$ at 43 GHz. This rules out ballistic motion of plasma components and implies that it is the jet nozzle that changes orientation. 

Many radio interferometric observations have shown that the blazar jets can change orientation in both space and time. They often present bends, sometimes as a part of a helical pattern, which can give rise to a wiggling structure that evolves in time, as can be seen from the curved trajectories of the (superluminal) radio features,
or from changes in the ridge lines, i.e.\ the lines that connect the points of the maximum intensity along the jet. 
This is the case e.g.\ of 
S5~0716+71 \citep{bach2005,kravchenko2020}, 
0735+178 \citep{gomez1999,britzen2010b}, 
0836+710 \citep{perucho2012}, 
OJ~287 \citep{vicente1996,zhao2022},
3C~273 \citep{baath1991}, 
PG~1553+113 \citep{lico2020}, 
3C~345 \citep{baath1992,zensus1995}, 
Mkn~501 \citep{conway1995}, 
1803+784 \citep{britzen2010a}, 
BL Lacertae \citep{cohen2015,kim2023}, 
CTA~102 \citep{fromm2013}, 
and 3C~454.3 \citep{jorstad2013},
and is also confirmed by sample studies from the MOJAVE survey \citep[e.g.][]{lister2021}.
A curved geometry implies differential Doppler beaming of the various parts of the jet and, if the jet structure changes in time, also the Doppler beaming will, leading to a possible geometric origin of at least some of the blazar variability we observe.

The extragalactic jet we know best is that of the close radiogalaxy M87 (Virgo A), which is actually alleged to be a misaligned BL Lac \citep{tsvetanov1998}.
Already in 1979, VLA observations revealed that the jet was not completely straight, but appears to wiggle \citep{owen1980}. The bending was ascribed to either Kelvin-Helmholtz (KH) helical instability \citep{hardee1982a,hardee1982b} or to precession. 
Subsequent VLBI and VLA observations showed complex structures, including oscillating patterns and helical filaments \citep{reid1989,owen1989}.
High-resolution observations at 8 and 15 GHz with the VLBA, VLA, and the 100-m Effelsberg telescope recently allowed \citet{nikonov2023} to recognize that the pc-scale structure of the M87 jet is dominated by three intertwining helical threads, likely originating from KH instability inside the plasma jet.

\begin{figure}[htbp]
\centering
\includegraphics[width=0.4\textwidth]{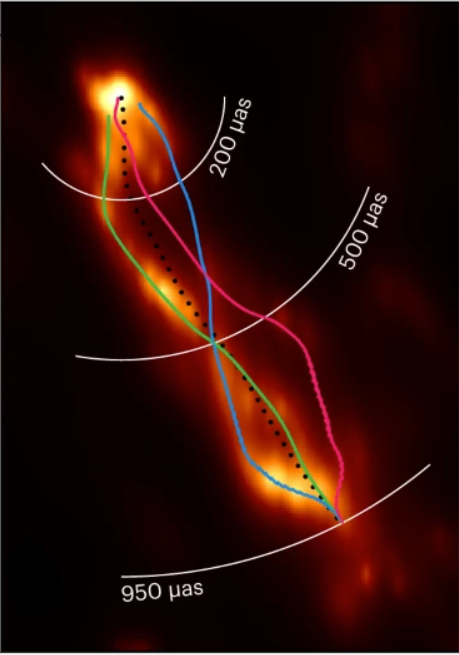}
\includegraphics[width=0.5\textwidth]{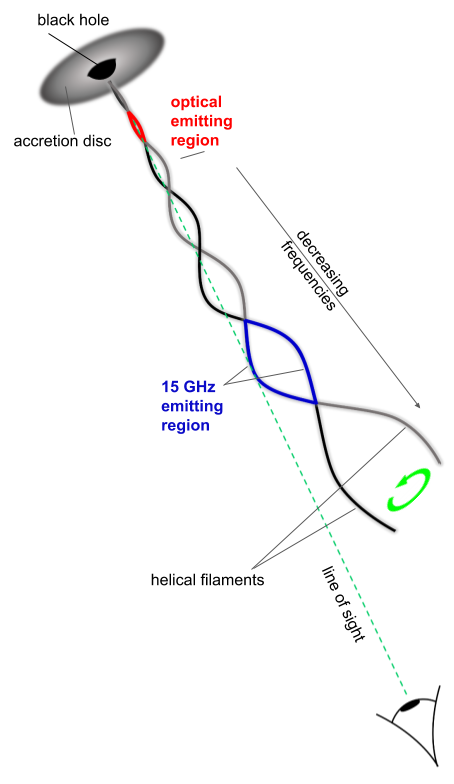}
\caption{Left: The filamentary structure of the 3C~279 jet reconstructed by \citet{fuentes2023}. The two main filaments are highlighted with green and blue lines, while the possible third filament is marked in red. They are likely produced by Kelvin--Helmholtz instabilities. From \citet{fuentes2023}. Right: The double-helix model proposed by \citet{raiteri2024} to explain the multifrequency variability of BL Lacertae. The rotating jet is made up of two interlaced plasma filaments and its axis is curved. The jet is inhomogeneous, with different regions in the jet emitting synchrotron radiation of different frequencies, with longer wavelengths coming from outer regions. Adapted from  \citet{raiteri2024}.}
\label{fig:filjet}
\end{figure}

A similar filamentary structure was identified in the jet  of the FSRQ 3C~279 by \citet{fuentes2023}. These authors analysed microarcsecond-scale angular resolution images at 22 GHz obtained with the space-VLBI mission RadioAstron
supported by 23 ground-based radio antennas. They found that the radio emission of the source comes from a core and up to three interlaced filaments, probably produced by KH instabilities (see Fig.~\ref{fig:filjet}).
The jet brightness along the filaments varies because of differential Doppler boosting, which can explain the moving components commonly observed.
As noted by \citet{fuentes2023}, emission from the filaments, whose cross section is much smaller than that of an entire jet, would also help to explain the shortest variability time scales observed (see Sect.~\ref{sec:vfv}).

Two filaments forming a double helical structure were identified in the jet of the FSRQ 3C~273 by \citet{lobanov2001} when examining radio data obtained by the space radio telescope HALCA
together with a series of ground-based radio telescopes.
Filaments and twisted structures were also recognized in the jet of 1308+326 \citep{britzen2017}.

\subsection{Jet structure and motion: theory}


Theoretical models have explained the observed twisted structure of extragalactic jets as the result of flow instabilities \citep{hardee1982a,hardee1982b}.
Numerical simulations have become increasingly more realistic, to the point of including a general (G) - relativistic (R) magneto (M) - hydrodynamical (HD) treatment of the plasma in three dimensions (3D).

3D-RHD simulations by \citet{hughes2002} showed that the impact of a jet on an oblique ambient density gradient can result in bending of the flow by a strong oblique shock. Moreover, if the flow undergoes precession, changes in its direction would be large enough to imply that significant Doppler boosting is observed. 
\citet{nakamura2001} performed 3D-MHD simulations to investigate the wiggled structure of AGN radio jets and found that the jet distortion can be produced by helical kink instabilities. 
The development of kink instabilities was also found by \citet{moll2008} in their 3D-MHD simulations when considering rotating magnetic fields.
\citet{mignone2010} stressed that the jet structure changes dramatically when passing from a 2D to a 3D description. They run high-resolution 3D-RMHD simulations with an important contribution of the toroidal component of the magnetic field, which seems to be needed to explain the jet acceleration and collimation. This led to the development of strong current-driven kink instabilities that produce jet wiggling. 
In the 3D-RMHD simulations of rotating jets by \citet{singh2016}, where kink instabilities are triggered by a precessional perturbation, magnetic reconnection occurs in filamentary current sheets associated with kink-unstable zones.
3D-GRMHD simulations of a rapidly spinning SMBH in gas threaded with magnetic field showed that the SMBH can launch relativistic jets that develop twists and bends \citep{lalakos2024}.

Accretion discs in AGN may be warped. This happens, e.g., when the rotation axis of the accretion disc and that of the SMBH are misaligned. In this case, the jet is expected to precess. The formation of precessing jets when considering tilted accretion discs has been confirmed by 3D-RMHD simulations \citep{liska2018}.

The theoretical results mentioned above suggest that jet orientation changes are likely to occur, and this would necessarily cause variability.

\subsection{Geometric models}
\label{sec:geo}

In light of the observational evidence and theoretical predictions mentioned in the previous sections, which argue for jet bends and twisting in both space and time, and due to the fact that the relativistic effects on the observable quantities depend on the viewing angle, it goes without saying that an exploration of the causes and consequences of jet bending and twisting is needed.
This issue has been addressed already many years ago and has received more attention in recent years, following the increasing observational and theoretical clues.

\citet{kaastra1992} proposed that jet wiggles observed in radio images can originate from orbital motion in a BBHS, because of the change in the velocity direction of the SMBH carrying the jet. The fact that the observations do not show a regular modulation was ascribed to additional effects, including the jet internal structure, instabilities, and interaction with the ambient medium.

An inhomogeneous helical jet that rotates around the helix axis likely because of orbital motion in a BBHS was proposed by \citet{villata1999} to explain the extreme spectral variability of Mkn~501 from the optical band to X-rays. Inhomogeneous means that the radiation at different frequencies that we observe come from distinct regions of the jet, the higher the emitted synchrotron frequency, the closer its production region to the SMBH. The jet curvature makes the jet emitting zones have different orientations and thus Doppler factors. Moreover, the rotation of the jet implies that such differential Doppler beaming changes in time. The success of the model to interpret the observational data was taken as supporting evidence of the presence of BBHS at the centre of AGNs.

The model by \citet{villata1999} was later elaborated 
and proposed to interpret the multiwavelength variability of a number of blazars. In particular, it explains the presence of both correlated and uncorrelated optical and radio flares in BL Lacertae as due, respectively, to similar or different alignment of the corresponding emitting regions with the line of sight \citep{villata2009a}.
Further elaboration of this model led to the twisting jet model by \citet{raiteri2017_nature}, which successfully reproduced the long-term optical and radio behaviour of the FSRQ CTA~102, including the extraordinary outburst observed in 2016--2017.
In this work, from the observed light curves the authors derived the behaviour of the Doppler factor and viewing angle in time for the emission at various frequencies. Then, by interpolating through frequencies, they obtained the behaviour of the Doppler factor and viewing angle along the jet at various epochs. This allowed them to reconstruct the SEDs of the source in different brightness states, which resulted in excellent agreement with the observations. The critical point they had to take into account is that if the Doppler factor is variable in time, from Eq.~\ref{eq:nu} it follows that the radiation that we observe at a certain wavelength was emitted at frequencies that change in time in the rest frame, and therefore come from different parts of the inhomogeneous jet. 
The model was further refined by \citet{raiteri2024} to explain the behaviour of BL Lacertae in 2019--2022, including a period of exceptional activity in 2021. The authors identified clues in favour of a double-helix structure of the jet, which led to the final picture of a wiggling filamentary jet. This is sketched in Fig.~\ref{fig:filjet}.

Another geometric model was developed by \citet{larionov2013}, who analysed optical (photometric and polarimetric) and $\gamma$-ray monitoring data of S5~0716+71 together with VLBA images. The behaviour in time of the optical flux, polarisation degree and angle was interpreted as due to shocks that propagate in the jet along a helical path.

Occurrence of the firehose instability was proposed by \citet{subramanian2012} to explain the fast variability observed at TeV energies. The instability  originates from pressure anisotropy and implies jet wiggling that can disrupt the large-scale structure of the jet and cause variability. The model can in principle be generalized to explain also variability on longer time scales by changing the values of the physical parameters involved (Subramanian, private communication).

Further geometric models were proposed to explain periodicities in blazar light curve; these will be discussed in Sect.~\ref{sec:period}.

\subsection{Periodic behaviour}
\label{sec:period}

The variability of blazars, like that of AGNs in general, is defined as ``aperiodic", ``random" and thus ``unpredictable". However, there is a wealth of published papers claiming evidence of (quasi)periodic behaviour at all wavelengths on a variety of time scales.
Often these periodicities are transient and sometimes they are revealed in particular brightness states of the source.
Most models for interpreting periodicities on long and intermediate time scales are geometric, implying variations in the orientation of the jet and thus in the Doppler beaming. 
The reason may be jet rotation, precession, or orbital motion in a BBHS.
Precession itself can arise in a BBHS \citep{begelman1980}.
Periodicity on short time scales has also been interpreted as due to instabilities inside the jet or the accretion disc.

\citet{rieger2004} estimated the time scales of periodicities resulting from motion along helical paths in the jet under various assumptions on the driven mechanism. 
For ballistic helical motion induced by precession, observed periods of several tens of years are predicted.
For nonballistic helical motion, the observed periods are shortened by a factor $\Gamma^2$ because of light-travel time effects. They are found to be less than about 10 d if the helical motion is driven by internal jet rotation, and more than 10 d for orbital motion in a BBHS, while precession would produce longer periods, more likely longer than 1 yr.

The most famous example of long-term (quasi)periodic behaviour is that of the BL Lac object OJ~287. 
\citet{sillanpaa1988} reconstructed the source optical light curve back to 1890 and were able to recognise a quasi-periodicity of $\sim 11.65 \, \rm yr$ of the main outbursts. They proposed that this was due to a BBHS, where outbursts occur when the disc of the larger SMBH ($\sim 5 \times 10^9 \, M_\odot$) undergoes tidal action by the companion ($\sim 2 \times 10^7 \, M_\odot$) during the periastron passage, which induces mass flows from the accretion disc into the SMBH. They supported their model with numerical simulations.
The next outburst was predicted for 1994, and in order to follow the source behaviour in detail to confirm its occurrence, the ``OJ-94" Project was founded as an international consortium of optical observers.
The 1994 outburst was indeed observed \citep{sillanpaa1996a}, and its double-peak structure confirmed \citep{sillanpaa1996b}. The separation between the two peaks of each outburst is 1--2 yr.

\begin{figure}[ht]
\centering
\includegraphics[width=\textwidth]{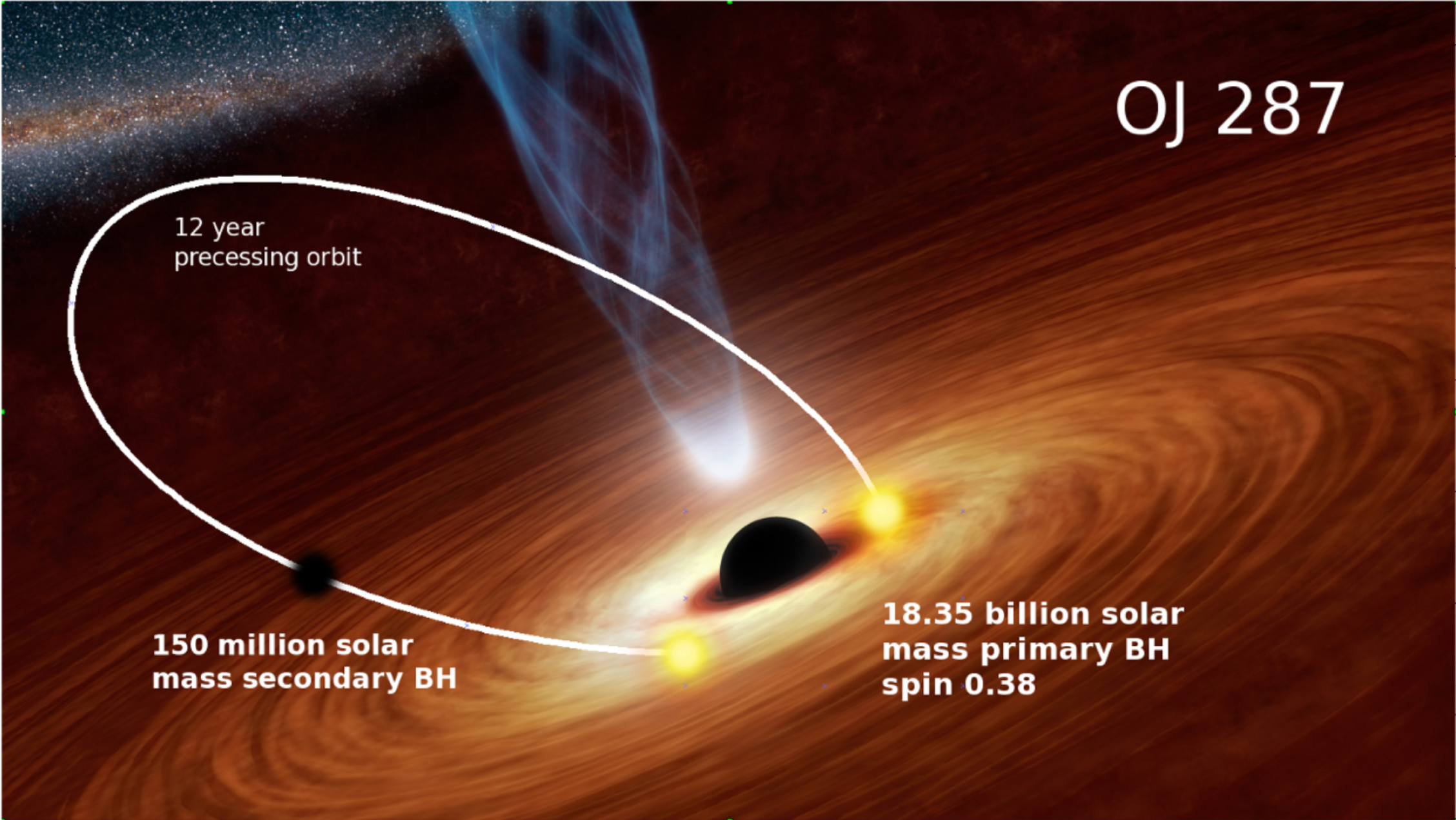}
\caption{The ``double impact" interpretation of the OJ~287 periodic outbursts \citep[see][and references therein]{valtonen2023}. In a BBHS, the secondary SMBH orbits around the primary SMBH on a precessing eccentric orbit with an observed period of $\sim 12 \, \rm yr$. The secondary impacts the accretion disc of the primary twice every orbit, causing double-peaked (quasi)periodic thermal outbursts. From \citet{dey2018}.}
\label{fig:oj287}
\end{figure}

According to other models, the secondary SMBH impacts the accretion disc of the primary SMBH, which is inclined with respect to the orbital plane.
Following one interpretation  \citep[][]{lehto1996,valtonen2023}, this happens twice every orbit, so that the two peaks of an outburst are thermal flares, and no counterparts at radio or X-rays are expected. However, matter ejected from the disc during the impact can end up in the jets and produce non-thermal flares.
A sketch of this model is shown in Fig.~\ref{fig:oj287}. 

In contrast, the analysis of the radio flux together with radio and optical polarisation data led \citet{valtaoja2000} to claim that while the first peak is due to a thermal flare, the second peak is produced by a synchrotron flare. They proposed that the impact occurs at the pericentre passage, leading to the optical thermal flare, which stimulates accretion into the primary SMBH and then the formation of shocks propagating down the jet, leading to a synchrotron flare about one year afterward. In this case, a radio counterpart and enhanced polarisation should be expected.

Furthermore, there is another class of models according to which the double-peaked outbursts of OJ~287 are the effect of variations in the Doppler beaming due to changes in the jet orientation.
In the model by \citet{villata1998}, both SMBHs in a BBHS have a jet that is precessing and is bent by the interaction with the ambient medium.
On the other side, the model by \citet{britzen2018} does not require a BBHS, and explains periodicity as due to jet precession, rotation and nutation. Moreover, these authors claim that the radio and optical variability shows a periodicity of $\sim 24 \, \rm yr$, which is roughly twice the classical period derived from the optical light curve.
Both these geometric models predict the occurrence of a nonthermal outburst when the jet emitting region reaches the minimum viewing angle.
In a recent paper, \citet{gopal-krishna2024} analysed the three major OJ~287 outbursts occurred in 1983, 2007 and 2015 using photo-polarimetric data and found inconsistencies with a thermal origin, favouring a synchrotron interpretation.


A $\sim 2.2 \, \rm yr$ quasi-periodicity was detected in the \textit{Fermi}-LAT $\gamma$-ray light curve of the BL Lac PG~1553+113, with some correlated behaviour in the optical 
\citep{ackermann2015}. 
An analysis of radio interferometric images at 15 GHz from the MOJAVE project led to the identification of seven superluminal jet components, the ejection time of which roughly corresponds to the epochs of the $\gamma$-ray flares \citep{caproni2017}. The fact that these components differ in both speed and direction of motion was interpreted as due to jet precession. However, \citet{lico2020} could not confirm periodicity in the radio band, but found evidence of jet wobbling in VLBA images at 15, 24 and 43 GHz.

\citet{sobacchi2017} explained both $\gamma$-ray and optical light curves, and optical--UV SED changes of PG~1553+113 through a BBHS, with one spine–sheath structured jet that precesses mostly because of orbital motion.
\citet{tavani2018} identified symmetric secondary peaks preceding and following the main periodic peaks in the $\gamma$-ray light curve, which would support the BBHS scenario with one or two jets.
\citet{gao2023} explained the periodicity in terms of a helical jet that rotates with constant angular velocity. They also noted an increase of the mean $\gamma$-ray brightness, which could possibly be due to an additional longer-term periodicity.

The periodicity of PG~1553+113 has recently been confirmed by \citet{abdollahi2024}, who identified 7 peaks repeating every $(2.1 \pm 0.2) \, \rm yr$ in the \textit{Fermi}-LAT $\gamma$-ray light curve covering 15 years. They found correlation with the optical flux as well as with the radio flux, with a radio delay of about 6 months. In contrast, no correlation was found with the X-ray flux. The authors discussed several possible astrophysical scenarios to explain the periodic behaviour.


Sinusoidal flux density variations were identified in the radio light curves of PKS $2131-021$ and $\rm J0805-0111$, with periods of $\sim 4.8 \, \rm yr$ and $\sim 3.6 \, \rm yr$, respectively \citep{kiehlmann2025,delaparra2025}. These sinusoidal oscillations are expected in case of a BBHS.

A mid-term quasi-periodicity of about one month was observed in the optical light curve of S4~0954+65 in 2019--2020, which was ascribed to changes in time in the pitch angle of a rotating inhomogeneous helical jet \citep{raiteri2021b}.

Periodic behaviour has been detected also on short time scales.
\citet{camenzind1992} proposed a lighthouse effect for the explanation of the intraday optical and radio variability observed in both FSRQs and BL Lac objects.
According to the lighthouse model, a rapid rotation of the SMBH implies a rapid rotation of emitting plasma blobs along the jet, producing quasi-periodic outbursts.

\begin{figure}[ht]
\centering
\includegraphics[width=\textwidth]{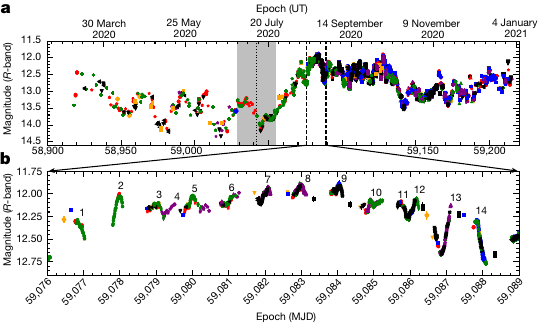}
\caption{a) Optical light curve of BL Lacertae from 2020 March 1 to December 31, showing a series of periodic oscillations during the first phase of the major outburst, enlarged in panel b). The vertical dotted line marks the ejection time of the superluminal knot, and the grey area indicates its uncertainty. Image reproduced with permission from \citet{jorstad2022}, copyright by the author(s)}
\label{fig:jorstad2022}
\end{figure}

Quasi-period oscillations (QPOs) with a time scale of $\sim 13 \, \rm h$ were observed in the optical and $\gamma$-ray light curves of BL Lacertae during an outburst in 2020 \citep{jorstad2022}. This transient quasi-periodicity was also recognised in the optical polarisation behaviour. The photometric and polarimetric information by the WEBT was complemented with VLBA images at 43 GHz at different epochs, which showed three stationary features and a superluminal knot travelling down the jet. The proposed interpretation was that the QPOs were produced by the growth of a kink instability \citep[see also][]{dong2020}, which was triggered when a perturbation (i.e.\ a shock) propagating off-axis crossed a standing recollimation shock. 
Fig.~\ref{fig:jorstad2022} shows the optical light curve analysed by \citet{jorstad2022} highlighting the quasi-periodic flux pulses.

\citet{espaillat2008} analysed the X-ray light curves of ten AGNs provided by \textit{XMM-Newton} using the wavelet technique. They found a QPO of slightly less than 1 hour in 3C~273, which was ascribed to oscillations in the accretion disk.

Besides the above-mentioned cases, a huge number of works have claimed the detection of QPOs in blazar light curves at various frequencies on a wide range of time scales. 
Sometimes the same object was found to exhibit different periodicities at different frequencies and/or different time intervals.
The analysis is often carried out on light curve segments, so, if real, these periodicities should likely come from transient phenomena.
However, as \citet{vaughan2016} demonstrated, stochastic processes characterised by red noise can produce false periodicities that last a few cycles.  Moreover,
\citet{covino2019} checked the reliability of some of the alleged periodicities 
in the \textit{Fermi}-LAT light curves and could not confirm any of them.
This calls for caution and requires the adoption of robust estimators of the period significance.

\subsection{Gravitational microlensing}

When we look at a distant source, a foreground galaxy may stand on the line of sight, and its stars (or other compact objects) can act as gravitational lenses and magnify the emission of the background source. This microlensing effect is important if the size of the source $R_S$ is less than the Einstein radius of the lensing star $R_E$:
\begin{equation}
R_E=\sqrt {\frac{4GM}{c^2} \, \frac{D_{\rm LS} D_{\rm OS}}{D_{\rm OL}}},
\end{equation}
where $G$ is the gravitational constant, $c$ is the speed of light, $M$ is the mass of the lens, $D_{\rm LS}$, $D_{\rm OS}$ and $D_{\rm OL}$ are the relative angular diameter distances between the source S, the lens L, and the observer O.
The relative motion of these three actors
causes variability, whose characteristic time scales are given by the Einstein radius crossing time $t_E=R_E/v$, and the source crossing time $t_S=R_S/v$, where $v$ is the effective transverse velocity of the source \citep{mosquera2011}. Microlensing events should typically produce symmetric flares in blazar light curves, since the duration of the rising and decaying phases is expected to be similar.

Magnification due to microlensing increases with decreasing size of the emitting region. Therefore, in the case of blazars, microlensing is expected to be chromatic if the jet radiation at different wavelengths comes from regions of different sizes. In particular, flux enhancement due to microlensing will be weaker for radio emission, which is likely coming from a jet region that is larger than those producing photons of higher frequencies. Microlensing will also affect less the flux of the broad emission lines, since the BLR is more extended than the jet emitting regions.
\citet{ostriker1985} proposed that a fraction of the order of 10\% of BL Lacs may actually be OVV quasars where the continuum emission is strongly amplified with respect to the line emission through microlensing by stars or black holes in intervening galaxies.

Gravitational microlensing by objects in a foreground galaxy was suggested as a possible explanation for the observed variability of some blazars, such as 0846+51W1 at $z=1.86$ \citep{nottale1986}, AO~0235+16 at $z=0.94$ \citep{stickel1988a}, and PKS~$0537-441$ at $z=0.894$ \citep{stickel1988b}.
\citet{gopal-krishna1991} showed that the effect of microlensing by stars in an intervening galaxy on superluminal knots propagating in the jet may produce IDV with time scales of about a hour in the optical and a day at radio centimetre wavelenghts. 

Microlensing was used to estimate the size of the $\gamma$-ray emitting region in the two spatially-resolved gravitationally-lensed blazars PKS~1830--211 \citep{neronov2015} and B0218+357 \citep{vovk2024}. In the former case, the analysis of flares in the \textit{Fermi}-LAT light curve led to the result that $\gamma$-ray photons at GeV energies come from a region of about $10^{15} \, \rm cm$ located within 10--100 Schwarzchild radii from the SMBH. In the latter case, by combining data from \textit{Fermi}-LAT and MAGIC, the authors constrained the dimension of the region to $10^{14}$--$10^{15} \, \rm cm$, and suggested a possible decrease of its size with increasing energy.

\section{Line variability}
\label{sec:lines}

In non-beamed AGN, variations in the continuum flux density are usually followed by variations in the flux of the broad emission lines. The time delay is ascribed to the light travel time from the source of the continuum radiation, i.e.\ the accretion disc, to the BLR zone where a given line is produced in response of excitation by the continuum photons. In this way, by measuring the time lag between the continuum and line variations, an estimate of the radius of the BLR can be inferred. 
Typically, the BLR radius is found to range from a few light days to several light months \citep[e.g][]{kaspi2000}. 

Under the assumption that the BLR is virialised, from the BLR radius and gas cloud velocity inferred from line broadening, an estimate of the SMBH mass can be derived. This technique is called ``Reverberation Mapping" 
\citep[e.g.][]{blandford1982_rm,kaspi2000}. 
The results of reverberation mapping have been used to derive scaling relations, which allow us to calculate the SMBH mass from single-epoch spectra using the line full-width-at-half-maximum (FWHM) together with line or continuum luminosity \citep[e.g.][and references therein]{shen2024}.

In the case of blazars, the continuum is usually dominated by jet emission, which is not the main source of ionising photons, but in principle may act as an additional photoionising radiation. 
If this were true, one should observe an enhancement of the line flux when the jet emission increases, which may even be simultaneous, if the jet dissipation zone is at the location of the BLR. 

 The literature on spectroscopic monitoring of blazars is not wide.
  In most cases, the variation of the line EW anticorrelates with the continuum flux, as expected if the BLR emission is not affected by jet radiation \citep[e.g.][]{corbett2000,carnerero2015,larionov2016a}. As mentioned in Sect.~\ref{sec:blatyp}, this can even cause a FSRQ to look like a BL Lac during outbursts.

   However, small but significant increases in the flux of broad emission lines were observed simultaneously with jet activity in 3C~454.3 \citep{leon2013,isler2013}. In particular, correlation between the line and the $\gamma$-ray flux would suggest that $\gamma$-rays are produced in the jet within the BLR, likely by an EC mechanism.

 Another interesting case is that of Ton~599 (4C~+29.45), which showed an increase in the flux of Mg\,II and Fe\,II emission lines several days after a flare observed in optical and $\gamma$-ray bands \citep{hallum2022}. The proposed scenario was that of polar BLR gas clouds ionised by jet radiation coming from regions located well upstream.

An increase in the broad emission line fluxes with brightness was also reported for 3C~279 by \citet{larionov2020}. This effect was found to be stronger for the Mg\,II red wing than for its blue wing, while the flux of the O\,II line remained essentially constant. The authors concluded that the variability of the red wing was probably due to infalling gas clouds located just outside the jet, at a distance of $\sim 0.6 \, \rm pc$ from the SMBH and within $\sim 10^{\circ}$ from the jet axis.

\section{Multiwavelength polarimetric behaviour}

One of the peculiar properties of blazars is that their emission is generally strongly polarised, and both the degree of polarisation ($P$) and the electric vector position angle (EVPA) can change on all time scales, as in the case of flux variability. This has been observed in the radio, optical, and near-infrared bands since the Sixties \citep[see e.g.][and references therein]{angel1980} and, more recently, also at X-rays (see Sect.~\ref{sec:ixpe}). 
High polarization has also been used as a tool to discover new blazars or confirm blazar candidates \citep{impey1988b,smith2007}.

Polarisation of blazar synchrotron radiation is produced by the relativistic motion of particles through magnetic fields. Therefore, polarimetry is expected to carry information on the magnetic field inside the jet emitting region. 
According to the theory of synchrotron radiation \citep{rybicki1979}, the linear polarisation degree in the case of radiating particles with a power-law energy distribution $N(E) \propto E^{-p}$, producing a power-law spectrum $F_\nu (\nu) \propto \nu^{-\alpha}$ with spectral index  $\alpha=(p-1)/2$, in a uniform magnetic field is
\begin{equation}
 P = (p + 1)/(p + 7/3)   = (\alpha+1)/(\alpha+5/3),
 \label{pola}
\end{equation}
which implies $P$ ranging from 60\% to 82\% when $p$ goes from 1 to 5 ($\alpha$ from 0 to 2).
Actually, the maximum $P$ observed is much lower,
    so there must be some mechanism that partly destroys the uniformity of the magnetic field\footnote{In the radio band also Faraday depolarization can decrease the observed polarization degree.}.
Moreover, one has to pay attention to the contributions by other unpolarised sources of radiation, such as the light of the host galaxy and the big blue bump (see Sect.~\ref{sec:blatyp}).
In this case, the degree of polarisation of the jet radiation $P_{\rm jet}$ can be found by correcting the observed value $P_{\rm obs}$ by the ``dilution" effect of the unpolarised emission contributions:
\begin{equation}
P_{\rm jet}=(P_{\rm obs} \times F_{\rm obs})/(F_{\rm obs}-F_{\rm unpol}),
\label{dilution}
\end{equation} 
where $F_{\rm obs}$ and $F_{\rm unpol}$ are the observed total flux density and the flux density of the unpolarised component, respectively. 
The contribution of the unpolarised component introduces a wavelength-dependence: starlight from the host galaxy dampens $P_{\rm obs}$ more towards the red, while the big blue bump decreases $P_{\rm obs}$ more towards the blue \citep{smith1988}. This wavelength-dependence can actually be used to separate the thermal from the non-thermal contributions in spectropolarimetric observations \citep[e.g][]{smith2016,otero2022}.
However, transient wavelength-dependent trends in polarisation have been detected in some BL Lacs in the optical and infrared bands, which cannot be explained by the contribution of unpolarised radiation, but  must correspond to intrinsic jet properties, suggesting the presence of two or more variable polarised components \citep{angel1980,smith2016}.


Rapid variations of $P$ on minute time scales have sometimes been detected in the optical band \citep{feigelson1986,covino2015}. Just as the radio flux varies more smoothly than the optical flux, usually this also occurs for the radio polarisation degree and EVPA.
However, rapid radio polarisation variability has also been detected \citep{quirrenbach1989b,bach2006a}.

In the radio-to-optical spectral range, HBLs are characterised by lower $P$ with respect to the values that LBL and FSRQs can reach \citep{jannuzi1994, angelakis2016,pushkarev2023}.
Moreover, BL Lacs, and in particular HBLs, tend to show a preferred EVPA, which suggests a main emission component with stable geometry or magnetic field 
\citep{jannuzi1994,hagen2002,angelakis2016,zobnina2023}. 
In general, $\gamma$-loud blazars, as well as blazars with stronger variability in the radio and optical bands, show higher polarisation in the optical band \citep{angelakis2016}.

 Sometimes changes in $P$ and EVPA appear to be correlated, or anticorrelated, with the flux behaviour. In contrast, in many cases the polarisation changes are found to be uncorrelated with flux, and this has been interpreted as the signature of a stochastic process, i.e.\ turbulence.

Some authors investigated the polarimetric signatures in the framework of jet models.
\citet{hughes1985} derived the polarimetric properties of jet emission in a shock-in-jet model, where a random magnetic field is compressed by the passage of a shock wave:
\begin{equation}
P \approx \frac {p+1}{p+7/3} \, \frac{(1-k^2) \cos^2\epsilon}{2-(1-k^2) \cos^2\epsilon}
\end{equation}
where $k<1$ represents the compression and $\epsilon$ is the viewing angle of the plane of compression in the rest frame, so that $\cos \epsilon = \sin \theta'$.
Polarimetric properties were also inferred for cylindrical relativistic jets characterised by helical magnetic fields with \citep{pariev2003} and without \citep{lyutikov2005} rotation.
The presence of helical magnetic fields has indeed been suggested on both theoretical and observational grounds.
The predictions of the above three models were compared in the context of a geometric interpretation of the long-term variability of the BL Lac object S4~0954+65 by \citet{raiteri2023b}, and allowed to reproduce the average behaviour of $P$ in time.

The shock-in-jet model by \citet{hughes1985} was able to explain the correlated behaviour of $F$ and $P$ during the radio outbursts of BL Lacertae in 1981--1982 and 1983. 
\citet{qian1991} interpreted the IDV of the radio flux density and polarisation observed in the FSRQ 0917+624, with anticorrelation between $F$ and $P$, in terms of a shock-in-jet model, where a shock moves across a jet characterised by small-scale inhomogeneities in the electron density and/or the magnetic field. 
According to the mildly wiggling jet model by \citet{gopal-krishna1992}, changes in the direction of propagation of a shock can lead to correlated variations of $F$ and $P$ when the viewing angle is greater than $1/\Gamma$, while they are anti-correlated for smaller $\theta$.
 The correlation between $P$ and $F$ during the 2006 outburst of AO~0235+16 was one of the source properties that were explained by \citet{hagen2008} in terms of the propagation of a transverse shock with variable plasma compression along a twisted jet.
 \citet{raiteri2013} showed that in a geometric model
 where the long-term flux variations are due to changes in the viewing angle, for a reasonable range of $\theta$ covered by the jet the relationship between $F$ and $P$ depends on the Lorentz factor of the emitting plasma. These quantities are correlated for high values of $\Gamma$, as found for the FSRQ 4C~38.41, and anti-correlated for low values of $\Gamma$, as in the case of BL Lacertae.
 
\subsection{Rotations of the polarisation angle}
As mentioned before, the EVPA can show wide rotations.
 One example was reported in \citet{ledden1979}. The authors monitored the source AO~0235+16 at 8 and 14.5 GHz at UMRAO and found a linear rotation of the EVPA during a flare. They suggested the presence of a rotating magnetic field structure.
  \citet{efimov2002} interpreted a large EVPA rotation in OJ~287 as due to a helical magnetic field in the jet.
It was shown that large swings in EVPA can be produced by the propagation of shocks in non-axisymmetric magnetic fields \citep{konigl1985}.
\citet{marscher2008} analysed the multiwavelength photometric and optical polarimetric behaviour of BL Lacertae together with high-resolution radio images. During a flare the degree of polarisation dropped to a minimum while the EVPA underwent a rotation of $240^{\circ}$. The authors proposed that this behaviour was the result of the propagation of a shock wave along a spiral path across the jet acceleration and collimation zone characterised by a helical magnetic field
(see Fig.~\ref{fig:marscher2008}). 
Numerical particle-in-cell (PIC) simulations show that large-amplitude EVPA swings can be produced by relativistic magnetic reconnection, 
 and can be correlated with multiwavelength flares \citep{zhang2022}.

\begin{figure}[htbp]
\centering
\includegraphics[width=0.9\textwidth]{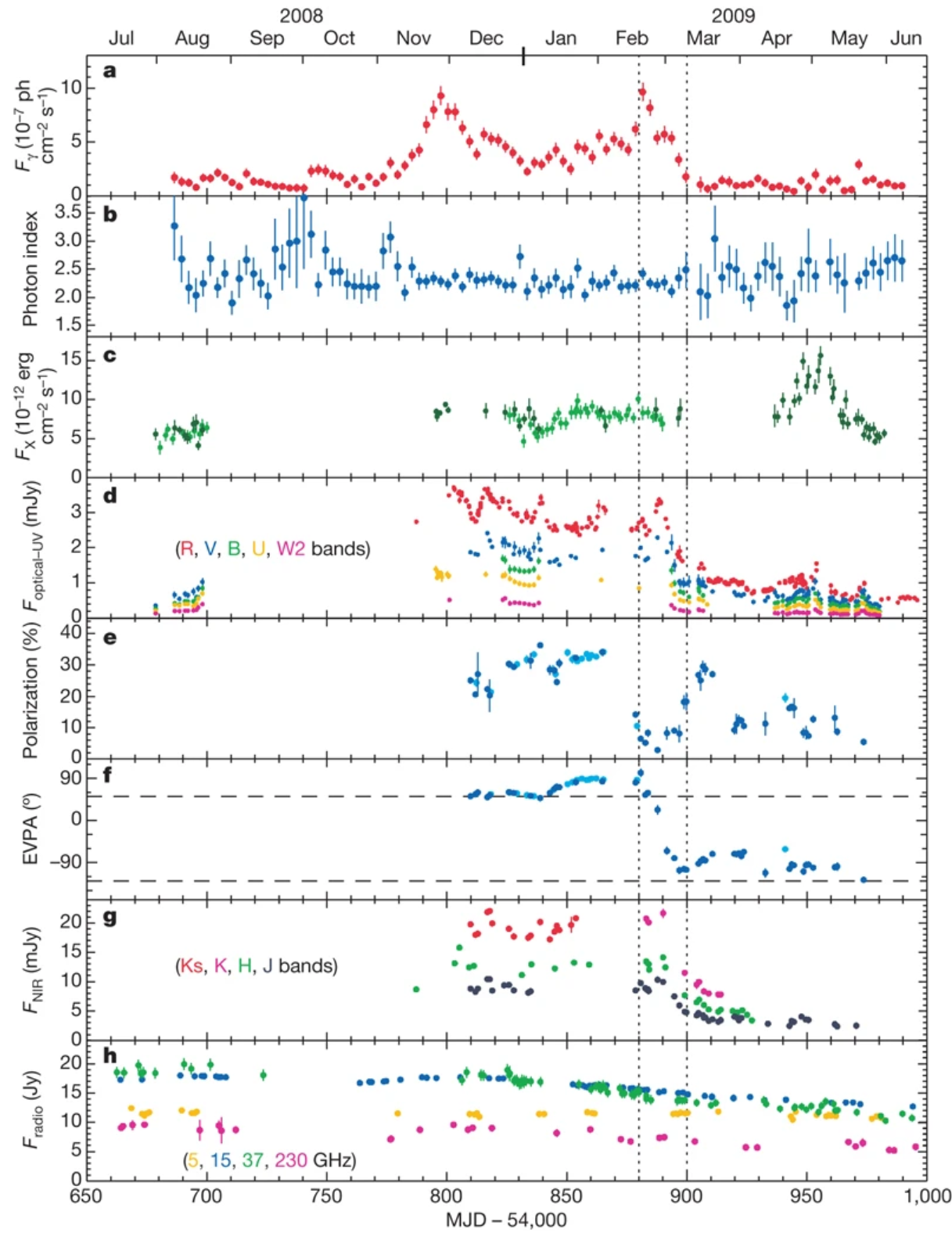}
\caption{Multiwavelength flux and polarimetric behaviour of the FSRQ~3C~279 in 2008--2009.
a, b, $\gamma$-ray flux and photon index measured by \textit{Fermi}-LAT. 
c, 2--10 keV X-ray flux from RXTE (light green) and \textit{Swift}-XRT (dark green).
d, Optical and UV flux densities in various bands from ground-based telescopes and \textit{Swift}-UVOT. 
e, f, Optical polarization degree and angle.
g, h, Near-IR and radio flux densities measured by ground-based telescopes.  
All UV, optical and near-IR data are corrected for Galactic absorption. From \citet{abdo2010}.}
\label{fig:abdo2010}
\end{figure}

Ample rotations of the optical EVPA have been observed simultaneously with $\gamma$-ray flares 
 \citep{abdo2010_nature,blinov2018}. 
  This suggests that the optical and $\gamma$-ray emission are cospatial, and favours a deterministic nature of the rotations, as due to the curved motion of emitting plasma in a jet with an ordered magnetic field.
  Fig.~\ref{fig:abdo2010} shows the multiwavelength and polarimetric behaviour of the FSRQ~3C~279 during a period of intense activity in 2008--2009 analysed by \citet{abdo2010_nature}, with a drop in the optical $P$ and an EVPA rotation at the time of a $\gamma$-ray flare.
  
However, in many cases wide EVPA rotations have been observed that are not correlated with flux and seem to have a stochastic nature due to turbulence \citep[e.g.][]{marscher2014,kiehlmann2017,raiteri2017_nature}.
The presence of turbulence in the jet emitting region would also explain EVPA rotations occurring both clockwise and anticlockwise in the same object.
Therefore, both deterministic and stochastic processes are likely to play a role in producing the observed EVPA rotations, with deterministic rotations likely occurring in a small jet region with a dominant helical magnetic field, and stochastic rotations probably produced in larger downstream jet regions with more turbulent magnetic field \citep{angelakis2016}. 

The point to be stressed is that the reconstruction of the EVPA rotation is sometimes tricky, as the EVPA measurement presents a $\pm \, n \times \pi$ uncertainty, so high-temporal coverage is needed to reconstruct the changes in EVPA in a reliable way. 
 A systematic shift in the temporal evolution of the normalised Stokes parameters $q$ and $u$ relative to each other was shown to be the signature of reliable EVPA rotations \citep{larionov2020}. 

\subsection{New results from IXPE}
\label{sec:ixpe}
Polarimetric information in the X-ray band became available with the launch of the space telescope Imaging X-ray Polarimetric Explorer (IXPE) in 2021. 
This opened the window to the possibility to test the proposed mechanisms for the production of high-energy photons and for the origin of the observed variability.

\subsubsection{HSP blazars}
Predictions on the X-ray polarimetric signatures of HBLs were derived by, e.g., \citet{krawczynski2012} and \citet{tavecchio2018}.
In HBL, the X-ray emission is synchrotron in nature and thus polarised, with the polarisation degree depending on the spectral index according to Eq.~\ref{pola} in case of uniform fields. Since the X-ray photons are emitted by the most energetic particles, which lose their energy faster than those emitting the optical radiation, their spectrum is softer (i.e., larger $\alpha$), and this would imply a higher $P$. Moreover, because of the fast cooling, the X-ray-emitting particles remain confined to a smaller jet region, with likely less turbulent magnetic field, and this also contributes to make $P_X$ larger than $P_{\rm optical}$ \citep{krawczynski2012}. 
In particular, if electrons are accelerated by shocks, the X-ray photons are expected to be emitted in a jet region close to the shock, where the compressed magnetic field has a high degree of ordering.
Low-energy photons would be emitted by electrons filling the post-shock region, which extends more and more downstream as the frequency decreases, and is characterized by an increasingly less ordered magnetic field \citep{angelakis2016,tavecchio2018}.
These models are commonly referred to as ``energy-stratified" models. They can also be applied to the synchrotron radiation of LSP blazars, when considering that the synchrotron peak in these objects does not occur in the X-ray band, but at infrared--optical frequencies, so it is at these frequencies that we can expect high polarisation, as observed. 

In line with the expectations, observations by IXPE found that in HBL the polarisation degree in X-rays is at least twice those at lower frequencies \citep{digesu2022,liodakis2022,middei2023b}. In the EHBL 1ES~0229+200, $P_X$ was observed to be more than seven times larger than $P_{\rm optical}$ \citep{ehlert2023}. Therefore, it seems that the higher the frequency of the synchrotron peak, the greater the ratio between $P_X$ and $P_{\rm optical}$, i.e.\  the chromaticity of the polarisation degree.

\citet{liodakis2022} analysed IXPE data on Mkn~501, and found an X-ray polarisation degree of $\approx 10\%$, which was about twice the host-corrected polarisation degree in the optical. The polarisation angle was parallel to the jet axis.
The authors favoured an energy-stratified model where particles are accelerated by shocks rather than by magnetic reconnection. 

As for the position angles, sometimes the X-ray EVPA was found to be in agreement with the low-frequency one \citep{liodakis2022,peirson2023}, while in other cases they were different \citep{ehlert2023}. 
Different EVPAs could further support the hypothesis of different production regions. However, the fact that the measurements are done at different epochs leaves the possibility of changes in the jet direction open.

In some cases, peculiar behaviour of the EVPA was found.
\citet{middei2023b} detected a smooth monotonic $\sim 125^{\circ}$ rotation of the optical EVPA in PG~1553+113, with no counterpart in the radio band or X-rays. This was interpreted as evidence that there should be a displacement in the corresponding emitting regions.

\begin{figure}[ht]
\centering
\includegraphics[width=0.9\textwidth]{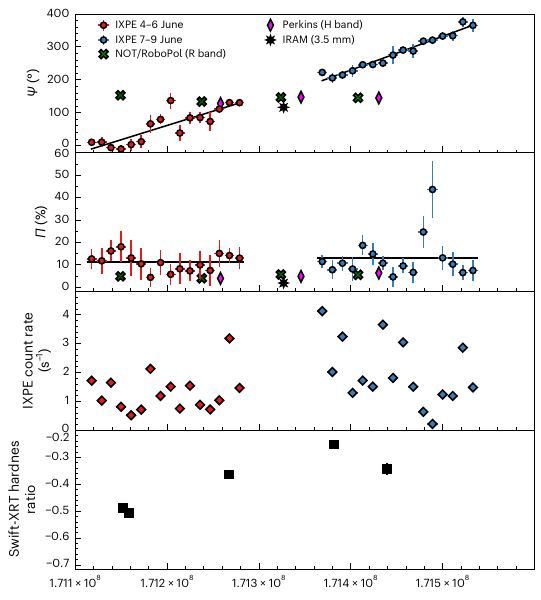}
\caption{Polarimetric behaviour of Mkn~421 during observations of IXPE in 2022.
From top to bottom: polarization angle, polarisation degree, IXPE photon count rate and \textit{Swift}-XRT hardness ratio as a function of time. Red and blue symbols refer to IXPE data acquired on June 4 and 7, respectively. \textit{Swift}-XRT hardness ratios are shown as black squares. Simultaneous radio-mm (black stars), near-IR (magenta diamonds) and optical (green crosses) polarisation data are shown. Image reproduced with permission from \citet{digesu2023}, copyright by the author(s).}
\label{fig:digesu2023}
\end{figure}

Vice versa, \citet{digesu2023} observed a rotation of about $360^{\circ}$ in the X-ray EVPA of Mkn~421, with nearly constant rotation rate, while the optical and near-infrared EVPAs did not change and were consistent with the radio one (see Fig.~\ref{fig:digesu2023}). The authors suggested that the X-rays were produced by an inner spine, while the optical photons came from a sheath surrounding it. 
The observed rotation would be caused by a localised shock propagating along the helical magnetic structure of the jet.

During IXPE observations of PKS~2155--304, \citet{kouch2024} detected a record value of $P_X=(30.7 \pm 2.0)\%$, which then dropped to half this value, while $P_{\rm optical}$ remained almost constant around 4\%.
In contrast, the optical EVPA rotated by about $50^{\circ}$ first in one direction and then in the opposite direction, while the X-ray EVPA remained stable. According to the authors, this behaviour would support the energy-stratified shock-acceleration scenario.

The amount of chromaticity of $P$ and the direction of EVPA in HBLs may also depend on the shape of the jet \citep{bolis2024a}. 

\subsubsection{LSP blazars}
Depending on the frequency of the synchrotron peak in the SED, the IXPE band in LSP blazars can still receive a minor contribution from the tail of the low-energy bump, which is synchrotron radiation with high polarisation, and a major contribution from the high-energy bump.
In the leptonic scenario, the high-energy emission contribution comes from inverse-Compton radiation, whose polarisation depends on the seed photons and physics of the emitting region. 

In the case of SSC, the polarisation degree of the synchrotron seed photons is reduced by scattering \citep{bonometto1973}, but $P_{\rm SSC}$ increases with the spectral index of the synchrotron radiation and with the angle between the direction of the magnetic field and the line of sight. It can reach values close to 50\% for $\alpha=1$ and a uniform and perpendicular magnetic field,
while it vanishes when the magnetic field is well aligned with the line of sight  \citep{krawczynski2012}. 
For inverse-Compton scattering on unpolarised seed photons (EC), $P_X$ is expected to be very low, likely smaller than 1\%. 
In the hadronic scenario, the high-energy contribution can come from synchrotron radiation from protons or charged secondary particles produced by hadronic cascades, which can be as highly polarised as the low-energy synchrotron component \citep{zhang2013}.
Unfortunately, this complex scenario makes a final statement on the nature of the high-energy emission difficult to achieve based on the $P_X$ value alone, since a low $P_X$ may indicate an EC process, but an SSC mechanism cannot be ruled out; likewise, a high $P_X$ can be the consequence of a hadronic process, but also of an SSC one.

\begin{figure}[ht]
\centering
\includegraphics[width=0.8\textwidth]{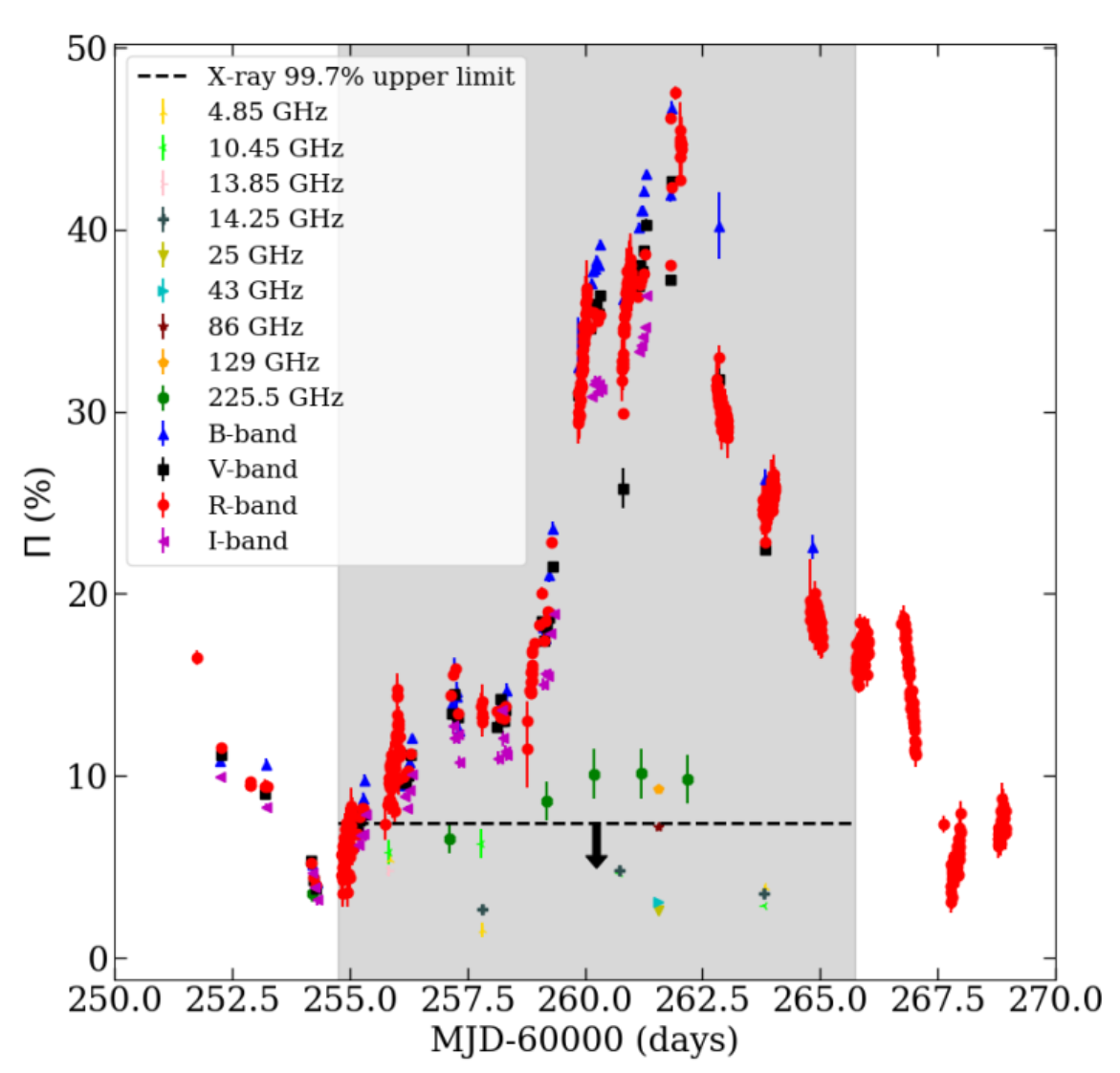}
\caption{Polarisation degree of BL Lacertae during the November 2023 optical polarisation exceptional event.
Different symbols and colours refer to different radio and optical bands. 
The grey shaded area indicates the time window of the IXPE observation.
The polarisation degree reached $\approx 48\%$ in optical and $\approx 10\%$ in radio, while in X-rays only an upper limit of 7.4\% was derived.  Image reproduced with permission from \citet{agudo2025}, copyright by the author(s).}
\label{fig:agudo2025}
\end{figure}

IXPE observations of BL Lacertae in May and July 2022, during a low state, yielded only upper limits $P_X \sim 14.2\%$ and $P_X \sim 12.6\%$, respectively, while the corresponding optical polarisation degrees were 6.8\% and 14.2\% \citep{middei2023a}.
Further IXPE observations of BL Lacertae in November 2022 during an outburst phase found $P_X \sim 22\%$, while $P_{\rm optical}$ varied in the range 10--18\%. The X-ray and optical EVPA were similar \citep{peirson2023}.
An exceptional event was observed in November 2023, when the polarisation degree of BL Lacertae reached $\approx 48\%$ in the optical band and $\approx 10\%$ in the mm band, but remained below the upper limit of 7.4\% in the X-rays \citep[][see Fig.~\ref{fig:agudo2025}]{agudo2025}.
These contrasting results are in line with the intense spectral variability shown by BL Lacertae, in particular in the X-ray domain (see Fig.~\ref{fig:seds}).

\citet{kouch2025} were able to put only an upper limit of $P_X \sim 10\%$ from the IXPE observations of S4~0954+65 during an optical and X-ray flare, when the optical polarisation degree was about 14\%.

Overall, the above results seem to favour inverse-Compton scattering as the dominant process for the origin of the high-energy emission.


\newpage
\section{Conclusions}
Notwithstanding decades of studies on blazars, these are still elusive sources.
A series of open questions, which are more or less connected to variability, are listed below:

\begin{itemize}

\item How many blazars are there in the sky?
We know a few thousand confirmed blazars and several tens of thousands of blazar candidates selected in different ways. Only a handful of high-redshift blazars have been detected so far. Deep monitoring surveys such as the incoming Rubin-LSST will help us to investigate blazar census also through variability, more efficiently with the help of machine learning techniques.

\item Are FSRQs and BL Lacs intrinsically different? The classical definition based on the EW of the emission lines makes FSRQs in outburst look like BL Lacs, and sometimes BL Lacs in faint states appear as FSRQs, including BL Lacertae itself. A more physical explanation of the divide between the two blazar types has been discussed several times in terms of accretion efficiency. However, the lack of a redshift determination for many BL Lacs prevents a robust physical interpretation of the blazar sequence. 
From the point of view of variability, there seems to be no compelling evidence that the two blazar types are different, once the jet emission is separated from the other possible contributions by the host galaxy, accretion disc and BLR, which affect the flux, spectral, and polarimetric behaviour in the infrared-to-UV frequency range.

\item What is the location of the emitting zone(s)? We know that most of the blazar emission comes from a very compact region inside the unresolved radio core, and this justified the one-zone models that were developed to interpret blazar variability. The distance of the dissipation zone from the SMBH is hard to estimate and may even vary in the same source at different epochs. In particular, the location with respect to the BLR or dusty torus can affect the nature of the seed photons for the production of high-energy photons through the EC process. 
Moreover, there is increasing evidence that blazar emission comes from multiple zones, with higher-frequency synchrotron (and their inverse-Compton) radiation being produced upstream with respect to the longer wavelengths. These models, where the position of the emitting zone in the jet is frequency-dependent, are called ``inhomogeneous" or ``energy-stratified". The difference between the two is that in inhomogeneous models the frequency-dependence is due to opacity reasons and occurs also for steady flows. In the case of energy-stratified models, the frequency-dependence is due to the energy-dependent cooling following particle acceleration.

\item What is the origin of the high-energy emission?
The mostly accepted scenario still involves a leptonic inverse-Compton origin, with an SSC process more likely in BL Lacs and an EC mechanisms required for FSRQs. Hadronic processes are needed to explain the production of high-energy neutrinos, if blazars really contribute to the population of cosmic neutrinos detected by IceCube (see below). Hadronic processes may also explain the uncorrelated variability between low- and high-energy emission.
However, results coming from the study of the particle content of blazar jets, of the correlation between the variability at low and high energies, and of the polarisation behaviour in different bands, seem to favour an inverse-Compton process as the typical mechanism to produce the high-energy photons.

\item Are blazars important sites for the production of high-energy neutrinos? The evidence for blazars as high-energy neutrino emitters is based on a few cases only, where neutrino detections were claimed to be strongly connected with blazar activity. However, it is not clear what type of blazars are the most likely neutrino emitters, and what would be the signature of neutrino emission in the multiwavelength light curves. LSP sources may be the best candidates, and $\gamma$-ray flares that are not necessarily correlated with low-energy ones may accompany neutrino emission.


\item What is the main particle acceleration mechanism? Many interpretations of blazar variability have been based on shocks travelling down the jet. Energy-stratified shock-acceleration models seem to be favoured by IXPE observations. However, magnetic reconnection can be a more efficient mechanism to accelerate particles and has been invoked to explain the observed blazar behaviour in several occasions. Turbulence is a necessary ingredient to account for fast erratic flux and polarization variability. 

\item How important are microlensing events to explain blazar flares? The stars (or other objects) of foreground galaxies on the line of sight towards distant blazars can produce microlensing effects that translate into observed flares, whose amplitude and duration depend on the mass of the lensing object, on the size of the emitting region, and on the geometric configuration. In principle, microlensing events should appear symmetric in the light curves, i.e~with similar duration of the rising and decaying phases, in contrast to intrinsic variability episodes, which are usually characterised by fast rise and slow decay. They should also be stronger for smaller emitting regions, so that we would expect faster and more pronounced flares in the optical than in the radio band. Indications of this phenomenon have been found in a few cases.

\item How important are orientation changes to explain blazar variability? 
There is robust observing evidence and theoretical support that blazar jets are bent, and that they twist in both space and time. The reasons can be both intrinsic, due to plasma instabilities developing in the jet, or extrinsic, due to jet rotation, precession, or orbital motion in a BBHS. In any case, changes in space in the viewing angle of the emitting regions along a curved jet imply differential Doppler boosting of the corresponding radiation observed in the various bands. If the orientation also changes in time, this will produce variability at all wavelengths. 

\item What is the fine structure of blazar jets? Is the jet made up of interlaced filaments like those observed in M87 and some blazars? Thin filaments would explain the small upper limits to the size of the jet emitting regions inferred from the very fast variability time scales at various wavelengths better than a full jet.

\item Are blazar light curves really periodic? Like the unbeamed AGNs hosted in massive elliptical galaxies, we would expect blazars to be powered by a BBHS and hence to show (quasi)periodic behaviour. Periodic behaviour would also be the result of jet rotation and precession. Yet, while there are countless claims of detected QPOs in blazar light curves at all frequencies and on all time scales, and sometimes multiple periods are found even in the same source, only a few blazars have shown persistent periodicities. 


\item What role can the jet play in photoionising the BLR? 
This should depend on the position of the dissipation zone in the jet with respect to the location of the BLR, but also on the BLR structure. 
The flux of the broad emission lines is generally seen not to respond to variations in the jet continuum, which is compatible with the idea that the lines are photoionised by the accretion disc radiation. However, in a few cases an increase in the line strength was observed during an outburst, suggesting a contribution by the jet.


\end{itemize}

\bmhead{Acknowledgements}

I am grateful to Massimo Villata and to the two anonymous referees for their careful reading of the manuscript and useful comments.
I thank the librarian of the INAF-Osservatorio Astrofisico di Torino, Tullia Carriero, for her help in finding articles that I did not have access to online.
I acknowledge financial support from the INAF Fundamental Research Funding Call 2023.
This research has made use of the Astrophysics Data System, funded by NASA under Cooperative Agreement 80NSSC21M00561.
Figure \ref{fig:seds} is based on archival data
provided by the Space Science Data Center - ASI.


\phantomsection
\addcontentsline{toc}{section}{References}
\bibliography{vables} 

@article{ice2018b,
	author = {Aartsen, Mark and Ackermann, Markus and Adams, Jenni and Aguilar, Juan Antonio and Ahlers, Markus and Ahrens, Maryon and Al Samarai, Imen and Altmann, David and Andeen, Karen and Anderson, Tyler and Ansseau, Isabelle and Anton, Gisela and Arg{\"u}elles, Carlos and Arsioli, Bruno and Auffenberg, Jan and Axani, Spencer and Bagherpour, Hadis and Bai, Xinhua and Barron, Jared and Barwick, Steve and Baum, Volker and Bay, Ryan and Beatty, James and Becker, Karl Heinz and Becker Tjus, Julia and BenZvi, Segev and Berley, David and Bernardini, Elisa and et al.},
	title = {Neutrino emission from the direction of the blazar TXS 0506+056 prior to the IceCube-170922A alert},
	volume = {361},
	number = {6398},
	pages = {147--151},
	year = {2018},
	doi = {10.1126/science.aat2890},
	journal = {Science}
}

@article{ice2018a,
	author = {Aartsen, Mark and Ackermann, Markus and Adams, Jenni and Aguilar, Juan Antonio and Ahlers, Markus and Ahrens, Maryon and Al Samarai, Imen and Altmann, David and Andeen, Karen and Anderson, Tyler and Ansseau, Isabelle and Anton, Gisela and Arg{\"u}elles, Carlos and Auffenberg, Jan and Axani, Spencer and Bagherpour, Hadis and Bai, Xinhua and Barron, Jared and Barwick, Steve and Baum, Volker and Bay, Ryan and Beatty, James and Becker, Karl Heinz and Tjus, Julia and BenZvi, Segev and Berley, David and Bernardini, Elisa and et al.},
	title = {Multimessenger observations of a flaring blazar coincident with high-energy neutrino IceCube-170922A},
	volume = {361},
	number = {6398},
	pages = {eaat1378},
	year = {2018},
	doi = {10.1126/science.aat1378},
	journal = {Science}
}

@ARTICLE{abdo2010_nature,
       author = {{Abdo}, A.~A. and {Ackermann}, M. and {Ajello}, M. and {Axelsson}, M. and {Baldini}, L. and {Ballet}, J. and {Barbiellini}, G. and {Bastieri}, D. and {Baughman}, B.~M. and {Bechtol}, K. and {Bellazzini}, R. and {Berenji}, B. and {Blandford}, R.~D. and {Bloom}, E.~D. and {Bock}, D.~C. -J. and {Bogart}, J.~R. and {Bonamente}, E. and {Borgland}, A.~W. and {Bouvier}, A. and {Bregeon}, J. and {Brez}, A. and {Brigida}, M. and {Bruel}, P. and {Burnett}, T.~H. and {Buson}, S. and {Caliandro}, G.~A. and {Cameron}, R.~A. and {Caraveo}, P.~A. and {Casandjian}, J.~M. and {Cavazzuti}, E. and {Cecchi}, C. and {{\c{C}}elik}, {\"O}. and {Chekhtman}, A. and {Cheung}, C.~C. and {Chiang}, J. and {Ciprini}, S. and {Claus}, R. and {Cohen-Tanugi}, J. and {Collmar}, W. and {Cominsky}, L.~R. and {Conrad}, J. and {Corbel}, S. and {Corbet}, R. and {Costamante}, L. and {Cutini}, S. and {Dermer}, C.~D. and {de Angelis}, A. and {de Palma}, F. and {Digel}, S.~W. and {Do Couto E Silva}, E. and {Drell}, P.~S. and {Dubois}, R. and {Dumora}, D. and {Farnier}, C. and {Favuzzi}, C. and {Fegan}, S.~J. and {Ferrara}, E.~C. and {Focke}, W.~B. and {Fortin}, P. and {Frailis}, M. and {Fuhrmann}, L. and {Fukazawa}, Y. and {Funk}, S. and {Fusco}, P. and {Gargano}, F. and {Gasparrini}, D. and {Gehrels}, N. and {Germani}, S. and {Giebels}, B. and {Giglietto}, N. and {Giommi}, P. and {Giordano}, F. and {Giroletti}, M. and {Glanzman}, T. and {Godfrey}, G. and {Grenier}, I.~A. and {Grove}, J.~E. and {Guillemot}, L. and {Guiriec}, S. and {Hanabata}, Y. and {Harding}, A.~K. and {Hayashida}, M. and {Hays}, E. and {Horan}, D. and {Hughes}, R.~E. and {Iafrate}, G. and {Itoh}, R. and {Jackson}, M.~S. and {J{\'o}hannesson}, G. and {Johnson}, A.~S. and {Johnson}, W.~N. and {Kadler}, M. and {Kamae}, T. and {Katagiri}, H. and {Kataoka}, J. and {Kawai}, N. and {Kerr}, M. and {Kn{\"o}dlseder}, J. and {Kocian}, M.~L. and {Kuss}, M. and {Lande}, J. and {Larsson}, S. and {Latronico}, L. and {Lemoine-Goumard}, M. and {Longo}, F. and {Loparco}, F. and {Lott}, B. and {Lovellette}, M.~N. and {Lubrano}, P. and {Macquart}, J. and {Madejski}, G.~M. and {Makeev}, A. and {Max-Moerbeck}, W. and {Mazziotta}, M.~N. and {McConville}, W. and {McEnery}, J.~E. and {McGlynn}, S. and {Meurer}, C. and {Michelson}, P.~F. and {Mitthumsiri}, W. and {Mizuno}, T. and {Moiseev}, A.~A. and {Monte}, C. and {Monzani}, M.~E. and {Morselli}, A. and {Moskalenko}, I.~V. and {Murgia}, S. and {Nestoras}, I. and {Nolan}, P.~L. and {Norris}, J.~P. and {Nuss}, E. and {Ohsugi}, T. and {Okumura}, A. and {Omodei}, N. and {Orlando}, E. and {Ormes}, J.~F. and {Paneque}, D. and {Panetta}, J.~H. and {Parent}, D. and {Pavlidou}, V. and {Pearson}, T.~J. and {Pelassa}, V. and {Pepe}, M. and {Pesce-Rollins}, M. and {Piron}, F. and {Porter}, T.~A. and {Rain{\`o}}, S. and {Rando}, R. and {Razzano}, M. and {Readhead}, A. and {Reimer}, A. and {Reimer}, O. and {Reposeur}, T. and {Reyes}, L.~C. and {Richards}, J.~L. and {Rochester}, L.~S. and {Rodriguez}, A.~Y. and {Roth}, M. and {Ryde}, F. and {Sadrozinski}, H.~F. -W. and {Sanchez}, D. and {Sander}, A. and {Saz Parkinson}, P.~M. and {Scargle}, J.~D. and {Sgr{\`o}}, C. and {Shaw}, M.~S. and {Shrader}, C. and {Siskind}, E.~J. and {Smith}, D.~A. and {Smith}, P.~D. and {Spandre}, G. and {Spinelli}, P. and {Stawarz}, L. and {Stevenson}, M. and {Strickman}, M.~S. and {Suson}, D.~J. and {Tajima}, H. and {Takahashi}, H. and {Takahashi}, T. and {Tanaka}, T. and {Taylor}, G.~B. and {Thayer}, J.~B. and {Thayer}, J.~G. and {Thompson}, D.~J. and {Tibaldo}, L. and {Torres}, D.~F. and {Tosti}, G. and {Tramacere}, A. and {Uchiyama}, Y. and {Usher}, T.~L. and {Vasileiou}, V. and {Vilchez}, N. and {Vitale}, V. and {Waite}, A.~P. and {Wang}, P. and {Wehrle}, A.~E. and {Winer}, B.~L. and {Wood}, K.~S. and {Ylinen}, T. and {Zensus}, J.~A. and {Uemura}, M. and {Ikejiri}, Y. and {Kawabata}, K.~S. and {Kino}, M. and {Sakimoto}, K. and {Sasada}, M. and {Sato}, S. and {Yamanaka}, M. and {Villata}, M. and {Raiteri}, C.~M. and {Agudo}, I. and {Aller}, H.~D. and {Aller}, M.~F. and {Angelakis}, E. and {Arkharov}, A.~A. and {Bach}, U. and {Ben{\'\i}tez}, E. and {Berdyugin}, A. and {Blinov}, D.~A. and {Boettcher}, M. and {Buemi}, C.~S. and {Chen}, W.~P. and {Dolci}, M. and {Dultzin}, D. and {Efimova}, N.~V. and {Gurwell}, M.~A. and {Gusbar}, C. and {G{\'o}mez}, J.~L. and {Heidt}, J. and {Hiriart}, D. and {Hovatta}, T. and {Jorstad}, S.~G. and {Konstantinova}, T.~S. and {Kopatskaya}, E.~N. and {Koptelova}, E. and {Kurtanidze}, O.~M. and {Lahteenmaki}, A. and {Larionov}, V.~M. and {Larionova}, E.~G. and {Leto}, P. and {Lin}, H.~C. and {Lindfors}, E. and {Marscher}, A.~P. and {McHardy}, I.~M. and {Melnichuk}, D.~A. and {Mommert}, M. and {Nilsson}, K. and {di Paola}, A. and {Reinthal}, R. and {Richter}, G.~M. and {Roca-Sogorb}, M. and {Roustazadeh}, P. and {Sigua}, L.~A. and {Takalo}, L.~O. and {Tornikoski}, M. and {Trigilio}, C. and {Troitsky}, I.~S. and {Umana}, G. and {Villforth}, C. and {Grainge}, K. and {Moderski}, R. and {Nalewajko}, K. and {Sikora}, M. and {Fermi LAT Collaboration} and {Members of the 3C Multi-Band Campaign}},
        title = "{A change in the optical polarization associated with a {\ensuremath{\gamma}}-ray flare in the blazar 3C279}",
      journal = {\nat},
         year = 2010,
        month = feb,
       volume = {463},
       number = {7283},
        pages = {919-923},
          doi = {10.1038/nature08841},
archivePrefix = {arXiv},
       eprint = {1004.3828},
 primaryClass = {astro-ph.CO},
       adsurl = {https://ui.adsabs.harvard.edu/abs/2010Natur.463..919A}
}

@ARTICLE{abdo2010,
       author = {{Abdo}, A.~A. and {Ackermann}, M. and {Agudo}, I. and {Ajello}, M. and {Allafort}, A. and {Aller}, H.~D. and {Aller}, M.~F. and {Antolini}, E. and {Arkharov}, A.~A. and {Axelsson}, M. and {Bach}, U. and {Baldini}, L. and {Ballet}, J. and {Barbiellini}, G. and {Bastieri}, D. and {Bechtol}, K. and {Bellazzini}, R. and {Berdyugin}, A. and {Berenji}, B. and {Blandford}, R.~D. and {Blinov}, D.~A. and {Bloom}, E.~D. and {Boettcher}, M. and {Bonamente}, E. and {Borgland}, A.~W. and {Bouvier}, A. and {Bregeon}, J. and {Brez}, A. and {Brigida}, M. and {Bruel}, P. and {Buehler}, R. and {Buemi}, C.~S. and {Burnett}, T.~H. and {Buson}, S. and {Caliandro}, G.~A. and {Cameron}, R.~A. and {Caraveo}, P.~A. and {Carosati}, D. and {Carrigan}, S. and {Casandjian}, J.~M. and {Cavazzuti}, E. and {Cecchi}, C. and {{\c{C}}elik}, {\"O}. and {Chekhtman}, A. and {Chen}, W.~P. and {Cheung}, C.~C. and {Chiang}, J. and {Ciprini}, S. and {Claus}, R. and {Cohen-Tanugi}, J. and {Conrad}, J. and {Corbel}, S. and {Costamante}, L. and {Dermer}, C.~D. and {de Angelis}, A. and {de Palma}, F. and {Donato}, D. and {Silva}, E. do Couto e. and {Drell}, P.~S. and {Dubois}, R. and {Dumora}, D. and {Farnier}, C. and {Favuzzi}, C. and {Fegan}, S.~J. and {Ferrara}, E.~C. and {Focke}, W.~B. and {Forn{\'e}}, E. and {Fortin}, P. and {Fukazawa}, Y. and {Funk}, S. and {Fusco}, P. and {Gargano}, F. and {Gasparrini}, D. and {Gehrels}, N. and {Germani}, S. and {Giebels}, B. and {Giglietto}, N. and {Giordano}, F. and {Giroletti}, M. and {Glanzman}, T. and {Godfrey}, G. and {Grenier}, I.~A. and {Grove}, J.~E. and {Guiriec}, S. and {Gurwell}, M.~A. and {Gusbar}, C. and {G{\'o}mez}, J.~L. and {Hadasch}, D. and {Hagen-Thorn}, V.~A. and {Hayashida}, M. and {Hays}, E. and {Horan}, D. and {Hughes}, R.~E. and {J{\'o}hannesson}, G. and {Johnson}, A.~S. and {Johnson}, W.~N. and {Kamae}, T. and {Katagiri}, H. and {Kataoka}, J. and {Kawai}, N. and {Kimeridze}, G. and {Kn{\"o}dlseder}, J. and {Konstantinova}, T.~S. and {Kopatskaya}, E.~N. and {Koptelova}, E. and {Kovalev}, Y.~Y. and {Kurtanidze}, O.~M. and {Kuss}, M. and {Lahteenmaki}, A. and {Lande}, J. and {Larionov}, V.~M. and {Larionova}, E.~G. and {Larionova}, L.~V. and {Larsson}, S. and {Latronico}, L. and {Lee}, S. -H. and {Leto}, P. and {Lister}, M.~L. and {Longo}, F. and {Loparco}, F. and {Lott}, B. and {Lovellette}, M.~N. and {Lubrano}, P. and {Madejski}, G.~M. and {Makeev}, A. and {Massaro}, E. and {Mazziotta}, M.~N. and {McConville}, W. and {McEnery}, J.~E. and {McHardy}, I.~M. and {Michelson}, P.~F. and {Mitthumsiri}, W. and {Mizuno}, T. and {Moiseev}, A.~A. and {Monte}, C. and {Monzani}, M.~E. and {Morozova}, D.~A. and {Morselli}, A. and {Moskalenko}, I.~V. and {Murgia}, S. and {Naumann-Godo}, M. and {Nikolashvili}, M.~G. and {Nolan}, P.~L. and {Norris}, J.~P. and {Nuss}, E. and {Ohno}, M. and {Ohsugi}, T. and {Okumura}, A. and {Omodei}, N. and {Orlando}, E. and {Ormes}, J.~F. and {Ozaki}, M. and {Paneque}, D. and {Panetta}, J.~H. and {Parent}, D. and {Pasanen}, M. and {Pelassa}, V. and {Pepe}, M. and {Pesce-Rollins}, M. and {Piron}, F. and {Porter}, T.~A. and {Pushkarev}, A.~B. and {Rain{\`o}}, S. and {Raiteri}, C.~M. and {Rando}, R. and {Razzano}, M. and {Reimer}, A. and {Reimer}, O. and {Reinthal}, R. and {Ripken}, J. and {Ritz}, S. and {Roca-Sogorb}, M. and {Rodriguez}, A.~Y. and {Roth}, M. and {Roustazadeh}, P. and {Ryde}, F. and {Sadrozinski}, H.~F. -W. and {Sander}, A. and {Scargle}, J.~D. and {Sgr{\`o}}, C. and {Sigua}, L.~A. and {Smith}, P.~D. and {Sokolovsky}, K. and {Spandre}, G. and {Spinelli}, P. and {Starck}, J. -L. and {Strickman}, M.~S. and {Suson}, D.~J. and {Takahashi}, H. and {Takahashi}, T. and {Takalo}, L.~O. and {Tanaka}, T. and {Taylor}, B. and {Thayer}, J.~B. and {Thayer}, J.~G. and {Thompson}, D.~J. and {Tibaldo}, L. and {Tornikoski}, M. and {Torres}, D.~F. and {Tosti}, G. and {Tramacere}, A. and {Trigilio}, C. and {Troitsky}, I.~S. and {Umana}, G. and {Usher}, T.~L. and {Vandenbroucke}, J. and {Vasileiou}, V. and {Vilchez}, N. and {Villata}, M. and {Vitale}, V. and {Waite}, A.~P. and {Wang}, P. and {Winer}, B.~L. and {Wood}, K.~S. and {Yang}, Z. and {Ylinen}, T. and {Ziegler}, M.},
        title = "{Fermi Large Area Telescope and Multi-wavelength Observations of the Flaring Activity of PKS 1510-089 between 2008 September and 2009 June}",
      journal = {\apj},
         year = 2010,
        month = oct,
       volume = {721},
       number = {2},
        pages = {1425-1447},
          doi = {10.1088/0004-637X/721/2/1425},
archivePrefix = {arXiv},
       eprint = {1007.1237},
 primaryClass = {astro-ph.CO},
       adsurl = {https://ui.adsabs.harvard.edu/abs/2010ApJ...721.1425A}
}

@ARTICLE{abdo2010_sed,
       author = {{Abdo}, A.~A. and {Ackermann}, M. and {Agudo}, I. and {Ajello}, M. and {Aller}, H.~D. and {Aller}, M.~F. and {Angelakis}, E. and {Arkharov}, A.~A. and {Axelsson}, M. and {Bach}, U. and {Baldini}, L. and {Ballet}, J. and {Barbiellini}, G. and {Bastieri}, D. and {Baughman}, B.~M. and {Bechtol}, K. and {Bellazzini}, R. and {Benitez}, E. and {Berdyugin}, A. and {Berenji}, B. and {Blandford}, R.~D. and {Bloom}, E.~D. and {Boettcher}, M. and {Bonamente}, E. and {Borgland}, A.~W. and {Bregeon}, J. and {Brez}, A. and {Brigida}, M. and {Bruel}, P. and {Burnett}, T.~H. and {Burrows}, D. and {Buson}, S. and {Caliandro}, G.~A. and {Calzoletti}, L. and {Cameron}, R.~A. and {Capalbi}, M. and {Caraveo}, P.~A. and {Carosati}, D. and {Casandjian}, J.~M. and {Cavazzuti}, E. and {Cecchi}, C. and {{\c{C}}elik}, {\"O}. and {Charles}, E. and {Chaty}, S. and {Chekhtman}, A. and {Chen}, W.~P. and {Chiang}, J. and {Chincarini}, G. and {Ciprini}, S. and {Claus}, R. and {Cohen-Tanugi}, J. and {Colafrancesco}, S. and {Cominsky}, L.~R. and {Conrad}, J. and {Costamante}, L. and {Cutini}, S. and {D'ammando}, F. and {Deitrick}, R. and {D'Elia}, V. and {Dermer}, C.~D. and {de Angelis}, A. and {de Palma}, F. and {Digel}, S.~W. and {Donnarumma}, I. and {Silva}, E. do Couto e. and {Drell}, P.~S. and {Dubois}, R. and {Dultzin}, D. and {Dumora}, D. and {Falcone}, A. and {Farnier}, C. and {Favuzzi}, C. and {Fegan}, S.~J. and {Focke}, W.~B. and {Forn{\'e}}, E. and {Fortin}, P. and {Frailis}, M. and {Fuhrmann}, L. and {Fukazawa}, Y. and {Funk}, S. and {Fusco}, P. and {G{\'o}mez}, J.~L. and {Gargano}, F. and {Gasparrini}, D. and {Gehrels}, N. and {Germani}, S. and {Giebels}, B. and {Giglietto}, N. and {Giommi}, P. and {Giordano}, F. and {Giuliani}, A. and {Glanzman}, T. and {Godfrey}, G. and {Grenier}, I.~A. and {Gronwall}, C. and {Grove}, J.~E. and {Guillemot}, L. and {Guiriec}, S. and {Gurwell}, M.~A. and {Hadasch}, D. and {Hanabata}, Y. and {Harding}, A.~K. and {Hayashida}, M. and {Hays}, E. and {Healey}, S.~E. and {Heidt}, J. and {Hiriart}, D. and {Horan}, D. and {Hoversten}, E.~A. and {Hughes}, R.~E. and {Itoh}, R. and {Jackson}, M.~S. and {J{\'o}hannesson}, G. and {Johnson}, A.~S. and {Johnson}, W.~N. and {Jorstad}, S.~G. and {Kadler}, M. and {Kamae}, T. and {Katagiri}, H. and {Kataoka}, J. and {Kawai}, N. and {Kennea}, J. and {Kerr}, M. and {Kimeridze}, G. and {Kn{\"o}dlseder}, J. and {Kocian}, M.~L. and {Kopatskaya}, E.~N. and {Koptelova}, E. and {Konstantinova}, T.~S. and {Kovalev}, Y.~Y. and {Kovalev}, Yu. A. and {Kurtanidze}, O.~M. and {Kuss}, M. and {Lande}, J. and {Larionov}, V.~M. and {Latronico}, L. and {Leto}, P. and {Lindfors}, E. and {Longo}, F. and {Loparco}, F. and {Lott}, B. and {Lovellette}, M.~N. and {Lubrano}, P. and {Madejski}, G.~M. and {Makeev}, A. and {Marchegiani}, P. and {Marscher}, A.~P. and {Marshall}, F. and {Max-Moerbeck}, W. and {Mazziotta}, M.~N. and {McConville}, W. and {McEnery}, J.~E. and {Meurer}, C. and {Michelson}, P.~F. and {Mitthumsiri}, W. and {Mizuno}, T. and {Moiseev}, A.~A. and {Monte}, C. and {Monzani}, M.~E. and {Morselli}, A. and {Moskalenko}, I.~V. and {Murgia}, S. and {Nestoras}, I. and {Nilsson}, K. and {Nizhelsky}, N.~A. and {Nolan}, P.~L. and {Norris}, J.~P. and {Nuss}, E. and {Ohsugi}, T. and {Ojha}, R. and {Omodei}, N. and {Orlando}, E. and {Ormes}, J.~F. and {Osborne}, J. and {Ozaki}, M. and {Pacciani}, L. and {Padovani}, P. and {Pagani}, C. and {Page}, K. and {Paneque}, D. and {Panetta}, J.~H. and {Parent}, D. and {Pasanen}, M. and {Pavlidou}, V. and {Pelassa}, V. and {Pepe}, M. and {Perri}, M. and {Pesce-Rollins}, M. and {Piranomonte}, S. and {Piron}, F. and {Pittori}, C. and {Porter}, T.~A. and {Puccetti}, S. and {Rahoui}, F. and {Rain{\`o}}, S. and {Raiteri}, C. and {Rando}, R. and {Razzano}, M. and {Reimer}, A. and {Reimer}, O.},
        title = "{The Spectral Energy Distribution of Fermi Bright Blazars}",
      journal = {\apj},
         year = 2010,
        month = jun,
       volume = {716},
       number = {1},
        pages = {30-70},
          doi = {10.1088/0004-637X/716/1/30},
archivePrefix = {arXiv},
       eprint = {0912.2040},
 primaryClass = {astro-ph.CO},
       adsurl = {https://ui.adsabs.harvard.edu/abs/2010ApJ...716...30A}
}

@ARTICLE{abdollahi2024,
       author = {{Abdollahi}, S. and {Baldini}, L. and {Barbiellini}, G. and {Bellazzini}, R. and {Berenji}, B. and {Bissaldi}, E. and {Blandford}, R.~D. and {Bonino}, R. and {Bruel}, P. and {Buson}, S. and {Cameron}, R.~A. and {Caraveo}, P.~A. and {Casaburo}, F. and {Cavazzuti}, E. and {Cheung}, C.~C. and {Chiaro}, G. and {Ciprini}, S. and {Cozzolongo}, G. and {Cristarella Orestano}, P. and {Cutini}, S. and {D'Ammando}, F. and {Di Lalla}, N. and {Dirirsa}, F. and {Di Venere}, L. and {Dom{\'\i}nguez}, A. and {Fegan}, S.~J. and {Ferrara}, E.~C. and {Fiori}, A. and {Fukazawa}, Y. and {Funk}, S. and {Fusco}, P. and {Gargano}, F. and {Garrappa}, S. and {Gasparrini}, D. and {Germani}, S. and {Giglietto}, N. and {Giordano}, F. and {Giroletti}, M. and {Green}, D. and {Grenier}, I.~A. and {Guiriec}, S. and {Hays}, E. and {Horan}, D. and {Kuss}, M. and {Larsson}, S. and {Laurenti}, M. and {Li}, J. and {Liodakis}, I. and {Longo}, F. and {Loparco}, F. and {Lott}, B. and {Lovellette}, M.~N. and {Lubrano}, P. and {Maldera}, S. and {Malyshev}, D. and {Manfreda}, A. and {Marcotulli}, L. and {Mart{\'\i}-Devesa}, G. and {Mazziotta}, M.~N. and {Mereu}, I. and {Michelson}, P.~F. and {Mitthumsiri}, W. and {Mizuno}, T. and {Monzani}, M.~E. and {Morselli}, A. and {Moskalenko}, I.~V. and {Negro}, M. and {Omodei}, N. and {Orienti}, M. and {Orlando}, E. and {Ormes}, J.~F. and {Paneque}, D. and {Perri}, M. and {Persic}, M. and {Pesce-Rollins}, M. and {Porter}, T.~A. and {Principe}, G. and {Rain{\`o}}, S. and {Rando}, R. and {Rani}, B. and {Razzano}, M. and {Reimer}, A. and {Reimer}, O. and {Saz Parkinson}, P.~M. and {Scotton}, L. and {Serini}, D. and {Sesana}, A. and {Sgr{\`o}}, C. and {Siskind}, E.~J. and {Spandre}, G. and {Spinelli}, P. and {Suson}, D.~J. and {Tajima}, H. and {Takahashi}, M.~N. and {Tak}, D. and {Thayer}, J.~B. and {Thompson}, D.~J. and {Torres}, D.~F. and {Valverde}, J. and {Verrecchia}, F. and {Zaharijas}, G.},
        title = "{Periodic Gamma-Ray Modulation of the Blazar PG 1553+113 Confirmed by Fermi-LAT and Multiwavelength Observations}",
      journal = {\apj},
         year = 2024,
        month = dec,
       volume = {976},
       number = {2},
          eid = {203},
        pages = {203},
          doi = {10.3847/1538-4357/ad64c5},
       adsurl = {https://ui.adsabs.harvard.edu/abs/2024ApJ...976..203A}
}

@ARTICLE{abe2025,
       author = {{Abe}, K. and {Abe}, S. and {Abhir}, J. and {Abhishek}, A. and {Acciari}, V.~A. and {Aguasca-Cabot}, A. and {Agudo}, I. and {Aniello}, T. and {Ansoldi}, S. and {Antonelli}, L.~A. and {Arbet Engels}, A. and {Arcaro}, C. and {Asano}, K. and {Baack}, D. and {Babi{\'c}}, A. and {Barres de Almeida}, U. and {Barrio}, J.~A. and {Batkovi{\'c}}, I. and {Bautista}, A. and {Baxter}, J. and {Becerra Gonz{\'a}lez}, J. and {Bednarek}, W. and {Bernardini}, E. and {Bernete}, J. and {Berti}, A. and {Besenrieder}, J. and {Bigongiari}, C. and {Biland}, A. and {Blanch}, O. and {Bonnoli}, G. and {Bo{\v{s}}njak}, {\v{Z}}. and {Bronzini}, E. and {Burelli}, I. and {Campoy-Ordaz}, A. and {Carosi}, R. and {Carretero-Castrillo}, M. and {Castro-Tirado}, A.~J. and {Cerasole}, D. and {Ceribella}, G. and {Chai}, Y. and {Cifuentes}, A. and {Colombo}, E. and {Contreras}, J.~L. and {Cortina}, J. and {Covino}, S. and {D'Amico}, G. and {D'Ammando}, F. and {D'Elia}, V. and {Da Vela}, P. and {Dazzi}, F. and {De Angelis}, A. and {De Lotto}, B. and {de Menezes}, R. and {Delfino}, M. and {Delgado}, J. and {Delgado Mendez}, C. and {Di Pierro}, F. and {Di Tria}, R. and {Di Venere}, L. and {Dominis Prester}, D. and {Donini}, A. and {Dorner}, D. and {Doro}, M. and {Eisenberger}, L. and {Elsaesser}, D. and {Escudero}, J. and {Fari{\~n}a}, L. and {Fattorini}, A. and {Foffano}, L. and {Font}, L. and {Fr{\"o}se}, S. and {Fukami}, S. and {Fukazawa}, Y. and {Garc{\'\i}a L{\'o}pez}, R.~J. and {Garczarczyk}, M. and {Gasparyan}, S. and {Gaug}, M. and {Giesbrecht Paiva}, J.~G. and {Giglietto}, N. and {Giordano}, F. and {Gliwny}, P. and {Godinovi{\'c}}, N. and {Gradetzke}, T. and {Grau}, R. and {Green}, D. and {Green}, J.~G. and {G{\"u}nther}, P. and {Hadasch}, D. and {Hahn}, A. and {Hassan}, T. and {Heckmann}, L. and {Herrera Llorente}, J. and {Hrupec}, D. and {Imazawa}, R. and {Ishio}, K. and {Jim{\'e}nez Mart{\'\i}nez}, I. and {Jormanainen}, J. and {Kankkunen}, S. and {Kayanoki}, T. and {Kerszberg}, D. and {Kluge}, G.~W. and {Kouch}, P.~M. and {Kubo}, H. and {Kushida}, J. and {L{\'a}inez}, M. and {Lamastra}, A. and {Leone}, F. and {Lindfors}, E. and {Lombardi}, S. and {Longo}, F. and {L{\'o}pez-Coto}, R. and {L{\'o}pez-Moya}, M. and {L{\'o}pez-Oramas}, A. and {Loporchio}, S. and {Lorini}, A. and {Majumdar}, P. and {Makariev}, M. and {Maneva}, G. and {Manganaro}, M. and {Mangano}, S. and {Mannheim}, K. and {Mariotti}, M. and {Mart{\'\i}nez}, M. and {Mart{\'\i}nez-Chicharro}, M. and {Mas-Aguilar}, A. and {Mazin}, D. and {Menchiari}, S. and {Mender}, S. and {Miceli}, D. and {Miener}, T. and {Miranda}, J.~M. and {Mirzoyan}, R. and {Molero Gonz{\'a}lez}, M. and {Molina}, E. and {Mondal}, H.~A. and {Moralejo}, A. and {Morcuende}, D. and {Nakamori}, T. and {Nanci}, C. and {Neustroev}, V. and {Nickel}, L. and {Nigro}, C. and {Nikoli{\'c}}, L. and {Nilsson}, K. and {Nishijima}, K. and {Njoh Ekoume}, T. and {Noda}, K. and {Nozaki}, S. and {Okumura}, A. and {Otero-Santos}, J. and {Paiano}, S. and {Paneque}, D. and {Paoletti}, R. and {Paredes}, J.~M. and {Peresano}, M. and {Persic}, M. and {Pihet}, M. and {Pirola}, G. and {Podobnik}, F. and {Prada Moroni}, P.~G. and {Prandini}, E. and {Principe}, G. and {Rhode}, W. and {Rib{\'o}}, M. and {Rico}, J. and {Righi}, C. and {Sahakyan}, N. and {Saito}, T. and {Saturni}, F.~G. and {Schmidt}, K. and {Schmuckermaier}, F. and {Schubert}, J.~L. and {Schweizer}, T. and {Sciaccaluga}, A. and {Silvestri}, G. and {Sitarek}, J. and {Sobczynska}, D. and {Stamerra}, A. and {Stri{\v{s}}kovi{\'c}}, J. and {Strom}, D. and {Suda}, Y. and {Tajima}, H. and {Takahashi}, M. and {Takeishi}, R. and {Tavecchio}, F. and {Temnikov}, P. and {Terauchi}, K. and {Terzi{\'c}}, T. and {Teshima}, M. and {Truzzi}, S. and {Tutone}, A. and {Ubach}, S. and {van Scherpenberg}, J. and {Ventura}, S. and {Verna}, G. and {Viale}, I. and {Vigorito}, C.~F. and {Vitale}, V. and {Vovk}, I. and {Walter}, R.},
        title = "{Characterization of Markarian 421 during its most violent year: Multiwavelength variability and correlations}",
      journal = {\aap},
         year = 2025,
        month = feb,
       volume = {694},
          eid = {A195},
        pages = {A195},
          doi = {10.1051/0004-6361/202451624},
archivePrefix = {arXiv},
       eprint = {2501.03831},
 primaryClass = {astro-ph.HE},
       adsurl = {https://ui.adsabs.harvard.edu/abs/2025A&A...694A.195A}
}

@ARTICLE{acciari2021,
       author = {{Acciari}, V.~A. and {Ansoldi}, S. and {Antonelli}, L.~A. and {Asano}, K. and {Babi{\'c}}, A. and {Banerjee}, B. and {Baquero}, A. and {de Almeida}, U. Barres and {Barrio}, J.~A. and {Becerra Gonz{\'a}lez}, J. and {Bednarek}, W. and {Bellizzi}, L. and {Bernardini}, E. and {Bernardos}, M. and {Berti}, A. and {Besenrieder}, J. and {Bhattacharyya}, W. and {Bigongiari}, C. and {Blanch}, O. and {Bonnoli}, G. and {Bo{\v{s}}njak}, {\v{Z}}. and {Busetto}, G. and {Carosi}, R. and {Ceribella}, G. and {Cerruti}, M. and {Chai}, Y. and {Chilingarian}, A. and {Cikota}, S. and {Colak}, S.~M. and {Colombo}, E. and {Contreras}, J.~L. and {Cortina}, J. and {Covino}, S. and {D'Amico}, G. and {D'Elia}, V. and {Da Vela}, P. and {Dazzi}, F. and {De Angelis}, A. and {De Lotto}, B. and {Delfino}, M. and {Delgado}, J. and {Delgado Mendez}, C. and {Depaoli}, D. and {Di Girolamo}, T. and {Di Pierro}, F. and {Di Venere}, L. and {Do Souto Espi{\~n}eira}, E. and {Dominis Prester}, D. and {Donini}, A. and {Doro}, M. and {Fallah Ramazani}, V. and {Fattorini}, A. and {Ferrara}, G. and {Foffano}, L. and {Fonseca}, M.~V. and {Font}, L. and {Fruck}, C. and {Fukami}, S. and {Garc{\'\i}a L{\'o}pez}, R.~J. and {Garczarczyk}, M. and {Gasparyan}, S. and {Gaug}, M. and {Giglietto}, N. and {Giordano}, F. and {Gliwny}, P. and {Godinovi{\'c}}, N. and {Green}, J.~G. and {Green}, D. and {Hadasch}, D. and {Hahn}, A. and {Heckmann}, L. and {Herrera}, J. and {Hoang}, J. and {Hrupec}, D. and {H{\"u}tten}, M. and {Inada}, T. and {Inoue}, S. and {Ishio}, K. and {Iwamura}, Y. and {Jormanainen}, J. and {Jouvin}, L. and {Kajiwara}, Y. and {Karjalainen}, M. and {Kerszberg}, D. and {Kobayashi}, Y. and {Kubo}, H. and {Kushida}, J. and {Lamastra}, A. and {Lelas}, D. and {Leone}, F. and {Lindfors}, E. and {Lombardi}, S. and {Longo}, F. and {L{\'o}pez}, M. and {L{\'o}pez-Coto}, R. and {L{\'o}pez-Oramas}, A. and {Loporchio}, S. and {Machado de Oliveira Fraga}, B. and {Maggio}, C. and {Majumdar}, P. and {Makariev}, M. and {Mallamaci}, M. and {Maneva}, G. and {Manganaro}, M. and {Maraschi}, L. and {Mariotti}, M. and {Mart{\'\i}nez}, M. and {Mazin}, D. and {Mender}, S. and {Mi{\'c}anovi{\'c}}, S. and {Miceli}, D. and {Miener}, T. and {Minev}, M. and {Miranda}, J.~M. and {Mirzoyan}, R. and {Molina}, E. and {Moralejo}, A. and {Morcuende}, D. and {Moreno}, V. and {Moretti}, E. and {Munar-Adrover}, P. and {Neustroev}, V. and {Nigro}, C. and {Nilsson}, K. and {Ninci}, D. and {Nishijima}, K. and {Noda}, K. and {Nozaki}, S. and {Ohtani}, Y. and {Oka}, T. and {Otero-Santos}, J. and {Palatiello}, M. and {Paneque}, D. and {Paoletti}, R. and {Paredes}, J.~M. and {Pavleti{\'c}}, L. and {Pe{\~n}il}, P. and {Perennes}, C. and {Persic}, M. and {Prada Moroni}, P.~G. and {Prandini}, E. and {Priyadarshi}, C. and {Puljak}, I. and {Rhode}, W. and {Rib{\'o}}, M. and {Rico}, J. and {Righi}, C. and {Rugliancich}, A. and {Saha}, L. and {Sahakyan}, N. and {Saito}, T. and {Sakurai}, S. and {Satalecka}, K. and {Schleicher}, B. and {Schmidt}, K. and {Schweizer}, T. and {Sitarek}, J. and {{\v{S}}nidari{\'c}}, I. and {Sobczynska}, D. and {Spolon}, A. and {Stamerra}, A. and {Strom}, D. and {Strzys}, M. and {Suda}, Y. and {Suri{\'c}}, T. and {Takahashi}, M. and {Tavecchio}, F. and {Temnikov}, P. and {Terzi{\'c}}, T. and {Teshima}, M. and {Torres-Alb{\`a}}, N. and {Tosti}, L. and {Truzzi}, S. and {van Scherpenberg}, J. and {Vanzo}, G. and {Vazquez Acosta}, M. and {Ventura}, S. and {Verguilov}, V. and {Vigorito}, C.~F. and {Vitale}, V. and {Vovk}, I. and {Will}, M. and {Zari{\'c}}, D. and {Arbet-Engels}, A. and {Baack}, D. and {Balbo}, M. and {Beck}, M. and {Biederbeck}, N. and {Biland}, A. and {Bretz}, T. and {Bruegge}, K. and {Buss}, J. and {Dorner}, D. and {Elsaesser}, D. and {Hildebrand}, D. and {Iotov}, R. and {Klinger}, M. and {Mannheim}, K. and {Neise}, D. and {Neronov}, A.},
        title = "{Multiwavelength variability and correlation studies of Mrk 421 during historically low X-ray and {\ensuremath{\gamma}}-ray activity in 2015-2016}",
      journal = {\mnras},
         year = 2021,
        month = jun,
       volume = {504},
       number = {1},
        pages = {1427-1451},
          doi = {10.1093/mnras/staa3727},
archivePrefix = {arXiv},
       eprint = {2012.01348},
 primaryClass = {astro-ph.HE},
       adsurl = {https://ui.adsabs.harvard.edu/abs/2021MNRAS.504.1427A}
}

@ARTICLE{ackermann2015,
       author = {{Ackermann}, M. and {Ajello}, M. and {Albert}, A. and {Atwood}, W.~B. and {Baldini}, L. and {Ballet}, J. and {Barbiellini}, G. and {Bastieri}, D. and {Becerra Gonzalez}, J. and {Bellazzini}, R. and {Bissaldi}, E. and {Blandford}, R.~D. and {Bloom}, E.~D. and {Bonino}, R. and {Bottacini}, E. and {Bregeon}, J. and {Bruel}, P. and {Buehler}, R. and {Buson}, S. and {Caliandro}, G.~A. and {Cameron}, R.~A. and {Caputo}, R. and {Caragiulo}, M. and {Caraveo}, P.~A. and {Cavazzuti}, E. and {Cecchi}, C. and {Chekhtman}, A. and {Chiang}, J. and {Chiaro}, G. and {Ciprini}, S. and {Cohen-Tanugi}, J. and {Conrad}, J. and {Cutini}, S. and {D'Ammando}, F. and {de Angelis}, A. and {de Palma}, F. and {Desiante}, R. and {Di Venere}, L. and {Dom{\'\i}nguez}, A. and {Drell}, P.~S. and {Favuzzi}, C. and {Fegan}, S.~J. and {Ferrara}, E.~C. and {Focke}, W.~B. and {Fuhrmann}, L. and {Fukazawa}, Y. and {Fusco}, P. and {Gargano}, F. and {Gasparrini}, D. and {Giglietto}, N. and {Giommi}, P. and {Giordano}, F. and {Giroletti}, M. and {Godfrey}, G. and {Green}, D. and {Grenier}, I.~A. and {Grove}, J.~E. and {Guiriec}, S. and {Harding}, A.~K. and {Hays}, E. and {Hewitt}, J.~W. and {Hill}, A.~B. and {Horan}, D. and {Jogler}, T. and {J{\'o}hannesson}, G. and {Johnson}, A.~S. and {Kamae}, T. and {Kuss}, M. and {Larsson}, S. and {Latronico}, L. and {Li}, J. and {Li}, L. and {Longo}, F. and {Loparco}, F. and {Lott}, B. and {Lovellette}, M.~N. and {Lubrano}, P. and {Magill}, J. and {Maldera}, S. and {Manfreda}, A. and {Max-Moerbeck}, W. and {Mayer}, M. and {Mazziotta}, M.~N. and {McEnery}, J.~E. and {Michelson}, P.~F. and {Mizuno}, T. and {Monzani}, M.~E. and {Morselli}, A. and {Moskalenko}, I.~V. and {Murgia}, S. and {Nuss}, E. and {Ohno}, M. and {Ohsugi}, T. and {Ojha}, R. and {Omodei}, N. and {Orlando}, E. and {Ormes}, J.~F. and {Paneque}, D. and {Pearson}, T.~J. and {Perkins}, J.~S. and {Perri}, M. and {Pesce-Rollins}, M. and {Petrosian}, V. and {Piron}, F. and {Pivato}, G. and {Porter}, T.~A. and {Rain{\`o}}, S. and {Rando}, R. and {Razzano}, M. and {Readhead}, A. and {Reimer}, A. and {Reimer}, O. and {Schulz}, A. and {Sgr{\`o}}, C. and {Siskind}, E.~J. and {Spada}, F. and {Spandre}, G. and {Spinelli}, P. and {Suson}, D.~J. and {Takahashi}, H. and {Thayer}, J.~B. and {Thompson}, D.~J. and {Tibaldo}, L. and {Torres}, D.~F. and {Tosti}, G. and {Troja}, E. and {Uchiyama}, Y. and {Vianello}, G. and {Wood}, K.~S. and {Wood}, M. and {Zimmer}, S. and {Berdyugin}, A. and {Corbet}, R.~H.~D. and {Hovatta}, T. and {Lindfors}, E. and {Nilsson}, K. and {Reinthal}, R. and {Sillanp{\"a}{\"a}}, A. and {Stamerra}, A. and {Takalo}, L.~O. and {Valtonen}, M.~J.},
        title = "{Multiwavelength Evidence for Quasi-periodic Modulation in the Gamma-Ray Blazar PG 1553+113}",
      journal = {\apjl},
         year = 2015,
        month = nov,
       volume = {813},
       number = {2},
          eid = {L41},
        pages = {L41},
          doi = {10.1088/2041-8205/813/2/L41},
archivePrefix = {arXiv},
       eprint = {1509.02063},
 primaryClass = {astro-ph.HE},
       adsurl = {https://ui.adsabs.harvard.edu/abs/2015ApJ...813L..41A}
}

@ARTICLE{agudo2025,
       author = {{Agudo}, Iv{\'a}n and {Liodakis}, Ioannis and {Otero-Santos}, Jorge and {Middei}, Riccardo and {Marscher}, Alan and {Jorstad}, Svetlana and {Zhang}, Haocheng and {Li}, Hui and {Di Gesu}, Laura and {Romani}, Roger W. and {Kim}, Dawoon E. and {Fenu}, Francesco and {Marshall}, Herman L. and {Pacciani}, Luigi and {Escudero Pedrosa}, Juan and {Aceituno}, Francisco Jos{\'e} and {Ag{\'\i}s-Gonz{\'a}lez}, Beatriz and {Bonnoli}, Giacomo and {Casanova}, V{\'\i}ctor and {Morcuende}, Daniel and {Piirola}, Vilppu and {Sota}, Alfredo and {Kouch}, Pouya M. and {Lindfors}, Elina and {McCall}, Callum and {Jermak}, Helen E. and {Steele}, Iain A. and {Borman}, George A. and {Grishina}, Tatiana S. and {Hagen-Thorn}, Vladimir A. and {Kopatskaya}, Evgenia N. and {Larionova}, Elena G. and {Morozova}, Daria A. and {Savchenko}, Sergey S. and {Shishkina}, Ekaterina V. and {Troitskiy}, Ivan S. and {Troitskaya}, Yulia V. and {Vasilyev}, Andrey A. and {Zhovtan}, Alexey V. and {Myserlis}, Ioannis and {Gurwell}, Mark and {Keating}, Garrett and {Rao}, Ramprasad and {Kang}, Sincheol and {Lee}, Sang-Sung and {Kim}, Sanghyun and {Cheong}, Whee Yeon and {Jeong}, Hyeon-Woo and {Angelakis}, Emmanouil and {Kraus}, Alexander and {Blinov}, Dmitry and {Maharana}, Siddharth and {Bachev}, Rumen and {Jormanainen}, Jenni and {Nilsson}, Kari and {Fallah Ramazani}, Vandad and {Casadio}, Carolina and {Fuentes}, Antonio and {Traianou}, Efthalia and {Thum}, Clemens and {G{\'o}mez}, Jos{\'e} L. and {Antonelli}, Lucio Angelo and {Bachetti}, Matteo and {Baldini}, Luca and {Baumgartner}, Wayne H. and {Bellazzini}, Ronaldo and {Bianchi}, Stefano and {Bongiorno}, Stephen D. and {Bonino}, Raffaella and {Brez}, Alessandro and {Bucciantini}, Niccol{\`o} and {Capitanio}, Fiamma and {Castellano}, Simone and {Cavazzuti}, Elisabetta and {Chen}, Chien-Ting and {Ciprini}, Stefano and {Costa}, Enrico and {De Rosa}, Alessandra and {Del Monte}, Ettore and {Di Lalla}, Niccol{\`o} and {Di Marco}, Alessandro and {Donnarumma}, Immacolata and {Doroshenko}, Victor and {Dov{\v{c}}iak}, Michal and {Ehlert}, Steven R. and {Enoto}, Teruaki and {Evangelista}, Yuri and {Fabiani}, Sergio and {Ferrazzoli}, Riccardo and {Garc{\'\i}a}, Javier A. and {Gunji}, Shuichi and {Hayashida}, Kiyoshi and {Heyl}, Jeremy and {Iwakiri}, Wataru and {Kaaret}, Philip and {Karas}, Vladimir and {Kislat}, Fabian and {Kitaguchi}, Takao and {Kolodziejczak}, Jeffery J. and {Krawczynski}, Henric and {La Monaca}, Fabio and {Latronico}, Luca and {Maldera}, Simone and {Manfreda}, Alberto and {Marin}, Fr{\'e}d{\'e}ric and {Marinucci}, Andrea and {Massaro}, Francesco and {Matt}, Giorgio and {Mitsuishi}, Ikuyuki and {Mizuno}, Tsunefumi and {Muleri}, Fabio and {Negro}, Michela and {Ng}, Chi-Yung and {O'Dell}, Stephen L. and {Omodei}, Nicola and {Oppedisano}, Chiara and {Papitto}, Alessandro and {Pavlov}, George G. and {Peirson}, Abel L. and {Perri}, Matteo and {Pesce-Rollins}, Melissa and {Petrucci}, Pierre-Olivier and {Pilia}, Maura and {Possenti}, Andrea and {Poutanen}, Juri and {Puccetti}, Simonetta and {Ramsey}, Brian D. and {Rankin}, John and {Ratheesh}, Ajay and {Roberts}, Oliver J. and {Sgr{\`o}}, Carmelo and {Slane}, Patrick and {Soffitta}, Paolo and {Spandre}, Gloria and {Swartz}, Douglas A. and {Tamagawa}, Toru and {Tavecchio}, Fabrizio and {Taverna}, Roberto and {Tawara}, Yuzuru and {Tennant}, Allyn F. and {Thomas}, Nicholas E. and {Tombesi}, Francesco and {Trois}, Alessio and {Tsygankov}, Sergey S. and {Turolla}, Roberto and {Vink}, Jacco and {Weisskopf}, Martin C. and {Wu}, Kinwah and {Xie}, Fei and {Zane}, Silvia},
        title = "{High Optical-to-X-Ray Polarization Ratio Reveals Compton Scattering in BL Lacertae's Jet}",
      journal = {\apjl},
         year = 2025,
        month = may,
       volume = {985},
       number = {1},
          eid = {L15},
        pages = {L15},
          doi = {10.3847/2041-8213/adc572},
archivePrefix = {arXiv},
       eprint = {2505.01832},
 primaryClass = {astro-ph.HE},
       adsurl = {https://ui.adsabs.harvard.edu/abs/2025ApJ...985L..15A}
}

@ARTICLE{agudo2018,
       author = {{Agudo}, Iv{\'a}n and {Thum}, Clemens and {Ramakrishnan}, Venkatessh and {Molina}, Sol N. and {Casadio}, Carolina and {G{\'o}mez}, Jos{\'e} L.},
        title = "{POLAMI: Polarimetric Monitoring of Active Galactic Nuclei at Millimetre Wavelengths - III. Characterization of total flux density and polarization variability of relativistic jets}",
      journal = {\mnras},
         year = 2018,
        month = jan,
       volume = {473},
       number = {2},
        pages = {1850-1867},
          doi = {10.1093/mnras/stx2437},
archivePrefix = {arXiv},
       eprint = {1709.08744},
 primaryClass = {astro-ph.GA},
       adsurl = {https://ui.adsabs.harvard.edu/abs/2018MNRAS.473.1850A}
}

@ARTICLE{aharonian2007,
       author = {{Aharonian}, F. and {Akhperjanian}, A.~G. and {Bazer-Bachi}, A.~R. and {Behera}, B. and {Beilicke}, M. and {Benbow}, W. and {Berge}, D. and {Bernl{\"o}hr}, K. and {Boisson}, C. and {Bolz}, O. and {Borrel}, V. and {Boutelier}, T. and {Braun}, I. and {Brion}, E. and {Brown}, A.~M. and {B{\"u}hler}, R. and {B{\"u}sching}, I. and {Bulik}, T. and {Carrigan}, S. and {Chadwick}, P.~M. and {Clapson}, A.~C. and {Chounet}, L. -M. and {Coignet}, G. and {Cornils}, R. and {Costamante}, L. and {Degrange}, B. and {Dickinson}, H.~J. and {Djannati-Ata{\"\i}}, A. and {Domainko}, W. and {Drury}, L. O'C. and {Dubus}, G. and {Dyks}, J. and {Egberts}, K. and {Emmanoulopoulos}, D. and {Espigat}, P. and {Farnier}, C. and {Feinstein}, F. and {Fiasson}, A. and {F{\"o}rster}, A. and {Fontaine}, G. and {Funk}, Seb. and {Funk}, S. and {F{\"u}{\ss}ling}, M. and {Gallant}, Y.~A. and {Giebels}, B. and {Glicenstein}, J.~F. and {Gl{\"u}ck}, B. and {Goret}, P. and {Hadjichristidis}, C. and {Hauser}, D. and {Hauser}, M. and {Heinzelmann}, G. and {Henri}, G. and {Hermann}, G. and {Hinton}, J.~A. and {Hoffmann}, A. and {Hofmann}, W. and {Holleran}, M. and {Hoppe}, S. and {Horns}, D. and {Jacholkowska}, A. and {de Jager}, O.~C. and {Kendziorra}, E. and {Kerschhaggl}, M. and {Kh{\'e}lifi}, B. and {Komin}, Nu. and {Kosack}, K. and {Lamanna}, G. and {Latham}, I.~J. and {Le Gallou}, R. and {Lemi{\`e}re}, A. and {Lemoine-Goumard}, M. and {Lenain}, J. -P. and {Lohse}, T. and {Martin}, J.~M. and {Martineau-Huynh}, O. and {Marcowith}, A. and {Masterson}, C. and {Maurin}, G. and {McComb}, T.~J.~L. and {Moderski}, R. and {Moulin}, E. and {de Naurois}, M. and {Nedbal}, D. and {Nolan}, S.~J. and {Olive}, J. -P. and {Orford}, K.~J. and {Osborne}, J.~L. and {Ostrowski}, M. and {Panter}, M. and {Pedaletti}, G. and {Pelletier}, G. and {Petrucci}, P. -O. and {Pita}, S. and {P{\"u}hlhofer}, G. and {Punch}, M. and {Ranchon}, S. and {Raubenheimer}, B.~C. and {Raue}, M. and {Rayner}, S.~M. and {Renaud}, M. and {Ripken}, J. and {Rob}, L. and {Rolland}, L. and {Rosier-Lees}, S. and {Rowell}, G. and {Rudak}, B. and {Ruppel}, J. and {Sahakian}, V. and {Santangelo}, A. and {Saug{\'e}}, L. and {Schlenker}, S. and {Schlickeiser}, R. and {Schr{\"o}der}, R. and {Schwanke}, U. and {Schwarzburg}, S. and {Schwemmer}, S. and {Shalchi}, A. and {Sol}, H. and {Spangler}, D. and {Stawarz}, {\L}. and {Steenkamp}, R. and {Stegmann}, C. and {Superina}, G. and {Tam}, P.~H. and {Tavernet}, J. -P. and {Terrier}, R. and {van Eldik}, C. and {Vasileiadis}, G. and {Venter}, C. and {Vialle}, J.~P. and {Vincent}, P. and {Vivier}, M. and {V{\"o}lk}, H.~J. and {Volpe}, F. and {Wagner}, S.~J. and {Ward}, M. and {Zdziarski}, A.~A.},
        title = "{An Exceptional Very High Energy Gamma-Ray Flare of PKS 2155-304}",
      journal = {\apjl},
         year = 2007,
        month = aug,
       volume = {664},
       number = {2},
        pages = {L71-L74},
          doi = {10.1086/520635},
archivePrefix = {arXiv},
       eprint = {0706.0797},
 primaryClass = {astro-ph},
       adsurl = {https://ui.adsabs.harvard.edu/abs/2007ApJ...664L..71A}
}

@ARTICLE{ahnen2016,
       author = {{Ahnen}, M.~L. and {Ansoldi}, S. and {Antonelli}, L.~A. and {Antoranz}, P. and {Babic}, A. and {Banerjee}, B. and {Bangale}, P. and {Barres de Almeida}, U. and {Barrio}, J.~A. and {Becerra Gonz{\'a}lez}, J. and {Bednarek}, W. and {Bernardini}, E. and {Biasuzzi}, B. and {Biland}, A. and {Blanch}, O. and {Bonnefoy}, S. and {Bonnoli}, G. and {Borracci}, F. and {Bretz}, T. and {Buson}, S. and {Carosi}, A. and {Chatterjee}, A. and {Clavero}, R. and {Colin}, P. and {Colombo}, E. and {Contreras}, J.~L. and {Cortina}, J. and {Covino}, S. and {Da Vela}, P. and {Dazzi}, F. and {De Angelis}, A. and {De Lotto}, B. and {de O{\~n}a Wilhelmi}, E. and {Di Pierro}, F. and {Dom{\'\i}nguez}, A. and {Dominis Prester}, D. and {Dorner}, D. and {Doro}, M. and {Einecke}, S. and {Eisenacher Glawion}, D. and {Elsaesser}, D. and {Fern{\'a}ndez-Barral}, A. and {Fidalgo}, D. and {Fonseca}, M.~V. and {Font}, L. and {Frantzen}, K. and {Fruck}, C. and {Galindo}, D. and {Garc{\'\i}a L{\'o}pez}, R.~J. and {Garczarczyk}, M. and {Garrido Terrats}, D. and {Gaug}, M. and {Giammaria}, P. and {Godinovi{\'c}}, N. and {Gonz{\'a}lez Mu{\~n}oz}, A. and {Gora}, D. and {Guberman}, D. and {Hadasch}, D. and {Hahn}, A. and {Hanabata}, Y. and {Hayashida}, M. and {Herrera}, J. and {Hose}, J. and {Hrupec}, D. and {Hughes}, G. and {Idec}, W. and {Kodani}, K. and {Konno}, Y. and {Kubo}, H. and {Kushida}, J. and {La Barbera}, A. and {Lelas}, D. and {Lindfors}, E. and {Lombardi}, S. and {Longo}, F. and {L{\'o}pez}, M. and {L{\'o}pez-Coto}, R. and {Majumdar}, P. and {Makariev}, M. and {Mallot}, K. and {Maneva}, G. and {Manganaro}, M. and {Mannheim}, K. and {Maraschi}, L. and {Marcote}, B. and {Mariotti}, M. and {Mart{\'\i}nez}, M. and {Mazin}, D. and {Menzel}, U. and {Miranda}, J.~M. and {Mirzoyan}, R. and {Moralejo}, A. and {Moretti}, E. and {Nakajima}, D. and {Neustroev}, V. and {Niedzwiecki}, A. and {Nievas Rosillo}, M. and {Nilsson}, K. and {Nishijima}, K. and {Noda}, K. and {Nogu{\'e}s}, L. and {Orito}, R. and {Overkemping}, A. and {Paiano}, S. and {Palacio}, J. and {Palatiello}, M. and {Paneque}, D. and {Paoletti}, R. and {Paredes}, J.~M. and {Paredes-Fortuny}, X. and {Pedaletti}, G. and {Perri}, L. and {Persic}, M. and {Poutanen}, J. and {Prada Moroni}, P.~G. and {Prandini}, E. and {Puljak}, I. and {Rhode}, W. and {Rib{\'o}}, M. and {Rico}, J. and {Rodriguez Garcia}, J. and {Saito}, T. and {Satalecka}, K. and {Schultz}, C. and {Schweizer}, T. and {Shore}, S.~N. and {Sillanp{\"a}{\"a}}, A. and {Sitarek}, J. and {Snidaric}, I. and {Sobczynska}, D. and {Stamerra}, A. and {Steinbring}, T. and {Strzys}, M. and {Takalo}, L. and {Takami}, H. and {Tavecchio}, F. and {Temnikov}, P. and {Terzi{\'c}}, T. and {Tescaro}, D. and {Teshima}, M. and {Thaele}, J. and {Torres}, D.~F. and {Toyama}, T. and {Treves}, A. and {Verguilov}, V. and {Vovk}, I. and {Ward}, J.~E. and {Will}, M. and {Wu}, M.~H. and {Zanin}, R. and {MAGIC Collaboration} and {Blinov}, D.~A. and {Chen}, W.~P. and {Efimova}, N.~V. and {Forn{\'e}}, E. and {Grishina}, T.~S. and {Hovatta}, T. and {Jordan}, B. and {Kimeridze}, G.~N. and {Kopatskaya}, E.~N. and {Koptelova}, E. and {Kurtanidze}, O.~M. and {Kurtanidze}, S.~O. and {L{\"a}hteenm{\"a}ki}, A. and {Larionov}, V.~M. and {Larionova}, E.~G. and {Larionova}, L.~V. and {Ligustri}, R. and {Lin}, H.~C. and {McBreen}, B. and {Morozova}, D.~A. and {Nikolashvili}, M.~G. and {Raiteri}, C.~M. and {Ros}, J.~A. and {Sadun}, A.~C. and {Sigua}, L.~A. and {Tornikoski}, M. and {Troitsky}, I.~S. and {Villata}, M.},
        title = "{Long-term multi-wavelength variability and correlation study of Markarian 421 from 2007 to 2009}",
      journal = {\aap},
         year = 2016,
        month = sep,
       volume = {593},
          eid = {A91},
        pages = {A91},
          doi = {10.1051/0004-6361/201628447},
archivePrefix = {arXiv},
       eprint = {1605.09017},
 primaryClass = {astro-ph.GA},
       adsurl = {https://ui.adsabs.harvard.edu/abs/2016A&A...593A..91A}
}

@ARTICLE{ajello2020,
       author = {{Ajello}, M. and {Angioni}, R. and {Axelsson}, M. and {Ballet}, J. and {Barbiellini}, G. and {Bastieri}, D. and {Becerra Gonzalez}, J. and {Bellazzini}, R. and {Bissaldi}, E. and {Bloom}, E.~D. and {Bonino}, R. and {Bottacini}, E. and {Bruel}, P. and {Buson}, S. and {Cafardo}, F. and {Cameron}, R.~A. and {Cavazzuti}, E. and {Chen}, S. and {Cheung}, C.~C. and {Ciprini}, S. and {Costantin}, D. and {Cutini}, S. and {D'Ammando}, F. and {de la Torre Luque}, P. and {de Menezes}, R. and {de Palma}, F. and {Desai}, A. and {Di Lalla}, N. and {Di Venere}, L. and {Dom{\'\i}nguez}, A. and {Dirirsa}, F. Fana and {Ferrara}, E.~C. and {Finke}, J. and {Franckowiak}, A. and {Fukazawa}, Y. and {Funk}, S. and {Fusco}, P. and {Gargano}, F. and {Garrappa}, S. and {Gasparrini}, D. and {Giglietto}, N. and {Giordano}, F. and {Giroletti}, M. and {Green}, D. and {Grenier}, I.~A. and {Guiriec}, S. and {Harita}, S. and {Hays}, E. and {Horan}, D. and {Itoh}, R. and {J{\'o}hannesson}, G. and {Kovac'evic'}, M. and {Krauss}, F. and {Kreter}, M. and {Kuss}, M. and {Larsson}, S. and {Leto}, C. and {Li}, J. and {Liodakis}, I. and {Longo}, F. and {Loparco}, F. and {Lott}, B. and {Lovellette}, M.~N. and {Lubrano}, P. and {Madejski}, G.~M. and {Maldera}, S. and {Manfreda}, A. and {Mart{\'\i}-Devesa}, G. and {Massaro}, F. and {Mazziotta}, M.~N. and {Mereu}, I. and {Meyer}, M. and {Migliori}, G. and {Mirabal}, N. and {Mizuno}, T. and {Monzani}, M.~E. and {Morselli}, A. and {Moskalenko}, I.~V. and {Negro}, M. and {Nemmen}, R. and {Nuss}, E. and {Ojha}, L.~S. and {Ojha}, R. and {Omodei}, N. and {Orienti}, M. and {Orlando}, E. and {Ormes}, J.~F. and {Paliya}, V.~S. and {Pei}, Z. and {Pe{\~n}a-Herazo}, H. and {Persic}, M. and {Pesce-Rollins}, M. and {Petrov}, L. and {Piron}, F. and {Poon}, H. and {Principe}, G. and {Rain{\`o}}, S. and {Rando}, R. and {Rani}, B. and {Razzano}, M. and {Razzaque}, S. and {Reimer}, A. and {Reimer}, O. and {Schinzel}, F.~K. and {Serini}, D. and {Sgr{\`o}}, C. and {Siskind}, E.~J. and {Spandre}, G. and {Spinelli}, P. and {Suson}, D.~J. and {Tachibana}, Y. and {Thompson}, D.~J. and {Torres}, D.~F. and {Torresi}, E. and {Troja}, E. and {Valverde}, J. and {van Zyl}, P. and {Yassine}, M.},
        title = "{The Fourth Catalog of Active Galactic Nuclei Detected by the Fermi Large Area Telescope}",
      journal = {\apj},
         year = 2020,
        month = apr,
       volume = {892},
       number = {2},
          eid = {105},
        pages = {105},
          doi = {10.3847/1538-4357/ab791e},
archivePrefix = {arXiv},
       eprint = {1905.10771},
 primaryClass = {astro-ph.HE},
       adsurl = {https://ui.adsabs.harvard.edu/abs/2020ApJ...892..105A}
}

@ARTICLE{albert2007,
       author = {{Albert}, J. and {Aliu}, E. and {Anderhub}, H. and {Antoranz}, P. and {Armada}, A. and {Baixeras}, C. and {Barrio}, J.~A. and {Bartko}, H. and {Bastieri}, D. and {Becker}, J.~K. and {Bednarek}, W. and {Berger}, K. and {Bigongiari}, C. and {Biland}, A. and {Bock}, R.~K. and {Bordas}, P. and {Bosch-Ramon}, V. and {Bretz}, T. and {Britvitch}, I. and {Camara}, M. and {Carmona}, E. and {Chilingarian}, A. and {Coarasa}, J.~A. and {Commichau}, S. and {Contreras}, J.~L. and {Cortina}, J. and {Costado}, M.~T. and {Curtef}, V. and {Danielyan}, V. and {Dazzi}, F. and {De Angelis}, A. and {Delgado}, C. and {de los Reyes}, R. and {De Lotto}, B. and {Domingo-Santamar{\'\i}a}, E. and {Dorner}, D. and {Doro}, M. and {Errando}, M. and {Fagiolini}, M. and {Ferenc}, D. and {Fern{\'a}ndez}, E. and {Firpo}, R. and {Flix}, J. and {Fonseca}, M.~V. and {Font}, L. and {Fuchs}, M. and {Galante}, N. and {Garc{\'\i}a-L{\'o}pez}, R.~J. and {Garczarczyk}, M. and {Gaug}, M. and {Giller}, M. and {Goebel}, F. and {Hakobyan}, D. and {Hayashida}, M. and {Hengstebeck}, T. and {Herrero}, A. and {H{\"o}hne}, D. and {Hose}, J. and {Hrupec}, D. and {Hsu}, C.~C. and {Jacon}, P. and {Jogler}, T. and {Kosyra}, R. and {Kranich}, D. and {Kritzer}, R. and {Laille}, A. and {Lindfors}, E. and {Lombardi}, S. and {Longo}, F. and {L{\'o}pez}, J. and {L{\'o}pez}, M. and {Lorenz}, E. and {Majumdar}, P. and {Maneva}, G. and {Mannheim}, K. and {Mansutti}, O. and {Mariotti}, M. and {Mart{\'\i}nez}, M. and {Mazin}, D. and {Merck}, C. and {Meucci}, M. and {Meyer}, M. and {Miranda}, J.~M. and {Mirzoyan}, R. and {Mizobuchi}, S. and {Moralejo}, A. and {Nieto}, D. and {Nilsson}, K. and {Ninkovic}, J. and {O{\~n}a-Wilhelmi}, E. and {Otte}, N. and {Oya}, I. and {Paneque}, D. and {Panniello}, M. and {Paoletti}, R. and {Paredes}, J.~M. and {Pasanen}, M. and {Pascoli}, D. and {Pauss}, F. and {Pegna}, R. and {Persic}, M. and {Peruzzo}, L. and {Piccioli}, A. and {Prandini}, E. and {Puchades}, N. and {Raymers}, A. and {Rhode}, W. and {Rib{\'o}}, M. and {Rico}, J. and {Rissi}, M. and {Robert}, A. and {R{\"u}gamer}, S. and {Saggion}, A. and {Saito}, T. and {S{\'a}nchez}, A. and {Sartori}, P. and {Scalzotto}, V. and {Scapin}, V. and {Schmitt}, R. and {Schweizer}, T. and {Shayduk}, M. and {Shinozaki}, K. and {Shore}, S.~N. and {Sidro}, N. and {Sillanp{\"a}{\"a}}, A. and {Sobczynska}, D. and {Stamerra}, A. and {Stark}, L.~S. and {Takalo}, L. and {Tavecchio}, F. and {Temnikov}, P. and {Tescaro}, D. and {Teshima}, M. and {Torres}, D.~F. and {Turini}, N. and {Vankov}, H. and {Vitale}, V. and {Wagner}, R.~M. and {Wibig}, T. and {Wittek}, W. and {Zandanel}, F. and {Zanin}, R. and {Zapatero}, J.},
        title = "{Variable Very High Energy {\ensuremath{\gamma}}-Ray Emission from Markarian 501}",
      journal = {\apj},
         year = 2007,
        month = nov,
       volume = {669},
       number = {2},
        pages = {862-883},
          doi = {10.1086/521382},
archivePrefix = {arXiv},
       eprint = {astro-ph/0702008},
 primaryClass = {astro-ph},
       adsurl = {https://ui.adsabs.harvard.edu/abs/2007ApJ...669..862A}
}

@INPROCEEDINGS{alexander1997,
       author = {{Alexander}, Tal},
        title = "{Is AGN Variability Correlated with Other AGN Properties? ZDCF Analysis of Small Samples of Sparse Light Curves}",
    booktitle = {Astronomical Time Series},
         year = 1997,
       editor = {{Maoz}, D. and {Sternberg}, A. and {Leibowitz}, E.~M.},
       series = {Astrophysics and Space Science Library},
       volume = {218},
        month = jan,
        pages = {163},
          doi = {10.1007/978-94-015-8941-3_14},
       adsurl = {https://ui.adsabs.harvard.edu/abs/1997ASSL..218..163A}
}

@ARTICLE{aller2017,
       author = {{Aller}, Margo F. and {Aller}, Hugh D. and {Hughes}, Philip A.},
        title = "{The University of Michigan Centimeter-Band All Stokes Blazar Monitoring Program: Single-Dish Polarimetry as a Probe of Parsec-Scale Magnetic Fields}",
      journal = {Galaxies},
         year = 2017,
        month = nov,
       volume = {5},
       number = {4},
          eid = {75},
        pages = {75},
          doi = {10.3390/galaxies5040075},
archivePrefix = {arXiv},
       eprint = {1711.05763},
 primaryClass = {astro-ph.HE},
       adsurl = {https://ui.adsabs.harvard.edu/abs/2017Galax...5...75A}
}

@ARTICLE{angel1980,
       author = {{Angel}, J.~R.~P. and {Stockman}, H.~S.},
        title = "{Optical and infrared polarization of active extragalactic objects}",
      journal = {\araa},
         year = 1980,
        month = jan,
       volume = {18},
        pages = {321-361},
          doi = {10.1146/annurev.aa.18.090180.001541},
       adsurl = {https://ui.adsabs.harvard.edu/abs/1980ARA&A..18..321A}
}

@ARTICLE{angelakis2019,
       author = {{Angelakis}, E. and {Fuhrmann}, L. and {Myserlis}, I. and {Zensus}, J.~A. and {Nestoras}, I. and {Karamanavis}, V. and {Marchili}, N. and {Krichbaum}, T.~P. and {Kraus}, A. and {Rachen}, J.~P.},
        title = "{F-GAMMA: Multi-frequency radio monitoring of Fermi blazars. The 2.64 to 43 GHz Effelsberg light curves from 2007-2015}",
      journal = {\aap},
         year = 2019,
        month = jun,
       volume = {626},
          eid = {A60},
        pages = {A60},
          doi = {10.1051/0004-6361/201834363},
archivePrefix = {arXiv},
       eprint = {1902.04404},
 primaryClass = {astro-ph.HE},
       adsurl = {https://ui.adsabs.harvard.edu/abs/2019A&A...626A..60A}
}

@ARTICLE{angelakis2016,
       author = {{Angelakis}, E. and {Hovatta}, T. and {Blinov}, D. and {Pavlidou}, V. and {Kiehlmann}, S. and {Myserlis}, I. and {B{\"o}ttcher}, M. and {Mao}, P. and {Panopoulou}, G.~V. and {Liodakis}, I. and {King}, O.~G. and {Balokovi{\'c}}, M. and {Kus}, A. and {Kylafis}, N. and {Mahabal}, A. and {Marecki}, A. and {Paleologou}, E. and {Papadakis}, I. and {Papamastorakis}, I. and {Pazderski}, E. and {Pearson}, T.~J. and {Prabhudesai}, S. and {Ramaprakash}, A.~N. and {Readhead}, A.~C.~S. and {Reig}, P. and {Tassis}, K. and {Urry}, M. and {Zensus}, J.~A.},
        title = "{RoboPol: the optical polarization of gamma-ray-loud and gamma-ray-quiet blazars}",
      journal = {\mnras},
         year = 2016,
        month = dec,
       volume = {463},
       number = {3},
        pages = {3365-3380},
          doi = {10.1093/mnras/stw2217},
archivePrefix = {arXiv},
       eprint = {1609.00640},
 primaryClass = {astro-ph.HE},
       adsurl = {https://ui.adsabs.harvard.edu/abs/2016MNRAS.463.3365A}
}

@ARTICLE{baath1992,
       author = {{Baath}, L.~B. and {Rogers}, A.~E.~E. and {Inoue}, M. and {Padin}, S. and {Wright}, M.~C.~H. and {Zensus}, A. and {Kus}, A.~J. and {Backer}, D.~C. and {Booth}, R.~S. and {Carlstrom}, J.~E. and {Dickman}, R.~L. and {Emerson}, D.~T. and {Hirabayashi}, H. and {Hodges}, M.~W. and {Kobayashi}, H. and {Lamb}, J. and {Moran}, J.~M. and {Morimoto}, M. and {Plambeck}, R.~L. and {Predmore}, C.~R. and {Ronnang}, B. and {Woody}, D.},
        title = "{VLBI observations of active galactic nuclei at 3 MM.}",
      journal = {\aap},
         year = 1992,
        month = apr,
       volume = {257},
        pages = {31-46},
       adsurl = {https://ui.adsabs.harvard.edu/abs/1992A&A...257...31B}
}

@ARTICLE{baath1991,
       author = {{Baath}, L.~B. and {Padin}, S. and {Woody}, D. and {Rogers}, A.~E.~E. and {Wright}, M.~C.~H. and {Zensus}, A. and {Kus}, A.~J. and {Backer}, D.~C. and {Booth}, R.~S. and {Carlstrom}, J.~E. and {Dickman}, R.~L. and {Emerson}, D.~T. and {Hirabayashi}, H. and {Hodges}, M.~W. and {Inoue}, M. and {Moran}, J.~M. and {Morimoto}, M. and {Payne}, J. and {Plambeck}, R.~L. and {Predmore}, C.~R. and {Ronnang}, B.},
        title = "{The microarcsecond structure of 3C 273 at 3 mm.}",
      journal = {\aap},
         year = 1991,
        month = jan,
       volume = {241},
        pages = {L1},
       adsurl = {https://ui.adsabs.harvard.edu/abs/1991A&A...241L...1B}
}

@ARTICLE{bach2007,
       author = {{Bach}, U. and {Raiteri}, C.~M. and {Villata}, M. and {Fuhrmann}, L. and {Buemi}, C.~S. and {Larionov}, V.~M. and {Letog}, P. and {Arkharov}, A.~A. and {Coloma}, J.~M. and {di Paola}, A. and {Dolci}, M. and {Efimova}, N. and {Forn{\'e}}, E. and {Ibrahimov}, M.~A. and {Hagen-Thorn}, V. and {Konstantinova}, T. and {Kopatskaya}, E. and {Lanteri}, L. and {Kurtanidze}, O.~M. and {Maccaferri}, G. and {Nikolashvili}, M.~G. and {Orlati}, A. and {Ros}, J.~A. and {Tosti}, G. and {Trigilio}, C. and {Umana}, G.},
        title = "{Multi-frequency monitoring of {\ensuremath{\gamma}}-ray loud blazars. I. Light curves and spectral energy distributions}",
      journal = {\aap},
         year = 2007,
        month = mar,
       volume = {464},
       number = {1},
        pages = {175-186},
          doi = {10.1051/0004-6361:20066561},
archivePrefix = {arXiv},
       eprint = {astro-ph/0612149},
 primaryClass = {astro-ph},
       adsurl = {https://ui.adsabs.harvard.edu/abs/2007A&A...464..175B}
}

@ARTICLE{bach2006a,
       author = {{Bach}, U. and {Krichbaum}, T.~P. and {Kraus}, A. and {Witzel}, A. and {Zensus}, J.~A.},
        title = "{Space-VLBI polarimetry of the BL Lacertae object S5 0716+714: rapid polarization variability in the VLBI core}",
      journal = {\aap},
         year = 2006,
        month = jun,
       volume = {452},
       number = {1},
        pages = {83-95},
          doi = {10.1051/0004-6361:20053943},
archivePrefix = {arXiv},
       eprint = {astro-ph/0511761},
 primaryClass = {astro-ph},
       adsurl = {https://ui.adsabs.harvard.edu/abs/2006A&A...452...83B}
}

@ARTICLE{bach2005,
       author = {{Bach}, U. and {Krichbaum}, T.~P. and {Ros}, E. and {Britzen}, S. and {Tian}, W.~W. and {Kraus}, A. and {Witzel}, A. and {Zensus}, J.~A.},
        title = "{Kinematic study of the blazar S5 0716+714}",
      journal = {\aap},
         year = 2005,
        month = apr,
       volume = {433},
       number = {3},
        pages = {815-825},
          doi = {10.1051/0004-6361:20040388},
archivePrefix = {arXiv},
       eprint = {astro-ph/0412406},
 primaryClass = {astro-ph},
       adsurl = {https://ui.adsabs.harvard.edu/abs/2005A&A...433..815B}
}

@ARTICLE{banados2025,
       author = {{Ba{\~n}ados}, Eduardo and {Momjian}, Emmanuel and {Connor}, Thomas and {Belladitta}, Silvia and {Decarli}, Roberto and {Mazzucchelli}, Chiara and {Venemans}, Bram P. and {Walter}, Fabian and {Wang}, Feige and {Xie}, Zhang-Liang and {Barth}, Aaron J. and {Eilers}, Anna-Christina and {Fan}, Xiaohui and {Khusanova}, Yana and {Schindler}, Jan-Torge and {Stern}, Daniel and {Yang}, Jinyi and {Andika}, Irham Taufik and {Carilli}, Christopher L. and {Farina}, Emanuele P. and {Fabian}, Andrew and {Hennawi}, Joseph F. and {Pensabene}, Antonio and {Rojas-Ruiz}, Sof{\'\i}a},
        title = "{A blazar in the epoch of reionization}",
      journal = {Nature Astronomy},
         year = 2025,
        month = feb,
       volume = {9},
        pages = {293-301},
          doi = {10.1038/s41550-024-02431-4},
       adsurl = {https://ui.adsabs.harvard.edu/abs/2025NatAs...9..293B}
}

@ARTICLE{begelman2008,
       author = {{Begelman}, Mitchell C. and {Fabian}, Andrew C. and {Rees}, Martin J.},
        title = "{Implications of very rapid TeV variability in blazars}",
      journal = {\mnras},
         year = 2008,
        month = feb,
       volume = {384},
       number = {1},
        pages = {L19-L23},
          doi = {10.1111/j.1745-3933.2007.00413.x},
archivePrefix = {arXiv},
       eprint = {0709.0540},
 primaryClass = {astro-ph},
       adsurl = {https://ui.adsabs.harvard.edu/abs/2008MNRAS.384L..19B}
}

@ARTICLE{begelman1980,
       author = {{Begelman}, M.~C. and {Blandford}, R.~D. and {Rees}, M.~J.},
        title = "{Massive black hole binaries in active galactic nuclei}",
      journal = {\nat},
         year = 1980,
        month = sep,
       volume = {287},
       number = {5780},
        pages = {307-309},
          doi = {10.1038/287307a0},
       adsurl = {https://ui.adsabs.harvard.edu/abs/1980Natur.287..307B}
}

@ARTICLE{bhatta2020,
       author = {{Bhatta}, Gopal and {Dhital}, Niraj},
        title = "{The Nature of {\ensuremath{\gamma}}-Ray Variability in Blazars}",
      journal = {\apj},
         year = 2020,
        month = mar,
       volume = {891},
       number = {2},
          eid = {120},
        pages = {120},
          doi = {10.3847/1538-4357/ab7455},
archivePrefix = {arXiv},
       eprint = {1911.08198},
 primaryClass = {astro-ph.HE},
       adsurl = {https://ui.adsabs.harvard.edu/abs/2020ApJ...891..120B}
}

@ARTICLE{blandford1982_rm,
       author = {{Blandford}, R.~D. and {McKee}, C.~F.},
        title = "{Reverberation mapping of the emission line regions of Seyfert galaxies and quasars.}",
      journal = {\apj},
         year = 1982,
        month = apr,
       volume = {255},
        pages = {419-439},
          doi = {10.1086/159843},
       adsurl = {https://ui.adsabs.harvard.edu/abs/1982ApJ...255..419B}
}

@ARTICLE{blandford1982,
       author = {{Blandford}, R.~D. and {Payne}, D.~G.},
        title = "{Hydromagnetic flows from accretion disks and the production of radio jets.}",
      journal = {\mnras},
         year = 1982,
        month = jun,
       volume = {199},
        pages = {883-903},
          doi = {10.1093/mnras/199.4.883},
       adsurl = {https://ui.adsabs.harvard.edu/abs/1982MNRAS.199..883B}
}

@ARTICLE{blandford1979,
       author = {{Blandford}, R.~D. and {K{\"o}nigl}, A.},
        title = "{Relativistic jets as compact radio sources.}",
      journal = {\apj},
         year = 1979,
        month = aug,
       volume = {232},
        pages = {34-48},
          doi = {10.1086/157262},
       adsurl = {https://ui.adsabs.harvard.edu/abs/1979ApJ...232...34B}
}

@ARTICLE{blandford1978,
       author = {{Blandford}, R.~D. and {Rees}, M.~J.},
        title = "{Extended and compact extragalactic radio sources: interpretation and theory.}",
      journal = {\physscr},
         year = 1978,
        month = mar,
       volume = {17},
        pages = {265-274},
          doi = {10.1088/0031-8949/17/3/020},
       adsurl = {https://ui.adsabs.harvard.edu/abs/1978PhyS...17..265B}
}

@ARTICLE{blandford1977,
       author = {{Blandford}, R.~D. and {Znajek}, R.~L.},
        title = "{Electromagnetic extraction of energy from Kerr black holes.}",
      journal = {\mnras},
         year = 1977,
        month = may,
       volume = {179},
        pages = {433-456},
          doi = {10.1093/mnras/179.3.433},
       adsurl = {https://ui.adsabs.harvard.edu/abs/1977MNRAS.179..433B}
}

@ARTICLE{blazejowski2000,
       author = {{B{\l}a{\.z}ejowski}, M. and {Sikora}, M. and {Moderski}, R. and {Madejski}, G.~M.},
        title = "{Comptonization of Infrared Radiation from Hot Dust by Relativistic Jets in Quasars}",
      journal = {\apj},
         year = 2000,
        month = dec,
       volume = {545},
       number = {1},
        pages = {107-116},
          doi = {10.1086/317791},
archivePrefix = {arXiv},
       eprint = {astro-ph/0008154},
 primaryClass = {astro-ph},
       adsurl = {https://ui.adsabs.harvard.edu/abs/2000ApJ...545..107B}
}

@ARTICLE{blinov2018,
       author = {{Blinov}, D. and {Pavlidou}, V. and {Papadakis}, I. and {Kiehlmann}, S. and {Liodakis}, I. and {Panopoulou}, G.~V. and {Angelakis}, E. and {Balokovi{\'c}}, M. and {Hovatta}, T. and {King}, O.~G. and {Kus}, A. and {Kylafis}, N. and {Mahabal}, A. and {Maharana}, S. and {Myserlis}, I. and {Paleologou}, E. and {Papamastorakis}, I. and {Pazderski}, E. and {Pearson}, T.~J. and {Ramaprakash}, A. and {Readhead}, A.~C.~S. and {Reig}, P. and {Tassis}, K. and {Zensus}, J.~A.},
        title = "{RoboPol: connection between optical polarization plane rotations and gamma-ray flares in blazars}",
      journal = {\mnras},
         year = 2018,
        month = feb,
       volume = {474},
       number = {1},
        pages = {1296-1306},
          doi = {10.1093/mnras/stx2786},
archivePrefix = {arXiv},
       eprint = {1710.08922},
 primaryClass = {astro-ph.HE},
       adsurl = {https://ui.adsabs.harvard.edu/abs/2018MNRAS.474.1296B}
}

@ARTICLE{boettcher2019,
       author = {{B{\"o}ttcher}, Markus and {Baring}, Matthew G.},
        title = "{Multi-wavelength Variability Signatures of Relativistic Shocks in Blazar Jets}",
      journal = {\apj},
         year = 2019,
        month = dec,
       volume = {887},
       number = {2},
          eid = {133},
        pages = {133},
          doi = {10.3847/1538-4357/ab552a},
archivePrefix = {arXiv},
       eprint = {1911.02834},
 primaryClass = {astro-ph.HE},
       adsurl = {https://ui.adsabs.harvard.edu/abs/2019ApJ...887..133B}
}

@ARTICLE{boettcher2002,
       author = {{B{\"o}ttcher}, Markus and {Chiang}, James},
        title = "{X-Ray Spectral Variability Signatures of Flares in BL Lacertae Objects}",
      journal = {\apj},
         year = 2002,
        month = dec,
       volume = {581},
       number = {1},
        pages = {127-142},
          doi = {10.1086/344155},
archivePrefix = {arXiv},
       eprint = {astro-ph/0208238},
 primaryClass = {astro-ph},
       adsurl = {https://ui.adsabs.harvard.edu/abs/2002ApJ...581..127B}
}

@ARTICLE{bolis2024a,
       author = {{Bolis}, F. and {Sobacchi}, E. and {Tavecchio}, F.},
        title = "{Multifrequency polarimetry of high-synchrotron peaked blazars probes the shape of their jets}",
      journal = {\aap},
         year = 2024,
        month = oct,
       volume = {690},
          eid = {A14},
        pages = {A14},
          doi = {10.1051/0004-6361/202450387},
archivePrefix = {arXiv},
       eprint = {2407.10578},
 primaryClass = {astro-ph.HE},
       adsurl = {https://ui.adsabs.harvard.edu/abs/2024A&A...690A..14B}
}

@ARTICLE{bonning2012,
       author = {{Bonning}, Erin and {Urry}, C. Megan and {Bailyn}, Charles and {Buxton}, Michelle and {Chatterjee}, Ritaban and {Coppi}, Paolo and {Fossati}, Giovanni and {Isler}, Jedidah and {Maraschi}, Laura},
        title = "{SMARTS Optical and Infrared Monitoring of 12 Gamma-Ray Bright Blazars}",
      journal = {\apj},
         year = 2012,
        month = sep,
       volume = {756},
       number = {1},
          eid = {13},
        pages = {13},
          doi = {10.1088/0004-637X/756/1/13},
archivePrefix = {arXiv},
       eprint = {1201.4380},
 primaryClass = {astro-ph.HE},
       adsurl = {https://ui.adsabs.harvard.edu/abs/2012ApJ...756...13B}
}

@ARTICLE{bonometto1973,
       author = {{Bonometto}, S. and {Saggion}, A.},
        title = "{Polarization in Inverse Compton Scattering of Synchrotron Radiation}",
      journal = {\aap},
         year = 1973,
        month = feb,
       volume = {23},
        pages = {9},
       adsurl = {https://ui.adsabs.harvard.edu/abs/1973A&A....23....9B}
}

@ARTICLE{bregman1990a,
       author = {{Bregman}, Joel N.},
        title = "{Continuum radiation from active galactic nuclei}",
      journal = {\aapr},
         year = 1990,
        month = jan,
       volume = {2},
       number = {2},
        pages = {125-166},
          doi = {10.1007/BF00872765},
       adsurl = {https://ui.adsabs.harvard.edu/abs/1990A&ARv...2..125B}
}

@ARTICLE{bregman1990b,
       author = {{Bregman}, Joel N. and {Glassgold}, A.~E. and {Huggins}, P.~J. and {Neugebauer}, G. and {Soifer}, B.~T. and {Matthews}, K. and {Elias}, J.~H. and {Webb}, J.~R. and {Pollock}, J.~T. and {Leacock}, R.~J. and {Smith}, A.~G. and {Aller}, H.~D. and {Aller}, M.~F. and {Hughes}, P.~A. and {Maccagni}, D. and {Garilli}, B. and {Giommi}, P. and {Miller}, J.~S. and {Stephens}, S. and {Balonek}, T.~J. and {Dent}, W.~A. and {Kinsel}, W. and {Wisniewski}, W.~Z. and {Williams}, P.~M. and {Brand}, P.~W.~J.~L. and {Ku}, W.~H. -M.},
        title = "{Multifrequency Observations of BL Lacertae}",
      journal = {\apj},
         year = 1990,
        month = apr,
       volume = {352},
        pages = {574},
          doi = {10.1086/168559},
       adsurl = {https://ui.adsabs.harvard.edu/abs/1990ApJ...352..574B}
}

@ARTICLE{bregman1986,
       author = {{Bregman}, Joel N. and {Glassgold}, A.~E. and {Huggins}, P.~J. and {Neugebauer}, G. and {Soifer}, B.~T. and {Matthews}, K. and {Elias}, J. and {Webb}, J. and {Pollock}, J.~T. and {Pica}, A.~J. and {Leacock}, R.~J. and {Smith}, A.~G. and {Aller}, H.~D. and {Aller}, M.~F. and {Hodge}, P.~E. and {Dent}, W.~A. and {Balonek}, T.~J. and {Barvainis}, R.~E. and {Roellig}, T.~P.~L. and {Wisniewski}, W.~Z. and {Rieke}, G.~H. and {Lebofsky}, M.~J. and {Wills}, B.~J. and {Wills}, D. and {Ku}, W.~H. -M. and {Bregman}, Jesse D. and {Witteborn}, F.~C. and {Lester}, D.~F. and {Impey}, C.~D. and {Hackwell}, J.~A.},
        title = "{Multifrequency Observations of the Superluminal Quasar 3C 345}",
      journal = {\apj},
         year = 1986,
        month = feb,
       volume = {301},
        pages = {708},
          doi = {10.1086/163938},
       adsurl = {https://ui.adsabs.harvard.edu/abs/1986ApJ...301..708B}
}

@ARTICLE{bregman1984,
       author = {{Bregman}, J.~N. and {Glassgold}, A.~E. and {Huggins}, P.~J. and {Aller}, H.~D. and {Aller}, M.~F. and {Hodge}, P.~E. and {Rieke}, G.~H. and {Lebofsky}, M.~J. and {Pollock}, J.~T. and {Pica}, A.~J. and {Leacock}, R.~J. and {Smith}, A.~G. and {Webb}, J. and {Balonek}, T.~J. and {Dent}, W.~A. and {O'Dea}, C.~P. and {Ku}, W.~H. -M. and {Schwartz}, D.~A. and {Miller}, J.~S. and {Rudy}, R.~J. and {Levan}, P.~D.},
        title = "{Multifrequency observations of the BL Lacertae object 0735 +178.}",
      journal = {\apj},
         year = 1984,
        month = jan,
       volume = {276},
        pages = {454-465},
          doi = {10.1086/161632},
       adsurl = {https://ui.adsabs.harvard.edu/abs/1984ApJ...276..454B}
}

@ARTICLE{brinkmann2005,
       author = {{Brinkmann}, W. and {Papadakis}, I.~E. and {Raeth}, C. and {Mimica}, P. and {Haberl}, F.},
        title = "{XMM-Newton timing mode observations of Mrk 421}",
      journal = {\aap},
         year = 2005,
        month = nov,
       volume = {443},
       number = {2},
        pages = {397-411},
          doi = {10.1051/0004-6361:20052767},
archivePrefix = {arXiv},
       eprint = {astro-ph/0508433},
 primaryClass = {astro-ph},
       adsurl = {https://ui.adsabs.harvard.edu/abs/2005A&A...443..397B}
}

@ARTICLE{britzen2018,
       author = {{Britzen}, S. and {Fendt}, C. and {Witzel}, G. and {Qian}, S. -J. and {Pashchenko}, I.~N. and {Kurtanidze}, O. and {Zajacek}, M. and {Martinez}, G. and {Karas}, V. and {Aller}, M. and {Aller}, H. and {Eckart}, A. and {Nilsson}, K. and {Ar{\'e}valo}, P. and {Cuadra}, J. and {Subroweit}, M. and {Witzel}, A.},
        title = "{OJ287: deciphering the `Rosetta stone of blazars}",
      journal = {\mnras},
         year = 2018,
        month = aug,
       volume = {478},
       number = {3},
        pages = {3199-3219},
          doi = {10.1093/mnras/sty1026},
       adsurl = {https://ui.adsabs.harvard.edu/abs/2018MNRAS.478.3199B}
}

@ARTICLE{britzen2017,
       author = {{Britzen}, S. and {Qian}, S. -J. and {Steffen}, W. and {Kun}, E. and {Karouzos}, M. and {Gergely}, L. and {Schmidt}, J. and {Aller}, M. and {Aller}, H. and {Krause}, M. and {Fendt}, C. and {B{\"o}ttcher}, M. and {Witzel}, A. and {Eckart}, A. and {Moser}, L.},
        title = "{A swirling jet in the quasar 1308+326}",
      journal = {\aap},
         year = 2017,
        month = jun,
       volume = {602},
          eid = {A29},
        pages = {A29},
          doi = {10.1051/0004-6361/201629999},
       adsurl = {https://ui.adsabs.harvard.edu/abs/2017A&A...602A..29B}
}

@ARTICLE{britzen2010b,
       author = {{Britzen}, S. and {Witzel}, A. and {Gong}, B.~P. and {Zhang}, J.~W. and {Gopal-Krishna} and {Goyal}, A. and {Aller}, M.~F. and {Aller}, H.~D. and {Zensus}, J.~A.},
        title = "{Understanding BL Lacertae objects. Structural and kinematic mode changes in the BL Lac object PKS 0735+178}",
      journal = {\aap},
         year = 2010,
        month = jun,
       volume = {515},
          eid = {A105},
        pages = {A105},
          doi = {10.1051/0004-6361/200913685},
archivePrefix = {arXiv},
       eprint = {1002.3531},
 primaryClass = {astro-ph.CO},
       adsurl = {https://ui.adsabs.harvard.edu/abs/2010A&A...515A.105B}
}

@ARTICLE{britzen2010a,
       author = {{Britzen}, S. and {Kudryavtseva}, N.~A. and {Witzel}, A. and {Campbell}, R.~M. and {Ros}, E. and {Karouzos}, M. and {Mehta}, A. and {Aller}, M.~F. and {Aller}, H.~D. and {Beckert}, T. and {Zensus}, J.~A.},
        title = "{The kinematics in the pc-scale jets of AGN. The case of S5 1803+784}",
      journal = {\aap},
         year = 2010,
        month = feb,
       volume = {511},
          eid = {A57},
        pages = {A57},
          doi = {10.1051/0004-6361/20079267},
archivePrefix = {arXiv},
       eprint = {1001.1973},
 primaryClass = {astro-ph.CO},
       adsurl = {https://ui.adsabs.harvard.edu/abs/2010A&A...511A..57B}
}

@ARTICLE{brown1989,
       author = {{Brown}, L.~M.~J. and {Robson}, E.~I. and {Gear}, W.~K. and {Smith}, M.~G.},
        title = "{Multifrequency Observations of Blazars. IV. The Variability of the Radio to Ultraviolet Continuum}",
      journal = {\apj},
         year = 1989,
        month = may,
       volume = {340},
        pages = {150},
          doi = {10.1086/167381},
       adsurl = {https://ui.adsabs.harvard.edu/abs/1989ApJ...340..150B}
}

@ARTICLE{caccianiga2004,
       author = {{Caccianiga}, A. and {March{\~a}}, M.~J.~M.},
        title = "{The CLASS blazar survey: testing the blazar sequence}",
      journal = {\mnras},
         year = 2004,
        month = mar,
       volume = {348},
       number = {3},
        pages = {937-954},
          doi = {10.1111/j.1365-2966.2004.07415.x},
archivePrefix = {arXiv},
       eprint = {astro-ph/0311384},
 primaryClass = {astro-ph},
       adsurl = {https://ui.adsabs.harvard.edu/abs/2004MNRAS.348..937C}
}

@ARTICLE{camenzind1992,
       author = {{Camenzind}, M. and {Krockenberger}, M.},
        title = "{The lighthouse effect of relativistic jets in blazars. A geometric originof intraday variability.}",
      journal = {\aap},
         year = 1992,
        month = feb,
       volume = {255},
        pages = {59-62},
       adsurl = {https://ui.adsabs.harvard.edu/abs/1992A&A...255...59C}
}

@ARTICLE{caproni2017,
       author = {{Caproni}, Anderson and {Abraham}, Zulema and {Motter}, Juliana Cristina and {Monteiro}, Hektor},
        title = "{Jet Precession Driven by a Supermassive Black Hole Binary System in the BL Lac Object PG 1553+113}",
      journal = {\apjl},
         year = 2017,
        month = dec,
       volume = {851},
       number = {2},
          eid = {L39},
        pages = {L39},
          doi = {10.3847/2041-8213/aa9fea},
archivePrefix = {arXiv},
       eprint = {1712.06881},
 primaryClass = {astro-ph.GA},
       adsurl = {https://ui.adsabs.harvard.edu/abs/2017ApJ...851L..39C}
}

@ARTICLE{carnerero2017,
       author = {{Carnerero}, M.~I. and {Raiteri}, C.~M. and {Villata}, M. and {Acosta-Pulido}, J.~A. and {Larionov}, V.~M. and {Smith}, P.~S. and {D'Ammando}, F. and {Agudo}, I. and {Ar{\'e}valo}, M.~J. and {Bachev}, R. and {Barnes}, J. and {Boeva}, S. and {Bozhilov}, V. and {Carosati}, D. and {Casadio}, C. and {Chen}, W.~P. and {Damljanovic}, G. and {Eswaraiah}, E. and {Forn{\'e}}, E. and {Gantchev}, G. and {G{\'o}mez}, J.~L. and {Gonz{\'a}lez-Morales}, P.~A. and {Gri{\~n}{\'o}n-Mar{\'\i}n}, A.~B. and {Grishina}, T.~S. and {Holden}, M. and {Ibryamov}, S. and {Joner}, M.~D. and {Jordan}, B. and {Jorstad}, S.~G. and {Joshi}, M. and {Kopatskaya}, E.~N. and {Koptelova}, E. and {Kurtanidze}, O.~M. and {Kurtanidze}, S.~O. and {Larionova}, E.~G. and {Larionova}, L.~V. and {Latev}, G. and {L{\'a}zaro}, C. and {Ligustri}, R. and {Lin}, H.~C. and {Marscher}, A.~P. and {Mart{\'\i}nez-Lombilla}, C. and {McBreen}, B. and {Mihov}, B. and {Molina}, S.~N. and {Moody}, J.~W. and {Morozova}, D.~A. and {Nikolashvili}, M.~G. and {Nilsson}, K. and {Ovcharov}, E. and {Pace}, C. and {Panwar}, N. and {Pastor Yabar}, A. and {Pearson}, R.~L. and {Pinna}, F. and {Protasio}, C. and {Rizzi}, N. and {Redondo-Lorenzo}, F.~J. and {Rodr{\'\i}guez-Coira}, G. and {Ros}, J.~A. and {Sadun}, A.~C. and {Savchenko}, S.~S. and {Semkov}, E. and {Slavcheva-Mihova}, L. and {Smith}, N. and {Strigachev}, A. and {Troitskaya}, Yu. V. and {Troitsky}, I.~S. and {Vasilyev}, A.~A. and {Vince}, O.},
        title = "{Dissecting the long-term emission behaviour of the BL Lac object Mrk 421}",
      journal = {\mnras},
         year = 2017,
        month = dec,
       volume = {472},
       number = {4},
        pages = {3789-3804},
          doi = {10.1093/mnras/stx2185},
archivePrefix = {arXiv},
       eprint = {1709.02237},
 primaryClass = {astro-ph.HE},
       adsurl = {https://ui.adsabs.harvard.edu/abs/2017MNRAS.472.3789C}
}

@ARTICLE{carnerero2015,
       author = {{Carnerero}, M.~I. and {Raiteri}, C.~M. and {Villata}, M. and {Acosta-Pulido}, J.~A. and {D'Ammando}, F. and {Smith}, P.~S. and {Larionov}, V.~M. and {Agudo}, I. and {Ar{\'e}valo}, M.~J. and {Arkharov}, A.~A. and {Bach}, U. and {Bachev}, R. and {Ben{\'\i}tez}, E. and {Blinov}, D.~A. and {Bozhilov}, V. and {Buemi}, C.~S. and {Bueno Bueno}, A. and {Carosati}, D. and {Casadio}, C. and {Chen}, W.~P. and {Damljanovic}, G. and {di Paola}, A. and {Efimova}, N.~V. and {Ehgamberdiev}, Sh. A. and {Giroletti}, M. and {G{\'o}mez}, J.~L. and {Gonz{\'a}lez-Morales}, P.~A. and {Grinon-Marin}, A.~B. and {Grishina}, T.~S. and {Gurwell}, M.~A. and {Hiriart}, D. and {Hsiao}, H.~Y. and {Ibryamov}, S. and {Jorstad}, S.~G. and {Joshi}, M. and {Kopatskaya}, E.~N. and {Kurtanidze}, O.~M. and {Kurtanidze}, S.~O. and {L{\"a}hteenm{\"a}ki}, A. and {Larionova}, E.~G. and {Larionova}, L.~V. and {L{\'a}zaro}, C. and {Leto}, P. and {Lin}, C.~S. and {Lin}, H.~C. and {Manilla-Robles}, A.~I. and {Marscher}, A.~P. and {McHardy}, I.~M. and {Metodieva}, Y. and {Mirzaqulov}, D.~O. and {Mokrushina}, A.~A. and {Molina}, S.~N. and {Morozova}, D.~A. and {Nikolashvili}, M.~G. and {Orienti}, M. and {Ovcharov}, E. and {Panwar}, N. and {Pastor Yabar}, A. and {Puerto Gim{\'e}nez}, I. and {Ramakrishnan}, V. and {Richter}, G.~M. and {Rossini}, M. and {Sigua}, L.~A. and {Strigachev}, A. and {Taylor}, B. and {Tornikoski}, M. and {Trigilio}, C. and {Troitskaya}, Yu. V. and {Troitsky}, I.~S. and {Umana}, G. and {Valcheva}, A. and {Velasco}, S. and {Vince}, O. and {Wehrle}, A.~E. and {Wiesemeyer}, H.},
        title = "{Multiwavelength behaviour of the blazar OJ 248 from radio to {\ensuremath{\gamma}}-rays}",
      journal = {\mnras},
         year = 2015,
        month = jul,
       volume = {450},
       number = {3},
        pages = {2677-2691},
          doi = {10.1093/mnras/stv823},
archivePrefix = {arXiv},
       eprint = {1505.00916},
 primaryClass = {astro-ph.HE},
       adsurl = {https://ui.adsabs.harvard.edu/abs/2015MNRAS.450.2677C}
}

@ARTICLE{chang2019,
       author = {{Chang}, Y. -L. and {Arsioli}, B. and {Giommi}, P. and {Padovani}, P. and {Brandt}, C.~H.},
        title = "{The 3HSP catalogue of extreme and high-synchrotron peaked blazars}",
      journal = {\aap},
         year = 2019,
        month = dec,
       volume = {632},
          eid = {A77},
        pages = {A77},
          doi = {10.1051/0004-6361/201834526},
archivePrefix = {arXiv},
       eprint = {1909.08279},
 primaryClass = {astro-ph.HE},
       adsurl = {https://ui.adsabs.harvard.edu/abs/2019A&A...632A..77C}
}

@ARTICLE{clements1995,
       author = {{Clements}, S.~D. and {Smith}, A.~G. and {Aller}, H.~D. and {Aller}, M.~F.},
        title = "{Correlation Analysis of Optical and Radio Light Curves for a Large Sample of Active Galactic Nuclei}",
      journal = {\aj},
         year = 1995,
        month = aug,
       volume = {110},
        pages = {529},
          doi = {10.1086/117540},
       adsurl = {https://ui.adsabs.harvard.edu/abs/1995AJ....110..529C}
}

@ARTICLE{cohen2015,
       author = {{Cohen}, M.~H. and {Meier}, D.~L. and {Arshakian}, T.~G. and {Clausen-Brown}, E. and {Homan}, D.~C. and {Hovatta}, T. and {Kovalev}, Y.~Y. and {Lister}, M.~L. and {Pushkarev}, A.~B. and {Richards}, J.~L. and {Savolainen}, T.},
        title = "{Studies of the Jet in Bl Lacertae. II. Superluminal Alfv{\'e}n Waves}",
      journal = {\apj},
         year = 2015,
        month = apr,
       volume = {803},
       number = {1},
          eid = {3},
        pages = {3},
          doi = {10.1088/0004-637X/803/1/3},
archivePrefix = {arXiv},
       eprint = {1409.3599},
 primaryClass = {astro-ph.HE},
       adsurl = {https://ui.adsabs.harvard.edu/abs/2015ApJ...803....3C}
}

@ARTICLE{cohen1976,
       author = {{Cohen}, M.~H. and {Moffet}, A.~T. and {Romney}, J.~D. and {Schilizzi}, R.~T. and {Seielstad}, G.~A. and {Kellermann}, K.~I. and {Purcell}, G.~H. and {Shaffer}, D.~B. and {Pauliny-Toth}, I.~I.~K. and {Preuss}, E. and {Witzel}, A. and {Rinehart}, R.},
        title = "{Rapid increase in the size of 3C 345.}",
      journal = {\apjl},
         year = 1976,
        month = may,
       volume = {206},
        pages = {L1-L3},
          doi = {10.1086/182119},
       adsurl = {https://ui.adsabs.harvard.edu/abs/1976ApJ...206L...1C}
}

@ARTICLE{conway1995,
       author = {{Conway}, J.~E. and {Wrobel}, J.~M.},
        title = "{A Helical Jet in the Orthogonally Misaligned BL Lacertae Object Markarian 501 (B1652+398)}",
      journal = {\apj},
         year = 1995,
        month = jan,
       volume = {439},
        pages = {98},
          doi = {10.1086/175155},
       adsurl = {https://ui.adsabs.harvard.edu/abs/1995ApJ...439...98C}
}

@ARTICLE{corbett2000,
       author = {{Corbett}, E.~A. and {Robinson}, A. and {Axon}, D.~J. and {Hough}, J.~H.},
        title = "{A Seyfert-like nucleus concealed in BL Lacertae?}",
      journal = {\mnras},
         year = 2000,
        month = jan,
       volume = {311},
       number = {3},
        pages = {485-492},
          doi = {10.1046/j.1365-8711.2000.03045.x},
       adsurl = {https://ui.adsabs.harvard.edu/abs/2000MNRAS.311..485C}
}

@ARTICLE{covino2019,
       author = {{Covino}, S. and {Sandrinelli}, A. and {Treves}, A.},
        title = "{Gamma-ray quasi-periodicities of blazars. A cautious approach}",
      journal = {\mnras},
         year = 2019,
        month = jan,
       volume = {482},
       number = {1},
        pages = {1270-1274},
          doi = {10.1093/mnras/sty2720},
archivePrefix = {arXiv},
       eprint = {1810.02409},
 primaryClass = {astro-ph.HE},
       adsurl = {https://ui.adsabs.harvard.edu/abs/2019MNRAS.482.1270C}
}

@ARTICLE{covino2015,
       author = {{Covino}, S. and {Baglio}, M.~C. and {Foschini}, L. and {Sandrinelli}, A. and {Tavecchio}, F. and {Treves}, A. and {Zhang}, H. and {Barres de Almeida}, U. and {Bonnoli}, G. and {B{\"o}ttcher}, M. and {Cecconi}, M. and {D'Ammando}, F. and {di Fabrizio}, L. and {Giarrusso}, M. and {Leone}, F. and {Lindfors}, E. and {Lorenzi}, V. and {Molinari}, E. and {Paiano}, S. and {Prandini}, E. and {Raiteri}, C.~M. and {Stamerra}, A. and {Tagliaferri}, G.},
        title = "{Short timescale photometric and polarimetric behavior of two BL Lacertae type objects}",
      journal = {\aap},
         year = 2015,
        month = jun,
       volume = {578},
          eid = {A68},
        pages = {A68},
          doi = {10.1051/0004-6361/201525674},
archivePrefix = {arXiv},
       eprint = {1504.03020},
 primaryClass = {astro-ph.HE},
       adsurl = {https://ui.adsabs.harvard.edu/abs/2015A&A...578A..68C}
}

@ARTICLE{dabrusco2019,
       author = {{D'Abrusco}, Raffaele and {{\'A}lvarez Crespo}, Nuria and {Massaro}, Francesco and {Campana}, Riccardo and {Chavushyan}, Vahram and {Landoni}, Marco and {La Franca}, Fabio and {Masetti}, Nicola and {Milisavljevic}, Dan and {Paggi}, Alessandro and {Ricci}, Federica and {Smith}, Howard A.},
        title = "{Two New Catalogs of Blazar Candidates in the WISE Infrared Sky}",
      journal = {\apjs},
         year = 2019,
        month = may,
       volume = {242},
       number = {1},
          eid = {4},
        pages = {4},
          doi = {10.3847/1538-4365/ab16f4},
archivePrefix = {arXiv},
       eprint = {1903.11124},
 primaryClass = {astro-ph.HE},
       adsurl = {https://ui.adsabs.harvard.edu/abs/2019ApJS..242....4D}
}

@ARTICLE{dammando2013,
       author = {{D'Ammando}, F. and {Antolini}, E. and {Tosti}, G. and {Finke}, J. and {Ciprini}, S. and {Larsson}, S. and {Ajello}, M. and {Covino}, S. and {Gasparrini}, D. and {Gurwell}, M. and {Hauser}, M. and {Romano}, P. and {Schinzel}, F. and {Wagner}, S.~J. and {Impiombato}, D. and {Perri}, M. and {Persic}, M. and {Pian}, E. and {Polenta}, G. and {Sbarufatti}, B. and {Treves}, A. and {Vercellone}, S. and {Wehrle}, A. and {Zook}, A.},
        title = "{Long-term monitoring of PKS 0537-441 with Fermi-LAT and multiwavelength observations}",
      journal = {\mnras},
         year = 2013,
        month = may,
       volume = {431},
       number = {3},
        pages = {2481-2492},
          doi = {10.1093/mnras/stt344},
archivePrefix = {arXiv},
       eprint = {1302.5439},
 primaryClass = {astro-ph.HE},
       adsurl = {https://ui.adsabs.harvard.edu/abs/2013MNRAS.431.2481D}
}

@ARTICLE{dediego1998,
       author = {{de Diego}, J.~A. and {Dultzin-Hacyan}, D. and {Ram{\'\i}rez}, A. and {Ben{\'\i}tez}, E.},
        title = "{A Comparative Study of the Microvariability Properties in Radio-loud and Radio-quiet Quasars}",
      journal = {\apj},
         year = 1998,
        month = jul,
       volume = {501},
       number = {1},
        pages = {69-81},
          doi = {10.1086/305817},
       adsurl = {https://ui.adsabs.harvard.edu/abs/1998ApJ...501...69D}
}

@ARTICLE{dejaeger2023,
       author = {{de Jaeger}, T. and {Shappee}, B.~J. and {Kochanek}, C.~S. and {Hinkle}, J.~T. and {Garrappa}, S. and {Liodakis}, I. and {Franckowiak}, A. and {Stanek}, K.~Z. and {Beacom}, J.~F. and {Prieto}, J.~L.},
        title = "{Optical/{\ensuremath{\gamma}}-ray blazar flare correlations: understanding the high-energy emission process using ASAS-SN and Fermi light curves}",
      journal = {\mnras},
         year = 2023,
        month = mar,
       volume = {519},
       number = {4},
        pages = {6349-6380},
          doi = {10.1093/mnras/stad060},
archivePrefix = {arXiv},
       eprint = {2210.16329},
 primaryClass = {astro-ph.HE},
       adsurl = {https://ui.adsabs.harvard.edu/abs/2023MNRAS.519.6349D}
}

@ARTICLE{delaparra2025,
       author = {{de la Parra}, P.~V. and {Kiehlmann}, S. and {Mr{\'o}z}, P. and {Readhead}, A.~C.~S. and {Synani}, A. and {Begelman}, M.~C. and {Blandford}, R.~D. and {Ding}, Y. and {Harrison}, F. and {Liodakis}, I. and {Max-Moerbeck}, W. and {Pavlidou}, V. and {Reeves}, R. and {Vallisneri}, M. and {Aller}, M.~F. and {Graham}, M.~J. and {Hovatta}, T. and {Lawrence}, C.~R. and {Lazio}, T.~J.~W. and {Mahabal}, A.~A. and {Molina}, B. and {O'Neill}, S. and {Pearson}, T.~J. and {Ravi}, V. and {Tassis}, K. and {Zensus}, J.~A.},
        title = "{PKS J0805-0111: A Second Owens Valley Radio Observatory Blazar Showing Highly Significant Sinusoidal Radio Variability{\textemdash}The Tip of the Iceberg}",
      journal = {\apj},
         year = 2025,
        month = jul,
       volume = {987},
       number = {2},
          eid = {191},
        pages = {191},
          doi = {10.3847/1538-4357/addc60},
archivePrefix = {arXiv},
       eprint = {2408.02645},
 primaryClass = {astro-ph.HE},
       adsurl = {https://ui.adsabs.harvard.edu/abs/2025ApJ...987..191D}
}

@ARTICLE{dermer1992,
       author = {{Dermer}, C.~D. and {Schlickeiser}, R. and {Mastichiadis}, A.},
        title = "{High-energy gamma radiation from extragalactic radio sources.}",
      journal = {\aap},
         year = 1992,
        month = mar,
       volume = {256},
        pages = {L27-L30},
       adsurl = {https://ui.adsabs.harvard.edu/abs/1992A&A...256L..27D}
}

@ARTICLE{dey2018,
       author = {{Dey}, Lankeswar and {Valtonen}, M.~J. and {Gopakumar}, A. and {Zola}, S. and {Hudec}, R. and {Pihajoki}, P. and {Ciprini}, S. and {Matsumoto}, K. and {Sadakane}, K. and {Kidger}, M. and {Nilsson}, K. and {Mikkola}, S. and {Sillanp{\"a}{\"a}}, A. and {Takalo}, L.~O. and {Lehto}, H.~J. and {Berdyugin}, A. and {Piirola}, V. and {Jermak}, H. and {Baliyan}, K.~S. and {Pursimo}, T. and {Caton}, D.~B. and {Alicavus}, F. and {Baransky}, A. and {Blay}, P. and {Boumis}, P. and {Boyd}, D. and {Campas Torrent}, M. and {Campos}, F. and {Carrillo G{\'o}mez}, J. and {Chandra}, S. and {Chavushyan}, V. and {Dalessio}, J. and {Debski}, B. and {Drozdz}, M. and {Er}, H. and {Erdem}, A. and {Escartin P{\'e}rez}, A. and {Fallah Ramazani}, V. and {Filippenko}, A.~V. and {Gafton}, E. and {Ganesh}, S. and {Garcia}, F. and {Gazeas}, K. and {Godunova}, V. and {G{\'o}mez Pinilla}, F. and {Gopinathan}, M. and {Haislip}, J.~B. and {Harmanen}, J. and {Hurst}, G. and {Jan{\'\i}k}, J. and {Jelinek}, M. and {Joshi}, A. and {Kagitani}, M. and {Karjalainen}, R. and {Kaur}, N. and {Keel}, W.~C. and {Kouprianov}, V.~V. and {Kundera}, T. and {Kurowski}, S. and {Kvammen}, A. and {LaCluyze}, A.~P. and {Lee}, B.~C. and {Liakos}, A. and {Lindfors}, E. and {Lozano de Haro}, J. and {Mugrauer}, M. and {Naves Nogues}, R. and {Neely}, A.~W. and {Nelson}, R.~H. and {Ogloza}, W. and {Okano}, S. and {Pajdosz-{\'S}mierciak}, U. and {Pandey}, J.~C. and {Perri}, M. and {Poyner}, G. and {Provencal}, J. and {Raj}, A. and {Reichart}, D.~E. and {Reinthal}, R. and {Reynolds}, T. and {Saario}, J. and {Sadegi}, S. and {Sakanoi}, T. and {Salto Gonz{\'a}lez}, J. -L. and {Sameer} and {Schweyer}, T. and {Simon}, A. and {Siwak}, M. and {Sold{\'a}n Alfaro}, F.~C. and {Sonbas}, E. and {Steele}, I. and {Stocke}, J.~T. and {Strobl}, J. and {Tomov}, T. and {Tremosa Espasa}, L. and {Valdes}, J.~R. and {Valero P{\'e}rez}, J. and {Verrecchia}, F. and {Vasylenko}, V. and {Webb}, J.~R. and {Yoneda}, M. and {Zejmo}, M. and {Zheng}, W. and {Zielinski}, P.},
        title = "{Authenticating the Presence of a Relativistic Massive Black Hole Binary in OJ 287 Using Its General Relativity Centenary Flare: Improved Orbital Parameters}",
      journal = {\apj},
         year = 2018,
        month = oct,
       volume = {866},
       number = {1},
          eid = {11},
        pages = {11},
          doi = {10.3847/1538-4357/aadd95},
archivePrefix = {arXiv},
       eprint = {1808.09309},
 primaryClass = {astro-ph.HE},
       adsurl = {https://ui.adsabs.harvard.edu/abs/2018ApJ...866...11D}
}

@ARTICLE{digesu2023,
       author = {{Di Gesu}, Laura and {Marshall}, Herman L. and {Ehlert}, Steven R. and {Kim}, Dawoon E. and {Donnarumma}, Immacolata and {Tavecchio}, Fabrizio and {Liodakis}, Ioannis and {Kiehlmann}, Sebastian and {Agudo}, Iv{\'a}n and {Jorstad}, Svetlana G. and {Muleri}, Fabio and {Marscher}, Alan P. and {Puccetti}, Simonetta and {Middei}, Riccardo and {Perri}, Matteo and {Pacciani}, Luigi and {Negro}, Michela and {Romani}, Roger W. and {Di Marco}, Alessandro and {Blinov}, Dmitry and {Bourbah}, Ioakeim G. and {Kontopodis}, Evangelos and {Mandarakas}, Nikos and {Romanopoulos}, Stylianos and {Skalidis}, Raphael and {Vervelaki}, Anna and {Casadio}, Carolina and {Escudero}, Juan and {Myserlis}, Ioannis and {Gurwell}, Mark A. and {Rao}, Ramprasad and {Keating}, Garrett K. and {Kouch}, Pouya M. and {Lindfors}, Elina and {Aceituno}, Francisco Jos{\'e} and {Bernardos}, Maria I. and {Bonnoli}, Giacomo and {Casanova}, V{\'\i}ctor and {Garc{\'\i}a-Comas}, Maya and {Ag{\'\i}s-Gonz{\'a}lez}, Beatriz and {Husillos}, C{\'e}sar and {Marchini}, Alessandro and {Sota}, Alfredo and {Imazawa}, Ryo and {Sasada}, Mahito and {Fukazawa}, Yasushi and {Kawabata}, Koji S. and {Uemura}, Makoto and {Mizuno}, Tsunefumi and {Nakaoka}, Tatsuya and {Akitaya}, Hiroshi and {Savchenko}, Sergey S. and {Vasilyev}, Andrey A. and {G{\'o}mez}, Jos{\'e} L. and {Antonelli}, Lucio A. and {Barnouin}, Thibault and {Bonino}, Raffaella and {Cavazzuti}, Elisabetta and {Costamante}, Luigi and {Chen}, Chien-Ting and {Cibrario}, Nicol{\`o} and {De Rosa}, Alessandra and {Di Pierro}, Federico and {Errando}, Manel and {Kaaret}, Philip and {Karas}, Vladimir and {Krawczynski}, Henric and {Lisalda}, Lindsey and {Madejski}, Grzegorz and {Malacaria}, Christian and {Marin}, Fr{\'e}d{\'e}ric and {Marinucci}, Andrea and {Massaro}, Francesco and {Matt}, Giorgio and {Mitsuishi}, Ikuyuki and {O'Dell}, Stephen L. and {Paggi}, Alessandro and {Peirson}, Abel L. and {Petrucci}, Pierre-Olivier and {Ramsey}, Brian D. and {Tennant}, Allyn F. and {Wu}, Kinwah and {Bachetti}, Matteo and {Baldini}, Luca and {Baumgartner}, Wayne H. and {Bellazzini}, Ronaldo and {Bianchi}, Stefano and {Bongiorno}, Stephen D. and {Brez}, Alessandro and {Bucciantini}, Niccol{\`o} and {Capitanio}, Fiamma and {Castellano}, Simone and {Ciprini}, Stefano and {Costa}, Enrico and {Del Monte}, Ettore and {Di Lalla}, Niccol{\`o} and {Doroshenko}, Victor and {Dov{\v{c}}iak}, Michal and {Enoto}, Teruaki and {Evangelista}, Yuri and {Fabiani}, Sergio and {Ferrazzoli}, Riccardo and {Garcia}, Javier A. and {Gunji}, Shuichi and {Hayashida}, Kiyoshi and {Heyl}, Jeremy and {Iwakiri}, Wataru and {Kislat}, Fabian and {Kitaguchi}, Takao and {Kolodziejczak}, Jeffery J. and {La Monaca}, Fabio and {Latronico}, Luca and {Maldera}, Simone and {Manfreda}, Alberto and {Ng}, C. -Y. and {Omodei}, Nicola and {Oppedisano}, Chiara and {Papitto}, Alessandro and {Pavlov}, George G. and {Pesce-Rollins}, Melissa and {Pilia}, Maura and {Possenti}, Andrea and {Poutanen}, Juri and {Rankin}, John and {Ratheesh}, Ajay and {Roberts}, Oliver J. and {Sgr{\`o}}, Carmelo and {Slane}, Patrick and {Soffitta}, Paolo and {Spandre}, Gloria and {Swartz}, Douglas A. and {Tamagawa}, Toru and {Taverna}, Roberto and {Tawara}, Yuzuru and {Thomas}, Nicholas E. and {Tombesi}, Francesco and {Trois}, Alessio and {Tsygankov}, Sergey S. and {Turolla}, Roberto and {Vink}, Jacco and {Weisskopf}, Martin C. and {Xie}, Fei and {Zane}, Silvia},
        title = "{Discovery of X-ray polarization angle rotation in the jet from blazar Mrk 421.}",
      journal = {Nature Astronomy},
         year = 2023,
        month = oct,
       volume = {7},
        pages = {1245-1258},
          doi = {10.1038/s41550-023-02032-7},
archivePrefix = {arXiv},
       eprint = {2305.13497},
 primaryClass = {astro-ph.HE},
       adsurl = {https://ui.adsabs.harvard.edu/abs/2023NatAs...7.1245D}
}

@ARTICLE{digesu2022,
       author = {{Di Gesu}, Laura and {Donnarumma}, Immacolata and {Tavecchio}, Fabrizio and {Agudo}, Iv{\'a}n and {Barnounin}, Thibault and {Cibrario}, Nicol{\`o} and {Di Lalla}, Niccol{\`o} and {Di Marco}, Alessandro and {Escudero}, Juan and {Errando}, Manel and {Jorstad}, Svetlana G. and {Kim}, Dawoon E. and {Kouch}, Pouya M. and {Liodakis}, Ioannis and {Lindfors}, Elina and {Madejski}, Grzegorz and {Marshall}, Herman L. and {Marscher}, Alan P. and {Middei}, Riccardo and {Muleri}, Fabio and {Myserlis}, Ioannis and {Negro}, Michela and {Omodei}, Nicola and {Pacciani}, Luigi and {Paggi}, Alessandro and {Perri}, Matteo and {Puccetti}, Simonetta and {Antonelli}, Lucio A. and {Bachetti}, Matteo and {Baldini}, Luca and {Baumgartner}, Wayne H. and {Bellazzini}, Ronaldo and {Bianchi}, Stefano and {Bongiorno}, Stephen D. and {Bonino}, Raffaella and {Brez}, Alessandro and {Bucciantini}, Niccol{\`o} and {Capitanio}, Fiamma and {Castellano}, Simone and {Cavazzuti}, Elisabetta and {Ciprini}, Stefano and {Costa}, Enrico and {De Rosa}, Alessandra and {Del Monte}, Ettore and {Doroshenko}, Victor and {Dov{\v{c}}iak}, Michal and {Ehlert}, Steven R. and {Enoto}, Teruaki and {Evangelista}, Yuri and {Fabiani}, Sergio and {Ferrazzoli}, Riccardo and {Garcia}, Javier A. and {Gunji}, Shuichi and {Hayashida}, Kiyoshi and {Heyl}, Jeremy and {Iwakiri}, Wataru and {Karas}, Vladimir and {Kitaguchi}, Takao and {Kolodziejczak}, Jeffery J. and {Krawczynski}, Henric and {La Monaca}, Fabio and {Latronico}, Luca and {Maldera}, Simone and {Manfreda}, Alberto and {Marin}, Fr{\'e}d{\'e}ric and {Marinucci}, Andrea and {Massaro}, Francesco and {Matt}, Giorgio and {Mitsuishi}, Ikuyuki and {Mizuno}, Tsunefumi and {Ng}, C. -Y. and {O'Dell}, Stephen L. and {Oppedisano}, Chiara and {Papitto}, Alessandro and {Pavlov}, George G. and {Peirson}, Abel L. and {Pesce-Rollins}, Melissa and {Petrucci}, Pierre-Olivier and {Pilia}, Maura and {Possenti}, Andrea and {Poutanen}, Juri and {Ramsey}, Brian D. and {Rankin}, John and {Ratheesh}, Ajay and {Romani}, Roger W. and {Sgr{\`o}}, Carmelo and {Slane}, Patrick and {Soffitta}, Paolo and {Spandre}, Gloria and {Tamagawa}, Toru and {Taverna}, Roberto and {Tawara}, Yuzuru and {Tennant}, Allyn F. and {Thomas}, Nicolas E. and {Tombesi}, Francesco and {Trois}, Alessio and {Tsygankov}, Sergey and {Turolla}, Roberto and {Vink}, Jacco and {Weisskopf}, Martin C. and {Wu}, Kinwah and {Xie}, Fei and {Zane}, Silvia},
        title = "{The X-Ray Polarization View of Mrk 421 in an Average Flux State as Observed by the Imaging X-Ray Polarimetry Explorer}",
      journal = {\apjl},
         year = 2022,
        month = oct,
       volume = {938},
       number = {1},
          eid = {L7},
        pages = {L7},
          doi = {10.3847/2041-8213/ac913a},
archivePrefix = {arXiv},
       eprint = {2209.07184},
 primaryClass = {astro-ph.HE},
       adsurl = {https://ui.adsabs.harvard.edu/abs/2022ApJ...938L...7D}
}

@ARTICLE{dingler2024,
       author = {{Dingler}, Ryne and {Smith}, Krista Lynne},
        title = "{Optical Variability Properties of Southern TESS Blazars}",
      journal = {\apj},
         year = 2024,
        month = sep,
       volume = {973},
       number = {1},
          eid = {10},
        pages = {10},
          doi = {10.3847/1538-4357/ad4f87},
archivePrefix = {arXiv},
       eprint = {2406.10346},
 primaryClass = {astro-ph.HE},
       adsurl = {https://ui.adsabs.harvard.edu/abs/2024ApJ...973...10D}
}

@ARTICLE{dong2020,
       author = {{Dong}, Lingyi and {Zhang}, Haocheng and {Giannios}, Dimitrios},
        title = "{Kink instabilities in relativistic jets can drive quasi-periodic radiation signatures}",
      journal = {\mnras},
         year = 2020,
        month = may,
       volume = {494},
       number = {2},
        pages = {1817-1825},
          doi = {10.1093/mnras/staa773},
archivePrefix = {arXiv},
       eprint = {2003.07765},
 primaryClass = {astro-ph.HE},
       adsurl = {https://ui.adsabs.harvard.edu/abs/2020MNRAS.494.1817D}
}

@ARTICLE{edelson2002,
       author = {{Edelson}, Rick and {Turner}, T.~J. and {Pounds}, Ken and {Vaughan}, Simon and {Markowitz}, Alex and {Marshall}, Herman and {Dobbie}, Paul and {Warwick}, Robert},
        title = "{X-Ray Spectral Variability and Rapid Variability of the Soft X-Ray Spectrum Seyfert 1 Galaxies Arakelian 564 and Ton S180}",
      journal = {\apj},
         year = 2002,
        month = apr,
       volume = {568},
       number = {2},
        pages = {610-626},
          doi = {10.1086/323779},
archivePrefix = {arXiv},
       eprint = {astro-ph/0108387},
 primaryClass = {astro-ph},
       adsurl = {https://ui.adsabs.harvard.edu/abs/2002ApJ...568..610E}
}

@ARTICLE{edelson1988,
       author = {{Edelson}, R.~A. and {Krolik}, J.~H.},
        title = "{The Discrete Correlation Function: A New Method for Analyzing Unevenly Sampled Variability Data}",
      journal = {\apj},
         year = 1988,
        month = oct,
       volume = {333},
        pages = {646},
          doi = {10.1086/166773},
       adsurl = {https://ui.adsabs.harvard.edu/abs/1988ApJ...333..646E}
}

@ARTICLE{efimov2002,
       author = {{Efimov}, Yu. S. and {Shakhovskoy}, N.~M. and {Takalo}, L.~O. and {Sillanp{\"a}{\"a}}, A.},
        title = "{Photopolarimetric monitoring of OJ 287 in 1994-1997}",
      journal = {\aap},
         year = 2002,
        month = jan,
       volume = {381},
        pages = {408-419},
          doi = {10.1051/0004-6361:20011515},
       adsurl = {https://ui.adsabs.harvard.edu/abs/2002A&A...381..408E}
}

@ARTICLE{ehlert2023,
       author = {{Ehlert}, Steven R. and {Liodakis}, Ioannis and {Middei}, Riccardo and {Marscher}, Alan P. and {Tavecchio}, Fabrizio and {Agudo}, Iv{\'a}n and {Kouch}, Pouya M. and {Lindfors}, Elina and {Nilsson}, Kari and {Myserlis}, Ioannis and {Gurwell}, Mark and {Rao}, Ramprasad and {Aceituno}, Francisco Jos{\'e} and {Bonnoli}, Giacomo and {Casanova}, V{\'\i}ctor and {Ag{\'\i}s-Gonz{\'a}lez}, Beatriz and {Escudero}, Juan and {Husillos}, C{\'e}sar and {Otero Santos}, Jorge and {Sota}, Alfredo and {Angelakis}, Emmanouil and {Kraus}, Alexander and {Keating}, Garrett K. and {Antonelli}, Lucio A. and {Bachetti}, Matteo and {Baldini}, Luca and {Baumgartner}, Wayne H. and {Bellazzini}, Ronaldo and {Bianchi}, Stefano and {Bongiorno}, Stephen D. and {Bonino}, Raffaella and {Brez}, Alessandro and {Bucciantini}, Niccol{\'o} and {Capitanio}, Fiamma and {Castellano}, Simone and {Cavazzuti}, Elisabetta and {Chen}, Chien-Ting and {Ciprini}, Stefano and {Costa}, Enrico and {De Rosa}, Alessandra and {Del Monte}, Ettore and {Di Gesu}, Laura and {Di Lalla}, Niccol{\'o} and {Di Marco}, Alessandro and {Donnarumma}, Immacolata and {Doroshenko}, Victor and {Dov{\v{c}}iak}, Michal and {Enoto}, Teruaki and {Evangelista}, Yuri and {Fabiani}, Sergio and {Ferrazzoli}, Riccardo and {Garcia}, Javier A. and {Gunji}, Shuichi and {Hayashida}, Kiyoshi and {Heyl}, Jeremy and {Iwakiri}, Wataru and {Jorstad}, Svetlana G. and {Kaaret}, Philip and {Karas}, Vladimir and {Kislat}, Fabian and {Kitaguchi}, Takao and {Kolodziejczak}, Jeffery J. and {Krawczynski}, Henric and {La Monaca}, Fabio and {Latronico}, Luca and {Maldera}, Simone and {Manfreda}, Alberto and {Marin}, Fr{\'e}d{\'e}ric and {Marinucci}, Andrea and {Marshall}, Herman L. and {Massaro}, Francesco and {Matt}, Giorgio and {Mitsuishi}, Ikuyuki and {Mizuno}, Tsunefumi and {Muleri}, Fabio and {Negro}, Michela and {Ng}, C. -Y. and {O'Dell}, Stephen L. and {Omodei}, Nicola and {Oppedisano}, Chiara and {Papitto}, Alessandro and {Pavlov}, George G. and {Peirson}, Abel L. and {Perri}, Matteo and {Pesce-Rollins}, Melissa and {Petrucci}, Pierre-Olivier and {Pilia}, Maura and {Possenti}, Andrea and {Poutanen}, Juri and {Puccetti}, Simonetta and {Ramsey}, Brian D. and {Rankin}, John and {Ratheesh}, Ajay and {Roberts}, Oliver J. and {Romani}, Roger W. and {Sgr{\'o}}, Carmelo and {Slane}, Patrick and {Soffitta}, Paolo and {Spandre}, Gloria and {Swartz}, Douglas A. and {Tamagawa}, Toru and {Taverna}, Roberto and {Tawara}, Yuzuru and {Tennant}, Allyn F. and {Thomas}, Nicholas E. and {Tombesi}, Francesco and {Trois}, Alessio and {Tsygankov}, Sergey S. and {Turolla}, Roberto and {Vink}, Jacco and {Weisskopf}, Martin C. and {Wu}, Kinwah and {Xie}, Fei and {Zane}, Silvia},
        title = "{X-Ray Polarization of the BL Lacertae Type Blazar 1ES 0229+200}",
      journal = {\apj},
         year = 2023,
        month = dec,
       volume = {959},
       number = {1},
          eid = {61},
        pages = {61},
          doi = {10.3847/1538-4357/ad05c4},
archivePrefix = {arXiv},
       eprint = {2310.01635},
 primaryClass = {astro-ph.HE},
       adsurl = {https://ui.adsabs.harvard.edu/abs/2023ApJ...959...61E}
}

@ARTICLE{emma2013,
       author = {{Emmanoulopoulos}, D. and {McHardy}, I.~M. and {Papadakis}, I.~E.},
        title = "{Generating artificial light curves: revisited and updated}",
      journal = {\mnras},
         year = 2013,
        month = aug,
       volume = {433},
       number = {2},
        pages = {907-927},
          doi = {10.1093/mnras/stt764},
archivePrefix = {arXiv},
       eprint = {1305.0304},
 primaryClass = {astro-ph.IM},
       adsurl = {https://ui.adsabs.harvard.edu/abs/2013MNRAS.433..907E}
}

@ARTICLE{emma2010,
       author = {{Emmanoulopoulos}, D. and {McHardy}, I.~M. and {Uttley}, P.},
        title = "{On the use of structure functions to study blazar variability: caveats and problems}",
      journal = {\mnras},
         year = 2010,
        month = may,
       volume = {404},
       number = {2},
        pages = {931-946},
          doi = {10.1111/j.1365-2966.2010.16328.x},
archivePrefix = {arXiv},
       eprint = {1001.2045},
 primaryClass = {astro-ph.CO},
       adsurl = {https://ui.adsabs.harvard.edu/abs/2010MNRAS.404..931E}
}

@ARTICLE{espaillat2008,
       author = {{Espaillat}, C. and {Bregman}, J. and {Hughes}, P. and {Lloyd-Davies}, E.},
        title = "{Wavelet Analysis of AGN X-Ray Time Series: A QPO in 3C 273?}",
      journal = {\apj},
         year = 2008,
        month = may,
       volume = {679},
       number = {1},
        pages = {182-193},
          doi = {10.1086/587023},
archivePrefix = {arXiv},
       eprint = {0805.4342},
 primaryClass = {astro-ph},
       adsurl = {https://ui.adsabs.harvard.edu/abs/2008ApJ...679..182E}
}

@ARTICLE{everett2020,
       author = {{Everett}, W.~B. and {Zhang}, L. and {Crawford}, T.~M. and {Vieira}, J.~D. and {Aravena}, M. and {Archipley}, M.~A. and {Austermann}, J.~E. and {Benson}, B.~A. and {Bleem}, L.~E. and {Carlstrom}, J.~E. and {Chang}, C.~L. and {Chapman}, S. and {Crites}, A.~T. and {de Haan}, T. and {Dobbs}, M.~A. and {George}, E.~M. and {Halverson}, N.~W. and {Harrington}, N. and {Holder}, G.~P. and {Holzapfel}, W.~L. and {Hrubes}, J.~D. and {Knox}, L. and {Lee}, A.~T. and {Luong-Van}, D. and {Mangian}, A.~C. and {Marrone}, D.~P. and {McMahon}, J.~J. and {Meyer}, S.~S. and {Mocanu}, L.~M. and {Mohr}, J.~J. and {Natoli}, T. and {Padin}, S. and {Pryke}, C. and {Reichardt}, C.~L. and {Reuter}, C.~A. and {Ruhl}, J.~E. and {Sayre}, J.~T. and {Schaffer}, K.~K. and {Shirokoff}, E. and {Spilker}, J.~S. and {Stalder}, B. and {Staniszewski}, Z. and {Stark}, A.~A. and {Story}, K.~T. and {Switzer}, E.~R. and {Vanderlinde}, K. and {Wei{\ss}}, A. and {Williamson}, R.},
        title = "{Millimeter-wave Point Sources from the 2500 Square Degree SPT-SZ Survey: Catalog and Population Statistics}",
      journal = {\apj},
         year = 2020,
        month = sep,
       volume = {900},
       number = {1},
          eid = {55},
        pages = {55},
          doi = {10.3847/1538-4357/ab9df7},
archivePrefix = {arXiv},
       eprint = {2003.03431},
 primaryClass = {astro-ph.IM},
       adsurl = {https://ui.adsabs.harvard.edu/abs/2020ApJ...900...55E}
}

@ARTICLE{feigelson1986,
       author = {{Feigelson}, E.~D. and {Bradt}, H. and {McClintock}, J. and {Remillard}, R. and {Urry}, C.~M. and {Tapia}, S. and {Geldzahler}, B. and {Johnston}, K. and {Romanishin}, W. and {Wehinger}, P.~A. and {Wyckoff}, S. and {Madejski}, G. and {Schwartz}, D.~A. and {Thorstensen}, J. and {Schaefer}, B.~E.},
        title = "{H0323+022: A New BL Lacertae Object with Extremely Rapid Variability}",
      journal = {\apj},
         year = 1986,
        month = mar,
       volume = {302},
        pages = {337},
          doi = {10.1086/163993},
       adsurl = {https://ui.adsabs.harvard.edu/abs/1986ApJ...302..337F}
}

@ARTICLE{foschini2022,
       author = {{Foschini}, Luigi and {Lister}, Matthew L. and {Andernach}, Heinz and {Ciroi}, Stefano and {Marziani}, Paola and {Ant{\'o}n}, Sonia and {Berton}, Marco and {Dalla Bont{\`a}}, Elena and {J{\"a}rvel{\"a}}, Emilia and {March{\~a}}, Maria J.~M. and {Romano}, Patrizia and {Tornikoski}, Merja and {Vercellone}, Stefano and {Vietri}, Amelia},
        title = "{A New Sample of Gamma-Ray Emitting Jetted Active Galactic Nuclei}",
      journal = {Universe},
         year = 2022,
        month = nov,
       volume = {8},
       number = {11},
          eid = {587},
        pages = {587},
          doi = {10.3390/universe8110587},
archivePrefix = {arXiv},
       eprint = {2211.03400},
 primaryClass = {astro-ph.HE},
       adsurl = {https://ui.adsabs.harvard.edu/abs/2022Univ....8..587F}
}

@ARTICLE{fossati1998,
       author = {{Fossati}, G. and {Maraschi}, L. and {Celotti}, A. and {Comastri}, A. and {Ghisellini}, G.},
        title = "{A unifying view of the spectral energy distributions of blazars}",
      journal = {\mnras},
         year = 1998,
        month = sep,
       volume = {299},
       number = {2},
        pages = {433-448},
          doi = {10.1046/j.1365-8711.1998.01828.x},
archivePrefix = {arXiv},
       eprint = {astro-ph/9804103},
 primaryClass = {astro-ph},
       adsurl = {https://ui.adsabs.harvard.edu/abs/1998MNRAS.299..433F}
}

@ARTICLE{franckowiak2020,
       author = {{Franckowiak}, A. and {Garrappa}, S. and {Paliya}, V. and {Shappee}, B. and {Stein}, R. and {Strotjohann}, N.~L. and {Kowalski}, M. and {Buson}, S. and {Kiehlmann}, S. and {Max-Moerbeck}, W. and {Angioni}, R.},
        title = "{Patterns in the Multiwavelength Behavior of Candidate Neutrino Blazars}",
      journal = {\apj},
         year = 2020,
        month = apr,
       volume = {893},
       number = {2},
          eid = {162},
        pages = {162},
          doi = {10.3847/1538-4357/ab8307},
archivePrefix = {arXiv},
       eprint = {2001.10232},
 primaryClass = {astro-ph.HE},
       adsurl = {https://ui.adsabs.harvard.edu/abs/2020ApJ...893..162F}
}

@ARTICLE{fromm2013,
       author = {{Fromm}, C.~M. and {Ros}, E. and {Perucho}, M. and {Savolainen}, T. and {Mimica}, P. and {Kadler}, M. and {Lobanov}, A.~P. and {Zensus}, J.~A.},
        title = "{Catching the radio flare in CTA 102. III. Core-shift and spectral analysis}",
      journal = {\aap},
         year = 2013,
        month = sep,
       volume = {557},
          eid = {A105},
        pages = {A105},
          doi = {10.1051/0004-6361/201321784},
archivePrefix = {arXiv},
       eprint = {1306.6208},
 primaryClass = {astro-ph.CO},
       adsurl = {https://ui.adsabs.harvard.edu/abs/2013A&A...557A.105F}
}

@ARTICLE{fuentes2023,
       author = {{Fuentes}, Antonio and {G{\'o}mez}, Jos{\'e} L. and {Mart{\'\i}}, Jos{\'e} M. and {Perucho}, Manel and {Zhao}, Guang-Yao and {Lico}, Rocco and {Lobanov}, Andrei P. and {Bruni}, Gabriele and {Kovalev}, Yuri Y. and {Chael}, Andrew and {Akiyama}, Kazunori and {Bouman}, Katherine L. and {Sun}, He and {Cho}, Ilje and {Traianou}, Efthalia and {Toscano}, Teresa and {Dahale}, Rohan and {Foschi}, Marianna and {Gurvits}, Leonid I. and {Jorstad}, Svetlana and {Kim}, Jae-Young and {Marscher}, Alan P. and {Mizuno}, Yosuke and {Ros}, Eduardo and {Savolainen}, Tuomas},
        title = "{Filamentary structures as the origin of blazar jet radio variability}",
      journal = {Nature Astronomy},
         year = 2023,
        month = nov,
       volume = {7},
        pages = {1359-1367},
          doi = {10.1038/s41550-023-02105-7},
archivePrefix = {arXiv},
       eprint = {2311.01861},
 primaryClass = {astro-ph.HE},
       adsurl = {https://ui.adsabs.harvard.edu/abs/2023NatAs...7.1359F}
}

@ARTICLE{gaidos1996,
       author = {{Gaidos}, J.~A. and {Akerlof}, C.~W. and {Biller}, S. and {Boyle}, P.~J. and {Breslin}, A.~C. and {Buckley}, J.~H. and {Carter-Lewis}, D.~A. and {Catanese}, M. and {Cawley}, M.~F. and {Fegan}, D.~J. and {Finley}, J.~P. and {Gordo}, J. Buss{\"o}ns and {Hillas}, A.~M. and {Krennrich}, F. and {Lamb}, R.~C. and {Lessard}, R.~W. and {McEnery}, J.~E. and {Masterson}, C. and {Mohanty}, G. and {Moriarty}, P. and {Quinn}, J. and {Rodgers}, A.~J. and {Rose}, H.~J. and {Samuelson}, F. and {Schubnell}, M.~S. and {Sembroski}, G.~H. and {Srinivasan}, R. and {Weekes}, T.~C. and {Wilson}, C.~L. and {Zweerink}, J.},
        title = "{Extremely rapid bursts of TeV photons from the active galaxy Markarian 421}",
      journal = {\nat},
         year = 1996,
        month = sep,
       volume = {383},
       number = {6598},
        pages = {319-320},
          doi = {10.1038/383319a0},
       adsurl = {https://ui.adsabs.harvard.edu/abs/1996Natur.383..319G}
}

@ARTICLE{gao2023,
       author = {{Gao}, Quan-Gui and {Lu}, Fang-Wu and {Qin}, Long-hua and {Gong}, Yun-Lu and {Yu}, Gong-ming and {Li}, Huai-zhen and {Yi}, Ting-feng},
        title = "{A Geometric Model to Interpret the {\ensuremath{\gamma}}-Ray Quasiperiodic Oscillation of PG 1553+113}",
      journal = {\apj},
         year = 2023,
        month = mar,
       volume = {945},
       number = {2},
          eid = {146},
        pages = {146},
          doi = {10.3847/1538-4357/acbe3e},
       adsurl = {https://ui.adsabs.harvard.edu/abs/2023ApJ...945..146G}
}

@ARTICLE{gear1986,
       author = {{Gear}, W.~K. and {Robson}, E.~I. and {Brown}, L.~M.~J.},
        title = "{Infrared variability of the BL lacertae object OJ287 since its outburst in 1983}",
      journal = {\nat},
         year = 1986,
        month = dec,
       volume = {324},
       number = {6097},
        pages = {546-547},
          doi = {10.1038/324546a0},
       adsurl = {https://ui.adsabs.harvard.edu/abs/1986Natur.324..546G}
}

@ARTICLE{ghisellini2017,
       author = {{Ghisellini}, G. and {Righi}, C. and {Costamante}, L. and {Tavecchio}, F.},
        title = "{The Fermi blazar sequence}",
      journal = {\mnras},
         year = 2017,
        month = jul,
       volume = {469},
       number = {1},
        pages = {255-266},
          doi = {10.1093/mnras/stx806},
archivePrefix = {arXiv},
       eprint = {1702.02571},
 primaryClass = {astro-ph.HE},
       adsurl = {https://ui.adsabs.harvard.edu/abs/2017MNRAS.469..255G}
}

@ARTICLE{ghisellini2011,
       author = {{Ghisellini}, G. and {Tavecchio}, F. and {Foschini}, L. and {Ghirlanda}, G.},
        title = "{The transition between BL Lac objects and flat spectrum radio quasars}",
      journal = {\mnras},
         year = 2011,
        month = jul,
       volume = {414},
       number = {3},
        pages = {2674-2689},
          doi = {10.1111/j.1365-2966.2011.18578.x},
archivePrefix = {arXiv},
       eprint = {1012.0308},
 primaryClass = {astro-ph.CO},
       adsurl = {https://ui.adsabs.harvard.edu/abs/2011MNRAS.414.2674G}
}

@ARTICLE{ghisellini2008,
       author = {{Ghisellini}, G. and {Tavecchio}, F.},
        title = "{The blazar sequence: a new perspective}",
      journal = {\mnras},
         year = 2008,
        month = jul,
       volume = {387},
       number = {4},
        pages = {1669-1680},
          doi = {10.1111/j.1365-2966.2008.13360.x},
archivePrefix = {arXiv},
       eprint = {0802.1918},
 primaryClass = {astro-ph},
       adsurl = {https://ui.adsabs.harvard.edu/abs/2008MNRAS.387.1669G}
}

@ARTICLE{ghisellini1996,
       author = {{Ghisellini}, Gabriele and {Madau}, Piero},
        title = "{On the origin of the gamma-ray emission in blazars}",
      journal = {\mnras},
         year = 1996,
        month = may,
       volume = {280},
       number = {1},
        pages = {67-76},
          doi = {10.1093/mnras/280.1.67},
       adsurl = {https://ui.adsabs.harvard.edu/abs/1996MNRAS.280...67G}
}

@ARTICLE{ghisellini1993,
       author = {{Ghisellini}, G. and {Padovani}, P. and {Celotti}, A. and {Maraschi}, L.},
        title = "{Relativistic Bulk Motion in Active Galactic Nuclei}",
      journal = {\apj},
         year = 1993,
        month = apr,
       volume = {407},
        pages = {65},
          doi = {10.1086/172493},
       adsurl = {https://ui.adsabs.harvard.edu/abs/1993ApJ...407...65G}
}

@ARTICLE{ghisellini1985,
       author = {{Ghisellini}, G. and {Maraschi}, L. and {Treves}, A.},
        title = "{Inhomogeneous synchrotron-self-compton models and the problem of relativistic beaming of BL Lac objects.}",
      journal = {\aap},
         year = 1985,
        month = may,
       volume = {146},
        pages = {204-212},
       adsurl = {https://ui.adsabs.harvard.edu/abs/1985A&A...146..204G}
}

@ARTICLE{giannios2013,
       author = {{Giannios}, Dimitrios},
        title = "{Reconnection-driven plasmoids in blazars: fast flares on a slow envelope}",
      journal = {\mnras},
         year = 2013,
        month = may,
       volume = {431},
       number = {1},
        pages = {355-363},
          doi = {10.1093/mnras/stt167},
archivePrefix = {arXiv},
       eprint = {1211.0296},
 primaryClass = {astro-ph.HE},
       adsurl = {https://ui.adsabs.harvard.edu/abs/2013MNRAS.431..355G}
}

@ARTICLE{giannios2009,
       author = {{Giannios}, Dimitrios and {Uzdensky}, Dmitri A. and {Begelman}, Mitchell C.},
        title = "{Fast TeV variability in blazars: jets in a jet}",
      journal = {\mnras},
         year = 2009,
        month = may,
       volume = {395},
       number = {1},
        pages = {L29-L33},
          doi = {10.1111/j.1745-3933.2009.00635.x},
archivePrefix = {arXiv},
       eprint = {0901.1877},
 primaryClass = {astro-ph.HE},
       adsurl = {https://ui.adsabs.harvard.edu/abs/2009MNRAS.395L..29G}
}

@ARTICLE{giebels2009,
       author = {{Giebels}, B. and {Degrange}, B.},
        title = "{Lognormal variability in BL Lacertae}",
      journal = {\aap},
         year = 2009,
        month = sep,
       volume = {503},
       number = {3},
        pages = {797-799},
          doi = {10.1051/0004-6361/200912303},
archivePrefix = {arXiv},
       eprint = {0907.2425},
 primaryClass = {astro-ph.CO},
       adsurl = {https://ui.adsabs.harvard.edu/abs/2009A&A...503..797G}
}

@ARTICLE{giommi2020,
       author = {{Giommi}, P. and {Glauch}, T. and {Padovani}, P. and {Resconi}, E. and {Turcati}, A. and {Chang}, Y.~L.},
        title = "{Dissecting the regions around IceCube high-energy neutrinos: growing evidence for the blazar connection}",
      journal = {\mnras},
         year = 2020,
        month = sep,
       volume = {497},
       number = {1},
        pages = {865-878},
          doi = {10.1093/mnras/staa2082},
archivePrefix = {arXiv},
       eprint = {2001.09355},
 primaryClass = {astro-ph.HE},
       adsurl = {https://ui.adsabs.harvard.edu/abs/2020MNRAS.497..865G}
}

@ARTICLE{giommi2012,
       author = {{Giommi}, P. and {Padovani}, P. and {Polenta}, G. and {Turriziani}, S. and {D'Elia}, V. and {Piranomonte}, S.},
        title = "{A simplified view of blazars: clearing the fog around long-standing selection effects}",
      journal = {\mnras},
         year = 2012,
        month = mar,
       volume = {420},
       number = {4},
        pages = {2899-2911},
          doi = {10.1111/j.1365-2966.2011.20044.x},
archivePrefix = {arXiv},
       eprint = {1110.4706},
 primaryClass = {astro-ph.CO},
       adsurl = {https://ui.adsabs.harvard.edu/abs/2012MNRAS.420.2899G}
}

@ARTICLE{gomez1999,
       author = {{G{\'o}mez}, Jos{\'e}-Luis and {Marscher}, Alan P. and {Alberdi}, Antonio and {Gabuzda}, Denise C.},
        title = "{The Twisted Parsec-Scale Structure of 0735+178}",
      journal = {\apj},
         year = 1999,
        month = jul,
       volume = {519},
       number = {2},
        pages = {642-646},
          doi = {10.1086/307410},
archivePrefix = {arXiv},
       eprint = {astro-ph/9901379},
 primaryClass = {astro-ph},
       adsurl = {https://ui.adsabs.harvard.edu/abs/1999ApJ...519..642G}
}

@ARTICLE{gopal-krishna2024,
       author = {{Gopal-Krishna}},
        title = "{Clues on the nature of the quasi-periodic optical outbursts of the blazar OJ 287}",
      journal = {\aap},
         year = 2024,
        month = aug,
       volume = {688},
          eid = {L16},
        pages = {L16},
          doi = {10.1051/0004-6361/202449409},
archivePrefix = {arXiv},
       eprint = {2407.09273},
 primaryClass = {astro-ph.HE},
       adsurl = {https://ui.adsabs.harvard.edu/abs/2024A&A...688L..16G}
}

@ARTICLE{gopal-krishna1992,
       author = {{Gopal-Krishna} and {Wiita}, Paul J.},
        title = "{Swinging jets and the variability of active nuclei.}",
      journal = {\aap},
         year = 1992,
        month = jun,
       volume = {259},
        pages = {109-117},
       adsurl = {https://ui.adsabs.harvard.edu/abs/1992A&A...259..109G}
}

@ARTICLE{gopal-krishna1991,
       author = {{Gopal-Krishna} and {Subramanian}, K.},
        title = "{Gravitational micro-lensing of the relativistic jets of quasars}",
      journal = {\nat},
         year = 1991,
        month = feb,
       volume = {349},
       number = {6312},
        pages = {766-768},
          doi = {10.1038/349766a0},
       adsurl = {https://ui.adsabs.harvard.edu/abs/1991Natur.349..766G}
}

@ARTICLE{graham2014,
       author = {{Graham}, Matthew J. and {Djorgovski}, S.~G. and {Drake}, Andrew J. and {Mahabal}, Ashish A. and {Chang}, Melissa and {Stern}, Daniel and {Donalek}, Ciro and {Glikman}, Eilat},
        title = "{A novel variability-based method for quasar selection: evidence for a rest-frame {\ensuremath{\sim}}54 d characteristic time-scale}",
      journal = {\mnras},
         year = 2014,
        month = mar,
       volume = {439},
       number = {1},
        pages = {703-718},
          doi = {10.1093/mnras/stt2499},
archivePrefix = {arXiv},
       eprint = {1401.1785},
 primaryClass = {astro-ph.CO},
       adsurl = {https://ui.adsabs.harvard.edu/abs/2014MNRAS.439..703G}
}

@INPROCEEDINGS{gurwell2007,
       author = {{Gurwell}, M.~A. and {Peck}, A.~B. and {Hostler}, S.~R. and {Darrah}, M.~R. and {Katz}, C.~A.},
        title = "{Monitoring Phase Calibrators at Submillimeter Wavelengths}",
    booktitle = {From Z-Machines to ALMA: (Sub)Millimeter Spectroscopy of Galaxies},
         year = 2007,
       editor = {{Baker}, A.~J. and {Glenn}, J. and {Harris}, A.~I. and {Mangum}, J.~G. and {Yun}, M.~S.},
       series = {Astronomical Society of the Pacific Conference Series},
       volume = {375},
        month = oct,
        pages = {234},
       adsurl = {https://ui.adsabs.harvard.edu/abs/2007ASPC..375..234G}
}

@ARTICLE{hagen2008,
       author = {{Hagen-Thorn}, V.~A. and {Larionov}, V.~M. and {Jorstad}, S.~G. and {Arkharov}, A.~A. and {Hagen-Thorn}, E.~I. and {Efimova}, N.~V. and {Larionova}, L.~V. and {Marscher}, A.~P.},
        title = "{The Outburst of the Blazar AO 0235+164 in 2006 December: Shock-in-Jet Interpretation}",
      journal = {\apj},
         year = 2008,
        month = jan,
       volume = {672},
       number = {1},
        pages = {40-47},
          doi = {10.1086/523841},
archivePrefix = {arXiv},
       eprint = {0709.3550},
 primaryClass = {astro-ph},
       adsurl = {https://ui.adsabs.harvard.edu/abs/2008ApJ...672...40H}
}

@ARTICLE{hagen2002,
       author = {{Hagen-Thorn}, V.~A. and {Larionova}, E.~G. and {Jorstad}, S.~G. and {Bj{\"o}rnsson}, C. -I. and {Larionov}, V.~M.},
        title = "{Analysis of the long-term polarization behaviour of BL Lac}",
      journal = {\aap},
         year = 2002,
        month = apr,
       volume = {385},
        pages = {55-61},
          doi = {10.1051/0004-6361:20020145},
       adsurl = {https://ui.adsabs.harvard.edu/abs/2002A&A...385...55H}
}

@ARTICLE{hallum2022,
       author = {{Hallum}, Melissa K. and {Jorstad}, Svetlana G. and {Larionov}, Valeri M. and {Marscher}, Alan P. and {Joshi}, Manasvita and {Weaver}, Zachary R. and {Williamson}, Karen E. and {Agudo}, Iv{\'a}n and {Borman}, George A. and {Casadio}, Carolina and {Fuentes}, Antonio and {Grishina}, Tatiana S. and {Kopatskaya}, Evgenia N. and {Larionova}, Elena G. and {Larionova}, Liyudmila V. and {Morozova}, Daria A. and {Nikiforova}, Anna A. and {Savchenko}, Sergey S. and {Troitsky}, Ivan S. and {Troitskaya}, Yulia V. and {Vasilyev}, Andrey A.},
        title = "{Emission-line Variability during a Nonthermal Outburst in the Gamma-Ray Bright Quasar 1156+295}",
      journal = {\apj},
         year = 2022,
        month = feb,
       volume = {926},
       number = {2},
          eid = {180},
        pages = {180},
          doi = {10.3847/1538-4357/ac4710},
archivePrefix = {arXiv},
       eprint = {2202.00061},
 primaryClass = {astro-ph.HE},
       adsurl = {https://ui.adsabs.harvard.edu/abs/2022ApJ...926..180H}
}

@ARTICLE{hardee1982b,
       author = {{Hardee}, P.~E.},
        title = "{The jet in M 87.}",
      journal = {\apj},
         year = 1982,
        month = oct,
       volume = {261},
        pages = {457-462},
          doi = {10.1086/160356},
       adsurl = {https://ui.adsabs.harvard.edu/abs/1982ApJ...261..457H}
}

@ARTICLE{hardee1982a,
       author = {{Hardee}, P.~E.},
        title = "{Helical and pinching instability of supersonic expanding jets in extragalactic radio sources}",
      journal = {\apj},
         year = 1982,
        month = jun,
       volume = {257},
        pages = {509-526},
          doi = {10.1086/160008},
       adsurl = {https://ui.adsabs.harvard.edu/abs/1982ApJ...257..509H}
}

@ARTICLE{hartman2001,
       author = {{Hartman}, R.~C. and {Villata}, M. and {Balonek}, T.~J. and {Bertsch}, D.~L. and {Bock}, H. and {B{\"o}ttcher}, M. and {Carini}, M.~T. and {Collmar}, W. and {De Francesco}, G. and {Ferrara}, E.~C. and {Heidt}, J. and {Kanbach}, G. and {Katajainen}, S. and {Koskimies}, M. and {Kurtanidze}, O.~M. and {Lanteri}, L. and {Lawson}, A. and {Lin}, Y.~C. and {Marscher}, A.~P. and {McFarland}, J.~P. and {McHardy}, I.~M. and {Miller}, H.~R. and {Nikolashvili}, M. and {Nilsson}, K. and {Noble}, J.~C. and {Nucciarelli}, G. and {Ostorero}, L. and {Pursimo}, T. and {Raiteri}, C.~M. and {Rekola}, R. and {Savolainen}, T. and {Sillanp{\"a}{\"a}}, A. and {Smale}, A. and {Sobrito}, G. and {Takalo}, L.~O. and {Thompson}, D.~J. and {Tosti}, G. and {Wagner}, S.~J. and {Wilson}, J.~W.},
        title = "{Day-Scale Variability of 3C 279 and Searches for Correlations in Gamma-Ray, X-Ray, and Optical Bands}",
      journal = {\apj},
         year = 2001,
        month = sep,
       volume = {558},
       number = {2},
        pages = {583-589},
          doi = {10.1086/322462},
archivePrefix = {arXiv},
       eprint = {astro-ph/0105247},
 primaryClass = {astro-ph},
       adsurl = {https://ui.adsabs.harvard.edu/abs/2001ApJ...558..583H}
}

@ARTICLE{hayashida2012,
       author = {{Hayashida}, M. and {Madejski}, G.~M. and {Nalewajko}, K. and {Sikora}, M. and {Wehrle}, A.~E. and {Ogle}, P. and {Collmar}, W. and {Larsson}, S. and {Fukazawa}, Y. and {Itoh}, R. and {Chiang}, J. and {Stawarz}, {\L}. and {Blandford}, R.~D. and {Richards}, J.~L. and {Max-Moerbeck}, W. and {Readhead}, A. and {Buehler}, R. and {Cavazzuti}, E. and {Ciprini}, S. and {Gehrels}, N. and {Reimer}, A. and {Szostek}, A. and {Tanaka}, T. and {Tosti}, G. and {Uchiyama}, Y. and {Kawabata}, K.~S. and {Kino}, M. and {Sakimoto}, K. and {Sasada}, M. and {Sato}, S. and {Uemura}, M. and {Yamanaka}, M. and {Greiner}, J. and {Kruehler}, T. and {Rossi}, A. and {Macquart}, J.~P. and {Bock}, D.~C. -J. and {Villata}, M. and {Raiteri}, C.~M. and {Agudo}, I. and {Aller}, H.~D. and {Aller}, M.~F. and {Arkharov}, A.~A. and {Bach}, U. and {Ben{\'\i}tez}, E. and {Berdyugin}, A. and {Blinov}, D.~A. and {Blumenthal}, K. and {B{\"o}ttcher}, M. and {Buemi}, C.~S. and {Carosati}, D. and {Chen}, W.~P. and {Di Paola}, A. and {Dolci}, M. and {Efimova}, N.~V. and {Forn{\'e}}, E. and {G{\'o}mez}, J.~L. and {Gurwell}, M.~A. and {Heidt}, J. and {Hiriart}, D. and {Jordan}, B. and {Jorstad}, S.~G. and {Joshi}, M. and {Kimeridze}, G. and {Konstantinova}, T.~S. and {Kopatskaya}, E.~N. and {Koptelova}, E. and {Kurtanidze}, O.~M. and {L{\"a}hteenm{\"a}ki}, A. and {Lamerato}, A. and {Larionov}, V.~M. and {Larionova}, E.~G. and {Larionova}, L.~V. and {Leto}, P. and {Lindfors}, E. and {Marscher}, A.~P. and {McHardy}, I.~M. and {Molina}, S.~N. and {Morozova}, D.~A. and {Nikolashvili}, M.~G. and {Nilsson}, K. and {Reinthal}, R. and {Roustazadeh}, P. and {Sakamoto}, T. and {Sigua}, L.~A. and {Sillanp{\"a}{\"a}}, A. and {Takalo}, L. and {Tammi}, J. and {Taylor}, B. and {Tornikoski}, M. and {Trigilio}, C. and {Troitsky}, I.~S. and {Umana}, G.},
        title = "{The Structure and Emission Model of the Relativistic Jet in the Quasar 3C 279 Inferred from Radio to High-energy {\ensuremath{\gamma}}-Ray Observations in 2008-2010}",
      journal = {\apj},
         year = 2012,
        month = aug,
       volume = {754},
       number = {2},
          eid = {114},
        pages = {114},
          doi = {10.1088/0004-637X/754/2/114},
archivePrefix = {arXiv},
       eprint = {1206.0745},
 primaryClass = {astro-ph.HE},
       adsurl = {https://ui.adsabs.harvard.edu/abs/2012ApJ...754..114H}
}

@ARTICLE{healey2007,
       author = {{Healey}, Stephen E. and {Romani}, Roger W. and {Taylor}, Gregory B. and {Sadler}, Elaine M. and {Ricci}, Roberto and {Murphy}, Tara and {Ulvestad}, James S. and {Winn}, Joshua N.},
        title = "{CRATES: An All-Sky Survey of Flat-Spectrum Radio Sources}",
      journal = {\apjs},
         year = 2007,
        month = jul,
       volume = {171},
       number = {1},
        pages = {61-71},
          doi = {10.1086/513742},
archivePrefix = {arXiv},
       eprint = {astro-ph/0702346},
 primaryClass = {astro-ph},
       adsurl = {https://ui.adsabs.harvard.edu/abs/2007ApJS..171...61H}
}

@ARTICLE{heidt1996,
       author = {{Heidt}, J. and {Wagner}, S.~J.},
        title = "{Statistics of optical intraday variability in a complete sample of radio-selected BL Lacertae objects.}",
      journal = {\aap},
         year = 1996,
        month = jan,
       volume = {305},
        pages = {42},
          doi = {10.48550/arXiv.astro-ph/9506032},
archivePrefix = {arXiv},
       eprint = {astro-ph/9506032},
 primaryClass = {astro-ph},
       adsurl = {https://ui.adsabs.harvard.edu/abs/1996A&A...305...42H}
}

@ARTICLE{homan2021,
       author = {{Homan}, D.~C. and {Cohen}, M.~H. and {Hovatta}, T. and {Kellermann}, K.~I. and {Kovalev}, Y.~Y. and {Lister}, M.~L. and {Popkov}, A.~V. and {Pushkarev}, A.~B. and {Ros}, E. and {Savolainen}, T.},
        title = "{MOJAVE. XIX. Brightness Temperatures and Intrinsic Properties of Blazar Jets}",
      journal = {\apj},
         year = 2021,
        month = dec,
       volume = {923},
       number = {1},
          eid = {67},
        pages = {67},
          doi = {10.3847/1538-4357/ac27af},
archivePrefix = {arXiv},
       eprint = {2109.04977},
 primaryClass = {astro-ph.HE},
       adsurl = {https://ui.adsabs.harvard.edu/abs/2021ApJ...923...67H}
}

@ARTICLE{hovatta2009,
       author = {{Hovatta}, T. and {Valtaoja}, E. and {Tornikoski}, M. and {L{\"a}hteenm{\"a}ki}, A.},
        title = "{Doppler factors, Lorentz factors and viewing angles for quasars, BL Lacertae objects and radio galaxies}",
      journal = {\aap},
         year = 2009,
        month = feb,
       volume = {494},
       number = {2},
        pages = {527-537},
          doi = {10.1051/0004-6361:200811150},
archivePrefix = {arXiv},
       eprint = {0811.4278},
 primaryClass = {astro-ph},
       adsurl = {https://ui.adsabs.harvard.edu/abs/2009A&A...494..527H}
}

@ARTICLE{hovatta2014,
       author = {{Hovatta}, T. and {Pavlidou}, V. and {King}, O.~G. and {Mahabal}, A. and {Sesar}, B. and {Dancikova}, R. and {Djorgovski}, S.~G. and {Drake}, A. and {Laher}, R. and {Levitan}, D. and {Max-Moerbeck}, W. and {Ofek}, E.~O. and {Pearson}, T.~J. and {Prince}, T.~A. and {Readhead}, A.~C.~S. and {Richards}, J.~L. and {Surace}, J.},
        title = "{Connection between optical and {\ensuremath{\gamma}}-ray variability in blazars}",
      journal = {\mnras},
         year = 2014,
        month = mar,
       volume = {439},
       number = {1},
        pages = {690-702},
          doi = {10.1093/mnras/stt2494},
archivePrefix = {arXiv},
       eprint = {1401.0538},
 primaryClass = {astro-ph.HE},
       adsurl = {https://ui.adsabs.harvard.edu/abs/2014MNRAS.439..690H}
}

@ARTICLE{hufnagel1992,
       author = {{Hufnagel}, Beth R. and {Bregman}, Joel N.},
        title = "{Optical and Radio Variability in Blazars}",
      journal = {\apj},
         year = 1992,
        month = feb,
       volume = {386},
        pages = {473},
          doi = {10.1086/171033},
       adsurl = {https://ui.adsabs.harvard.edu/abs/1992ApJ...386..473H}
}

@ARTICLE{hughes2011,
       author = {{Hughes}, Philip A. and {Aller}, Margo F. and {Aller}, Hugh D.},
        title = "{Oblique Shocks as the Origin of Radio to Gamma-ray Variability in Active Galactic Nuclei}",
      journal = {\apj},
         year = 2011,
        month = jul,
       volume = {735},
       number = {2},
          eid = {81},
        pages = {81},
          doi = {10.1088/0004-637X/735/2/81},
archivePrefix = {arXiv},
       eprint = {1104.4256},
 primaryClass = {astro-ph.HE},
       adsurl = {https://ui.adsabs.harvard.edu/abs/2011ApJ...735...81H}
}

@ARTICLE{hughes2002,
       author = {{Hughes}, Philip A. and {Miller}, Mark A. and {Duncan}, G. Comer},
        title = "{Three-dimensional Hydrodynamic Simulations of Relativistic Extragalactic Jets}",
      journal = {\apj},
         year = 2002,
        month = jun,
       volume = {572},
       number = {2},
        pages = {713-728},
          doi = {10.1086/340382},
archivePrefix = {arXiv},
       eprint = {astro-ph/0202402},
 primaryClass = {astro-ph},
       adsurl = {https://ui.adsabs.harvard.edu/abs/2002ApJ...572..713H}
}

@ARTICLE{hughes1992,
       author = {{Hughes}, P.~A. and {Aller}, H.~D. and {Aller}, M.~F.},
        title = "{The University of Michigan Radio Astronomy Data Base. I. Structure Function Analysis and the Relation between BL Lacertae Objects and Quasi-stellar Objects}",
      journal = {\apj},
         year = 1992,
        month = sep,
       volume = {396},
        pages = {469},
          doi = {10.1086/171734},
       adsurl = {https://ui.adsabs.harvard.edu/abs/1992ApJ...396..469H}
}

@ARTICLE{hughes1985,
       author = {{Hughes}, P.~A. and {Aller}, H.~D. and {Aller}, M.~F.},
        title = "{Polarized radio outbursts in BL Lacertae. II. The flux and polarization of a piston-driven shock.}",
      journal = {\apj},
         year = 1985,
        month = nov,
       volume = {298},
        pages = {301-315},
          doi = {10.1086/163611},
       adsurl = {https://ui.adsabs.harvard.edu/abs/1985ApJ...298..301H}
}

@ARTICLE{impey1988b,
       author = {{Impey}, C.~D. and {Tapia}, S.},
        title = "{New Blazars Discovered by Polarimetry}",
      journal = {\apj},
         year = 1988,
        month = oct,
       volume = {333},
        pages = {666},
          doi = {10.1086/166775},
       adsurl = {https://ui.adsabs.harvard.edu/abs/1988ApJ...333..666I}
}

@ARTICLE{isler2013,
       author = {{Isler}, Jedidah C. and {Urry}, C.~M. and {Coppi}, P. and {Bailyn}, C. and {Chatterjee}, R. and {Fossati}, G. and {Bonning}, E.~W. and {Maraschi}, L. and {Buxton}, M.},
        title = "{A Time-resolved Study of the Broad-line Region in Blazar 3C 454.3}",
      journal = {\apj},
         year = 2013,
        month = dec,
       volume = {779},
       number = {2},
          eid = {100},
        pages = {100},
          doi = {10.1088/0004-637X/779/2/100},
archivePrefix = {arXiv},
       eprint = {1310.0817},
 primaryClass = {astro-ph.CO},
       adsurl = {https://ui.adsabs.harvard.edu/abs/2013ApJ...779..100I}
}

@ARTICLE{itoh2020,
       author = {{Itoh}, Ryosuke and {Utsumi}, Yousuke and {Inoue}, Yoshiyuki and {Ohta}, Kouji and {Doi}, Akihiro and {Morokuma}, Tomoki and {Kawabata}, Koji S. and {Tanaka}, Yasuyuki T.},
        title = "{Blazar Radio and Optical Survey (BROS): A Catalog of Blazar Candidates Showing Flat Radio Spectrum and Their Optical Identification in Pan-STARRS1 Surveys}",
      journal = {\apj},
         year = 2020,
        month = sep,
       volume = {901},
       number = {1},
          eid = {3},
        pages = {3},
          doi = {10.3847/1538-4357/abab07},
archivePrefix = {arXiv},
       eprint = {2008.00038},
 primaryClass = {astro-ph.HE},
       adsurl = {https://ui.adsabs.harvard.edu/abs/2020ApJ...901....3I}
}

@ARTICLE{jannuzi1994,
       author = {{Jannuzi}, Buell T. and {Smith}, Paul S. and {Elston}, Richard},
        title = "{The Optical Polarization Properties of X-Ray--selected BL Lacertae Objects}",
      journal = {\apj},
         year = 1994,
        month = jun,
       volume = {428},
        pages = {130},
          doi = {10.1086/174226},
       adsurl = {https://ui.adsabs.harvard.edu/abs/1994ApJ...428..130J}
}

@ARTICLE{jorstad2022,
       author = {{Jorstad}, S.~G. and {Marscher}, A.~P. and {Raiteri}, C.~M. and {Villata}, M. and {Weaver}, Z.~R. and {Zhang}, H. and {Dong}, L. and {G{\'o}mez}, J.~L. and {Perel}, M.~V. and {Savchenko}, S.~S. and {Larionov}, V.~M. and {Carosati}, D. and {Chen}, W.~P. and {Kurtanidze}, O.~M. and {Marchini}, A. and {Matsumoto}, K. and {Mortari}, F. and {Aceti}, P. and {Acosta-Pulido}, J.~A. and {Andreeva}, T. and {Apolonio}, G. and {Arena}, C. and {Arkharov}, A. and {Bachev}, R. and {Banfi}, M. and {Bonnoli}, G. and {Borman}, G.~A. and {Bozhilov}, V. and {Carnerero}, M.~I. and {Damljanovic}, G. and {Ehgamberdiev}, S.~A. and {Els{\"a}sser}, D. and {Frasca}, A. and {Gabellini}, D. and {Grishina}, T.~S. and {Gupta}, A.~C. and {Hagen-Thorn}, V.~A. and {Hallum}, M.~K. and {Hart}, M. and {Hasuda}, K. and {Hemrich}, F. and {Hsiao}, H.~Y. and {Ibryamov}, S. and {Irsmambetova}, T.~R. and {Ivanov}, D.~V. and {Joner}, M.~D. and {Kimeridze}, G.~N. and {Klimanov}, S.~A. and {Kn{\"o}tt}, J. and {Kopatskaya}, E.~N. and {Kurtanidze}, S.~O. and {Kurtenkov}, A. and {Kuutma}, T. and {Larionova}, E.~G. and {Leonini}, S. and {Lin}, H.~C. and {Lorey}, C. and {Mannheim}, K. and {Marino}, G. and {Minev}, M. and {Mirzaqulov}, D.~O. and {Morozova}, D.~A. and {Nikiforova}, A.~A. and {Nikolashvili}, M.~G. and {Ovcharov}, E. and {Papini}, R. and {Pursimo}, T. and {Rahimov}, I. and {Reinhart}, D. and {Sakamoto}, T. and {Salvaggio}, F. and {Semkov}, E. and {Shakhovskoy}, D.~N. and {Sigua}, L.~A. and {Steineke}, R. and {Stojanovic}, M. and {Strigachev}, A. and {Troitskaya}, Y.~V. and {Troitskiy}, I.~S. and {Tsai}, A. and {Valcheva}, A. and {Vasilyev}, A.~A. and {Vince}, O. and {Waller}, L. and {Zaharieva}, E. and {Chatterjee}, R.},
        title = "{Rapid quasi-periodic oscillations in the relativistic jet of BL Lacertae}",
      journal = {\nat},
         year = 2022,
        month = sep,
       volume = {609},
       number = {7926},
        pages = {265-268},
          doi = {10.1038/s41586-022-05038-9},
       adsurl = {https://ui.adsabs.harvard.edu/abs/2022Natur.609..265J}
}

@ARTICLE{jorstad2017,
       author = {{Jorstad}, Svetlana G. and {Marscher}, Alan P. and {Morozova}, Daria A. and {Troitsky}, Ivan S. and {Agudo}, Iv{\'a}n and {Casadio}, Carolina and {Foord}, Adi and {G{\'o}mez}, Jos{\'e} L. and {MacDonald}, Nicholas R. and {Molina}, Sol N. and {L{\"a}hteenm{\"a}ki}, Anne and {Tammi}, Joni and {Tornikoski}, Merja},
        title = "{Kinematics of Parsec-scale Jets of Gamma-Ray Blazars at 43 GHz within the VLBA-BU-BLAZAR Program}",
      journal = {\apj},
         year = 2017,
        month = sep,
       volume = {846},
       number = {2},
          eid = {98},
        pages = {98},
          doi = {10.3847/1538-4357/aa8407},
archivePrefix = {arXiv},
       eprint = {1711.03983},
 primaryClass = {astro-ph.GA},
       adsurl = {https://ui.adsabs.harvard.edu/abs/2017ApJ...846...98J}
}

@ARTICLE{jorstad2013,
       author = {{Jorstad}, Svetlana G. and {Marscher}, Alan P. and {Smith}, Paul S. and {Larionov}, Valeri M. and {Agudo}, Iv{\'a}n and {Gurwell}, Mark and {Wehrle}, Ann E. and {L{\"a}hteenm{\"a}ki}, Anne and {Nikolashvili}, Maria G. and {Schmidt}, Gary D. and {Arkharov}, Arkady A. and {Blinov}, Dmitry A. and {Blumenthal}, Kelly and {Casadio}, Carolina and {Chigladze}, Revaz A. and {Efimova}, Natalia V. and {Eggen}, Joseph R. and {G{\'o}mez}, Jos{\'e} L. and {Grupe}, Dirk and {Hagen-Thorn}, Vladimir A. and {Joshi}, Manasvita and {Kimeridze}, Givi N. and {Konstantinova}, Tatiana S. and {Kopatskaya}, Evgenia N. and {Kurtanidze}, Omar M. and {Kurtanidze}, Sofia O. and {Larionova}, Elena G. and {Larionova}, Liudmilla V. and {Sigua}, Lorand A. and {MacDonald}, Nicholas R. and {Maune}, Jeremy D. and {McHardy}, Ian M. and {Miller}, H. Richard and {Molina}, Sol N. and {Morozova}, Daria A. and {Scott}, Terri and {Taylor}, Brian W. and {Tornikoski}, Merja and {Troitsky}, Ivan S. and {Thum}, Clemens and {Walker}, Gary and {Williamson}, Karen E. and {Sallum}, Stephanie and {Consiglio}, Santina and {Strelnitski}, Vladimir},
        title = "{A Tight Connection between Gamma-Ray Outbursts and Parsec-scale Jet Activity in the Quasar 3C 454.3}",
      journal = {\apj},
         year = 2013,
        month = aug,
       volume = {773},
       number = {2},
          eid = {147},
        pages = {147},
          doi = {10.1088/0004-637X/773/2/147},
archivePrefix = {arXiv},
       eprint = {1307.2522},
 primaryClass = {astro-ph.HE},
       adsurl = {https://ui.adsabs.harvard.edu/abs/2013ApJ...773..147J}
}

@ARTICLE{kaastra1992,
       author = {{Kaastra}, J.~S. and {Roos}, N.},
        title = "{Massive binary black holes and wiggling jets.}",
      journal = {\aap},
         year = 1992,
        month = feb,
       volume = {254},
        pages = {96},
       adsurl = {https://ui.adsabs.harvard.edu/abs/1992A&A...254...96K}
}

@ARTICLE{kaspi2000,
       author = {{Kaspi}, Shai and {Smith}, Paul S. and {Netzer}, Hagai and {Maoz}, Dan and {Jannuzi}, Buell T. and {Giveon}, Uriel},
        title = "{Reverberation Measurements for 17 Quasars and the Size-Mass-Luminosity Relations in Active Galactic Nuclei}",
      journal = {\apj},
         year = 2000,
        month = apr,
       volume = {533},
       number = {2},
        pages = {631-649},
          doi = {10.1086/308704},
archivePrefix = {arXiv},
       eprint = {astro-ph/9911476},
 primaryClass = {astro-ph},
       adsurl = {https://ui.adsabs.harvard.edu/abs/2000ApJ...533..631K}
}

@ARTICLE{kellermann2007,
       author = {{Kellermann}, K.~I. and {Kovalev}, Y.~Y. and {Lister}, M.~L. and {Homan}, D.~C. and {Kadler}, M. and {Cohen}, M.~H. and {Ros}, E. and {Zensus}, J.~A. and {Vermeulen}, R.~C. and {Aller}, M.~F. and {Aller}, H.~D.},
        title = "{Doppler boosting, superluminal motion, and the kinematics of AGN jets}",
      journal = {\apss},
         year = 2007,
        month = oct,
       volume = {311},
       number = {1-3},
        pages = {231-239},
          doi = {10.1007/s10509-007-9622-5},
archivePrefix = {arXiv},
       eprint = {0708.3219},
 primaryClass = {astro-ph},
       adsurl = {https://ui.adsabs.harvard.edu/abs/2007Ap&SS.311..231K}
}

@ARTICLE{kellermann1969,
       author = {{Kellermann}, K.~I. and {Pauliny-Toth}, I.~I.~K.},
        title = "{The Spectra of Opaque Radio Sources}",
      journal = {\apjl},
         year = 1969,
        month = feb,
       volume = {155},
        pages = {L71},
          doi = {10.1086/180305},
       adsurl = {https://ui.adsabs.harvard.edu/abs/1969ApJ...155L..71K}
}

@ARTICLE{kiehlmann2017,
       author = {{Kiehlmann}, S. and {Blinov}, D. and {Pearson}, T.~J. and {Liodakis}, I.},
        title = "{Optical EVPA rotations in blazars: testing a stochastic variability model with RoboPol data}",
      journal = {\mnras},
         year = 2017,
        month = dec,
       volume = {472},
       number = {3},
        pages = {3589-3604},
          doi = {10.1093/mnras/stx2167},
archivePrefix = {arXiv},
       eprint = {1708.06777},
 primaryClass = {astro-ph.HE},
       adsurl = {https://ui.adsabs.harvard.edu/abs/2017MNRAS.472.3589K}
}

\end{document}